\documentclass[sw]{iosart2c}
\usepackage[titletoc,title]{appendix}
\usepackage{pifont}
\usepackage{marvosym}
\usepackage{times}%
\usepackage[T1]{fontenc}
\usepackage[table]{xcolor}
\usepackage[caption=false]{subfig}
\usepackage{color}
\usepackage{multirow}
\usepackage{graphicx}
\usepackage[linesnumbered, vlined, ruled,algo2e]{algorithm2e} 
\usepackage{booktabs}
\usepackage{amssymb}
\usepackage{amsfonts}
\usepackage{amsmath}
\usepackage{mathtools}
\usepackage{amsthm}
\usepackage{soul}
\usepackage{etex}
\usepackage{relsize}
\usepackage{epstopdf}
\usepackage{multirow}
\usepackage{url}
\usepackage{tablefootnote}
 \usepackage{hyperref}  
\hypersetup{ 
    colorlinks,%
    citecolor=black,%
    filecolor=black,%
    linkcolor=black,%
    urlcolor=black,%
pdftitle={A Hierarchical Aggregation Framework for Efficient Multilevel Visual Exploration and Analytics over Large Datasets},
pdfauthor={Nikos Bikakis, George Papastefanatos, Melina Skourla, Timos Sellis},
pdfsubject={A Hierarchical Aggregation Framework for Efficient Multilevel Visual Exploration and Analytics over Large Datasets},
pdfproducer={pdfbikTeX-1.1},
pdfcreator={bikTeX},
pdfkeywords={visual analytics, big data, progressive visualization, incremental visualization, multiresolution visualization, multiscale visualization, multilevel visualization, hierarchical visualization, incremental indexing, adaptive visualization, data summarization, data abstraction, adaptive indexing, binning, binning aggregation, data reduction, visualization recommendations, tree reconstruction, dynamic tree, Steerable Exploration, visualization preferences, scalable visualizations, visual aggregation, visualizing very large datasets, personalized exploration, semantic web, hierarchical navigation, personalized navigation, pay as you go visualization, RDF visualization, linked data visualization, web of data, SPARQL, hierarchical aggregation, OWL, scalable visual analytics, incremental visual exploration, hierarchical exploration, Steerable visualization, scalable exploration, multiresolution exploration, multiscale exploration, multilevel exploration, incremental exploration, personalized visual exploration, multiresolution navigation, multilevel navigation, incremental navigation, multiscale navigation, personalized visualization, visualization tree, personalized visual analytics, exploration user preferences, multiresolution visual analytics, hierarchical charts, RDF Facets, HETree, Faceted search, multilevel temporal visualization, LOD, rdf:SynopsViz, hierarchical timeline, linked open data, multilevel timeline, hierarchical histogram, tree visualization structure, numeric exploration, LD, multilevel charts, RDF Charts, Data exploration, RDF Statistics, temporal exploration, synopsviz, scalable navigation, personalization, large data visualization, multilevel numeric visualization, multi-scale, multi-level, pay-as-you-go visualization, multi-resolution}
}

\newcommand{\ico}{{ICO}\xspace}
\newcommand{\ada}{{ADA}\xspace}
\newcommand{\hetR}{\mbox{HETree-R}\xspace}
\newcommand{\hetC}{\mbox{HETree-C}\xspace}
\newcommand{\het}{\mbox{HETree}\xspace}

\newcommand{\app}[1]{
}

\newcommand{\eat}[1]{}

\newcommand{\stitle}[1]{\vspace{0.2cm}\noindent\textbf{#1}}
\newcommand{\sstitle}[1]{\vspace{0.2cm}\noindent\textit{#1}}

\definecolor{dark-gray}{gray}{0.2}

\DeclarePairedDelimiter\ceil{\lceil}{\rceil}
\DeclarePairedDelimiter\floor{\lfloor}{\rfloor}

\newcommand{\eoe} {}

\renewcommand{\H}{\ensuremath\mathcal{H}}
\newcommand{\T}{\ensuremath\mathcal{T}}

\newcommand{\subscript}[1]{\ensuremath{_{\textrm{#1}}}}

\newcommand{\myFontC}[1]{{\fontfamily{pcr}\selectfont #1}} 
\newcommand{\myFontP}[1]{{\fontfamily{ppl}\selectfont #1}} 

\newcommand*\yes{\ding{51}}
\newcommand*\no{}
\newcommand{\mystar}{\scalebox{0.8}{\tiny$\bigstar$}}
\newcommand{\mydiamond}{ \scalebox{0.8}{\tiny\ding{117}}}
 
\newcommand{\specialcell}[2][c]{%
  \begin{tabular}[#1]{@{}c@{}}#2\end{tabular}}

\newcommand{\tline}  {\specialrule{0.8 pt}{0pt}{1pt}}		 
\newcommand{\bline}  {\specialrule{0.8 pt}{1pt}{0pt}}		
\newcommand{\dlineB}  {\specialrule{0.8 pt}{1pt}{2pt} \specialrule{0.8 pt}{0pt}{0pt}}		

\newcommand{\boldline}  {\specialrule{0.8 pt}{0pt}{0pt}}		

\newcolumntype{'}{!{\vrule width 0.8pt}}

\newcounter{remarkcnt}
\newenvironment{myremark}
{\refstepcounter{remarkcnt}\smallskip\setlength{\leftskip}{5pt}\setlength{\rightskip}{0pt}\par\noindent\ignorespaces 
   \textbf{Remark~\theremarkcnt.}}
{\smallskip\par}

\newcounter{lemmcnt}

\newcounter{theorcnt}
\newenvironment{mytheor}
{\refstepcounter{theorcnt}\smallskip\setlength{\leftskip}{5pt}\setlength{\rightskip}{5pt}\par\noindent\ignorespaces 
   \textbf{Theorem~\thetheorcnt.}}
{\smallskip\par}

\newcounter{proposcnt}
\newenvironment{myproposition}
{\refstepcounter{proposcnt}\smallskip\setlength{\leftskip}{5pt}\setlength{\rightskip}{5pt}\par\noindent\ignorespaces 
   \textbf{Proposition~\theproposcnt.}}
{\smallskip\par}

\newcounter{ex}
\newenvironment{myExample}
{\refstepcounter{ex}\smallskip\setlength{\leftskip}{8pt}\setlength{\rightskip}{0pt}\par\noindent\ignorespaces 
   \textbf{Example~\theex.}}
{\medskip\par}

\SetKw{Logand}{and}	
\SetKw{Logor}{or}
\SetKw{Lognot}{not}	

\SetKw{mybreak}{break}	
\SetKw{mycontinue}{continue}	

\SetKwInput{KwVar}{Variables}
\SetKwInput{KwPar}{Parameters}

\SetKwData{myEOF}{EOF}	
\SetKwData{myfalse}{false}
\SetKwData{mytrue}{true}
\SetKwData{mynull}{null}
\SetKwData{myempty}{empty}
\SetKwBlock{myblock}{beginnikos}{end}

 \SetAlgoInsideSkip{} 
\SetAlgoSkip{} 
\SetAlgoCaptionSeparator{.} 

\newcommand{\mycomment} [1] 	{\tiny{\textcolor{dark-gray}{\myFontP{{#1}}}}} 
\SetKwComment{Comment}{\mycomment{/\!\!/}}{}
 \DontPrintSemicolon
\SetAlgoCaptionLayout{small}
\SetProcNameSty{small}
\SetProcArgSty{small}

\pubyear{2015}

\begin{document}

\begin{frontmatter}



\title{A Hierarchical Aggregation Framework for Efficient Multilevel Visual Exploration and Analysis\thanks{To appear in Semantic Web Journal (SWJ), 2016}
}

\runningtitle{A Hierarchical Aggregation Framework for Efficient Multilevel Visual Exploration and Analysis } 


\author[A,B]{\fnms{Nikos} \snm{Bikakis}},
\author[B]{\fnms{George} \snm{Papastefanatos}},
\author[A]{\fnms{Melina} \snm{Skourla}} and
\author[C]{\fnms{Timos} \snm{Sellis}}
\runningauthor{N. Bikakis et al.}
\address[A]{National Technical University of Athens, Greece}
\address[B]{ATHENA Research Center, Greece}
\address[C]{Swinburne University of Technology, Australia}

\begin{abstract}
Data exploration and visualization
systems are of great importance in the Big Data era, in
which the volume and heterogeneity of available information make
it difficult for humans to manually explore and analyse data.
Most traditional systems operate in an 
offline way, limited to accessing  preprocessed (static) sets of data. 
They also restrict themselves to dealing with small dataset sizes,
which can be easily handled with conventional techniques. 
However, the {Big Data era} has realized the availability of a great amount
and variety of big   datasets that are dynamic in nature;
most of them offer API or query endpoints for online access, 
or the data is received in a stream fashion.
Therefore, modern systems must address the challenge of on-the-fly scalable visualizations
over large dynamic sets of data,  offering efficient exploration techniques, 
as well as mechanisms for information abstraction and summarization.
Further, they must take into account different user-defined exploration scenarios and user preferences. 
%
In this work,  we present a generic model for
personalized multilevel exploration and analysis 
over large dynamic sets of numeric and temporal data. 
Our model is built on top of a     lightweight tree-based structure
which can be  efficiently constructed on-the-fly for a  given set of data.
This tree structure aggregates input objects into a hierarchical multiscale model. 
We define two versions of this structure, that adopt different
data organization approaches, well-suited to exploration and analysis context. 
In the proposed  structure, statistical computations can be efficiently performed on-the-fly.
Considering different exploration scenarios over large datasets, 
the proposed model enables efficient multilevel exploration, 
offering incremental construction and prefetching via user interaction, and dynamic adaptation of the hierarchies based on user preferences. 
 {A thorough theoretical  analysis is presented, illustrating the efficiency of the proposed methods.} 
The presented model is realized  in a web-based prototype tool,
called  {SynopsViz} that offers multilevel visual exploration and analysis over Linked Data datasets. 
Finally, we provide a  performance evaluation and a empirical user study employing  real datasets.\\

\end{abstract}

\end{frontmatter}





\section{Introduction}
\label{sec:intro}

Exploring, visualizing and analysing data is a core
task for data scientists and analysts in numerous applications.
Data exploration and visualization
enable users to identify interesting patterns, infer correlations and causalities, and support sense-making  activities 
over data that are not always possible with traditional data mining techniques  \cite{IdreosPC15,DadzieLP09}.  
This is of great importance in the  {Big Data era}, 
where the volume and heterogeneity of available information make it difficult for humans 
to manually explore and analyse large datasets.




One of the major challenges in visual exploration is
related to the  \textit{large size} that characterizes many datasets nowadays. 
Considering the visual information seeking mantra: ``\textit{overview first, zoom and filter, then details on demand}'' \cite{Shneiderman96}, gaining overview is a crucial task in the visual exploration scenario. 
However, offering an overview of  a large dataset is an extremely challenging task.
\textit{Information overloading} is a common issue in  large dataset visualization;
a basic requirement for the proposed approaches is to  offer mechanisms for information abstraction and summarization.

The above challenges can be overcome by adopting \textit{hierarchical aggregation} approaches  (for simplicity we also refer to them as  hierarchical) \cite{EF10}. 
Hierarchical approaches 
allow the visual exploration of very large datasets in a multilevel fashion, offering  overview of a dataset, as well as an  intuitive and  usable  way for finding specific  parts within a dataset.   
Particularly, in hierarchical approaches, the user first obtains an overview 
of the dataset (both structure and a summary of its content) before proceeding
to data exploration operations, such as roll-up and drill-down, filtering out a specific
part of it and finally retrieving details about the data. 
Therefore, hierarchical approaches directly support the visual information seeking mantra.
Also, hierarchical approaches can effectively address the problem of information overloading 
as it provides information abstraction and summarization.

A second challenge is related to the   availability of  API  and query endpoints (e.g., SPARQL) for online data access, as well as the cases where that data is received in a stream fashion.
 The latter pose the challenge of
handling large sets of data in a dynamic setting, and as a result, a
preprocessing phase, such as traditional indexing, is prevented.
In this respect,  modern techniques must offer scalability and efficient processing 
for on-the-fly analysis and visualization of dynamic datasets.

Finally, the requirement for on-the-fly  visualization must be coupled with the diversity of 
preferences and requirements posed by different users and tasks. 
Therefore, the proposed approaches should provide the user with the ability to customize the exploration experience, 
allowing users to organize data into different ways according to 
the type of information or the level of details she wishes to explore.


 
Considering the general problem of exploring big data \cite{Shneiderman08,MortonBGM14,bs16,IdreosPC15,HeerK12b,GGL15}, most approaches aim at   providing appropriate summaries and abstractions over the enormous number of available data objects. 
In this respect, a large number of  systems adopt \textit{approximation techniques} (a.k.a.\ \textit{data reduction} techniques)
in which  partial results are computed. 
Existing approaches are mostly based on: 
(1) sampling and filtering  \cite{FisherPDs12,ParkCM15,AgarwalMPMMS13,ImVM13,BattleSC13} 
and/or (2) aggregation (e.g., binning, clustering) \cite{EF10,JugelJM15, JugelJHM14a,LiuJH13,hw13,bcs15,LinsKS13,AbelloHK06,RodriguesTPTTF13}. 
Similarly,  some modern database-oriented systems adopt approximation techniques 
using query-based approaches (e.g., query translation, query rewriting) \cite{BattleSC13,JugelJM15,JugelJHM14a,VartakMPP14,WuBM14}.
Recently,  incremental approximation techniques are adopted; in these approaches approximate answers are computed over progressively larger samples of the data \cite{FisherPDs12,AgarwalMPMMS13,ImVM13}. 
 In a different context, an adaptive indexing approach is used in \cite{ZoumpatianosIP14}, where the indexes are created incrementally and adaptively throughout exploration. 
Further, in order to improve performance many systems exploit caching and prefetching techniques 
 \cite{TauheedHSMA12,KalininCZ14,JayachandranTKN14,bcs15,ChanXGH08,KhanSA14,DoshiRW03}. 
Finally, in other approaches,  parallel architectures are adopted \cite{EMJ16,KamatJTN14, KalininCZ15,ImVM13}.

Addressing the aforementioned challenges, in this work, we introduce a generic model that
combines personalized multilevel exploration with online analysis of numeric and temporal data. 
At the core lies a lightweight hierarchical  aggregation model, constructed \mbox{on-the-fly} for a given set of data. 
The proposed model is a tree-based structure that aggregates data objects into multiple 
levels of hierarchically related groups based on numerical or temporal values of the objects.
Our model also enriches groups (i.e., aggregations/summaries) with statistical information regarding their content, 
offering richer overviews and insights into the detailed data.
An additional feature is that it allows users to organize data exploration
in different ways, by parameterizing the number of groups, the range and 
cardinality of their contents, the number of hierarchy levels, and so on. 
On top of this model, we propose three  user exploration scenarios and present two methods for efficient exploration over large datasets: the first one achieves the incremental construction of the model based on  user interaction, whereas the second one enables dynamic and efficient adaptation of the model to the user's preferences. 
 {The efficiency of the proposed model is illustrated through a thorough theoretical  analysis, as well as an experimental evaluation.} 
Finally, the proposed model is realized  in a web-based tool,
called \textit{SynopsViz} that offers a variety of visualization techniques (e.g., charts, timelines) for multilevel visual exploration and analysis over Linked Data (LD) datasets.

 \stitle{Contributions.}
The main contributions of this work are summarized as follows.
\begin{itemize}

\item
We introduce a generic model for organizing, exploring, and analysing
numeric and temporal data in a multilevel fashion. 

\item
We implement our model as a lightweight, main memory tree-based structure, which can be  efficiently constructed on-the-fly.
 
\item
We propose two tree structure versions, 
which adopt different approaches for the data organization. 

\item
We describe a simple method to estimate the  tree construction parameters, when no user preferences are available.

\item
 We define different exploration scenarios assuming various user exploration preferences.

\item
We introduce  a method that incrementally constructs and prefetches the hierarchy tree via user interaction.

 \item
 {We propose an efficient method that dynamically adapts an existing hierarchy to a new, considering user's preferences.
}

\item
 {We present a thorough theoretical  analysis, illustrating the efficiency of the proposed model.
}
 
\item
We develop a prototype system  which implements the presented model, offering multilevel visual exploration and analysis over LD. 

\item
We conduct a thorough performance evaluation and an empirical  user study, using  the DBpedia 2014 dataset.

\end{itemize}
 
\stitle{Outline.}
The remainder of this paper is organized as follows. 
Section~\ref{sec:model} presents the proposed hierarchical model, and
Section~\ref{sec:ext} provides the exploration scenarios and methods for efficient hierarchical exploration.
Then,  Section~\ref{sec:system} presents the SynopsViz  tool  and demonstrate the basic functionality.
The evaluation of our system is presented in Section~\ref{sec:eval}.
Section~\ref{sec:related} reviews related work, while
Section~\ref{sec:concl} concludes this paper.

\vspace{-2mm}
\section{The \het Model}
\label{sec:model}
In this section we present {\het} ({\textbf{H}ierarchical  \textbf{E}xploration \textbf{Tree}}), 
a generic model for organizing, exploring, and analysing
numeric and temporal data in a multilevel fashion. 
Particularly, {\het}  is defined in the context of multilevel (visual) exploration and analysis. 
{The proposed model hierarchically organize arbitrary numeric and temporal data, 
without requiring it to be described by an hierarchical scheme. 
We should note that, our model is not bound to any specific type of visualization; rather it can be adopted by several "flat" visualization techniques (e.g., charts,  timeline), 
offering scalable and multilevel exploration over non-hierarchical data. 
} 

In what follows, we present  some basic aspects of our working scenario 
(i.e., visual exploration and analysis scenario) and highlight the main assumptions 
and requirements employed in the construction of our model.
First,  the input data in our scenario can be retrieved directly from a database, but also produced dynamically; 
e.g., from a query or from data filtering (e.g., faceted browsing). 
Thus, we consider that data  visualization is performed online; 
i.e., we do not assume an offline preprocessing phase in the construction of the visualization model. 
Second, users can specify different requirements or preferences with respect to the data organization.  
For example, a user prefers to 
organize the data as a deep hierarchy for a specific task, while for another task a 
flat hierarchical organization is more appropriate.
Therefore, even if the data is not dynamically produced, 
the data organization is dynamically adapted to the user preferences. 
The same also holds for any additional information (e.g., statistical information)  that is computed for each group of objects. 
This information must be recomputed when the groups of objects (i.e., data organization) are modified. 

From the above, a basic requirement is that the model must be constructed 
on-the-fly for any given data and users preferences.
Therefore, we implement our model as a lightweight, main memory tree structure, 
which can be efficiently constructed on-the-fly. We define two versions of this tree structure, following data organization approaches well-suited to visual 
exploration and analysis context: the first version considers fixed-range groups of data objects, whereas the second considers fixed-size groups.
Finally, our structure allows efficient on-the-fly statistical computations, 
which are extremely valuable for the hierarchical exploration and analysis scenario.

The basic idea of our model is to hierarchically group data objects based on values of one of their properties. 
Input data objects   are stored at the  leaves, while internal nodes aggregate 
their child nodes.
The root of the tree represents (i.e., aggregates) the whole dataset.
The basic concepts of our model can be considered similar to a simplified version  of a static 1D R-Tree \cite{Guttman84}.

Regarding the visual representation of the model and data exploration,
we consider that both data objects sets (leaf nodes contents) 
and entities representing groups of objects (leaf or internal nodes) 
are visually represented enabling the user to explore the data in a hierarchical manner. 
Note that our tree structure organizes data in a  hierarchical model, without setting any
constraints on the way the user interacts with these hierarchies. 
As such, it is possible that different strategies can be adopted, regarding the traversal policy, 
as well as the nodes of the tree that are rendered in each visualization stage.

In the rest of this section,  preliminaries are presented in Section~\ref{sec:prel}. 
In Section~\ref{sec:HETree}, we introduce the proposed tree structure. 
  Sections~\ref{sec:HETree-C} and \ref{sec:HETree-R} 
present the two versions of the  structure.   
Finally, Section~\ref{sec:param} discusses the specification of the parameters 
required for the tree construction,  and  Section~\ref{sec:stat} presents how
statistics computations can be performed over the tree.

\subsection{Preliminaries}
\label{sec:prel}

In this work we formalize data objects as RDF triples. 
However, the presented methods are generic and can be applied to any data objects with numeric or temporal attributes. Hence, in the 
following, the terms triple and (data) object will be used interchangeably.

We consider  an \textit{RDF dataset} $R$  consisting of a set of \textit{RDF triples}.
As \textit{input data}, we assume a set of {RDF triples} $D$, 
where $D \subseteq R$ and triples in $D$ have as objects either numeric (e.g.,  integer, decimal) or temporal values (e.g., date, time).
Let $tr$ be an {RDF triple}, $tr.s$, $tr.p$ and $tr.o$ represent, respectively, 
the \textit{subject},  \textit{predicate} and \textit{object} of the RDF triple $tr$.

Given input data  $D$,
$S$ is an \textit{ordered set} of RDF triples, produced from $D$, 
where triples are sorted based on objects' values, in ascending order. 
Assume that $S[i]$ denotes the $i$-th triple, with $S[1]$ the first triple. 
Then, for each $i < j$, we have that  ${S[i].o \leq S[j].o}$.
Also, $D=S$, i.e., for each $tr, tr \in D$ iff $tr \in S$.

Figure~\ref{fig:data} presents a set of 10 RDF triples, representing persons and their ages. 
In Figure~\ref{fig:data}, we assume that the subjects $p0$-$p9$ are instances of a class \textit{Person} and the predicate \textit{age} is a datatype property with integer range.

\begin{figure}[!ht] 
 \vspace{-1mm}
\centering
\small
\setlength{\tabcolsep}{20pt} 
\begin{tabular}{cc}
 {$p0 \textit{ age }   35$} &  {$p5 \textit{ age }    35$} \\
 {$p1 \textit{ age }  100$} &  {$p6 \textit{ age }    45$} \\
 {$p2 \textit{ age }  55$} &   {$p7 \textit{ age }    80$} \\
 {$p3 \textit{ age }  37$} &   {$p8 \textit{ age }     20$} \\
 {$p4 \textit{ age }  30$} &    {$p9 \textit{ age }    50$}  \\ 
\end{tabular} 
\vspace{-1mm}
\caption{Running example input data (data objects)}
 \label{fig:data}
\end{figure}

\begin{myExample}
\label{ex:data}
In Figure~\ref{fig:data}, given the RDF triple $tr = p0 \textit{ age } 35$, we have that $tr.s=p0$, $tr.p=age$ and $tr.o=35$. 
Also, given that all triples comprise the input data $D$ and $S$ is the ordered set of $D$ based on the object values, in ascending order;  we have that
$S[1] =  {p8 \textit{ age } 20}$ and 
$S[10] =  {p1 \textit{ age }  100}$. 
\eoe
\end{myExample}

Assume an \textit{interval} $I=[a,b]$, where $a, b \in \mathbb{R}$;
then, $I = \{ k \in \mathbb{R} \mid  a \leq k \leq b \}$.
Similarly, for $I=[a,b)$, we have that $I = \{ k \in \mathbb{R} \mid  a \leq k < b \}$.
Let $I^-$ and $I^+$ denote the lower and upper bound of the interval $I$, respectively.
That is, given $I=[a, b]$, then $I^-=a$ and $I^+=b$.  
The \textit{length} of an interval $I$ is defined as $|I^+ - I^-|$.
 
In this work we assume {rooted trees}.
The number of the children of a node is its \textit{degree}.
Nodes with degree $0$ are called \textit{leaf nodes}.
Moreover, any non-leaf node is called \textit{internal node}.
\textit{Sibling nodes} are the nodes that have the same parent.
The \textit{level of a node} is defined by letting the root node be at level zero. 
Additionally, the \textit{height of a node} is the length of the longest path from the node to a leaf. A leaf node has a height of $0$.

The \textit{height of a tree} is the maximum level of any node in the tree.
The \textit{degree of a tree} is the maximum degree of a node in the tree.
An \textit{ordered tree} is a tree where the children of each node are ordered.
A tree is called an $m$-\textit{ary tree} if every internal node has no more than $m$ children.
A  \textit{full} $m$-\textit{ary tree} is a tree where every internal node has exactly $m$ children.  
A \textit{perfect} $m$-\textit{ary tree} is a full  $m$-ary tree in which all leaves are at the same level.


\subsection{The \het Structure}
\label{sec:HETree}

In this section, we present in more detail the 
\textit{\mbox{\het}} structure. \mbox{\het} hierarchically organizes numeric and temporal data into groups;
 intervals  are used to represents these groups.%
 \footnote{Note that our structure handles numeric and temporal data in a similar manner. 
Also, other  types of one-dimensional data may be supported,
with the requirement that a total order can be defined over the data.}
\het is defined by the tree degree and the number of leaf nodes.%
\footnote{Note that following a similar approach, the \het can also 
be defined by specifying the tree height  instead of degree or number of leaves.}
Essentially, the number of leaf nodes corresponds to the number of groups where input data objects are organized. 
The tree degree corresponds to the (maximum) number of groups
where a group is split in the lower level.

Given a set of data objects (RDF triples) $D$, a positive integer $\ell$ denoting the number of leaf nodes; 
and a positive integer $d$ denoting the tree degree; 
an \textit{\het} $(D, \ell, d)$ is an \textit{ordered   d-ary tree}, 
with the following  basic properties.

\begin{itemize}
\item 
 The tree has exactly $\ell$ number of leaf nodes.

\item 
All leaf nodes appear in the same level.

\item 
Each leaf node contains a set of data objects,
sorted in ascending order based on their values. 
Given a leaf node $n$, $n.data$ denote the data objects contained in $n$.
 
\item 
Each internal node has at most \textit{d} children nodes. Let $n$ be  an internal node,  $n.c_i$ denotes the $i$-th child for 
the node $n$, with $n.c_1$ be the leftmost child.

\item 
Each node corresponds to an interval.
Given a node $n$, $n.I$ denotes the interval for the node $n$.

\item 
At each level, all nodes are sorted based on the lower bounds of their intervals. That is, let $n$ be an internal node, 
for any $i<j$,  we have that $n.c_i.I^- \leq  n.c_j.I^-$.

\item
For a leaf node, its interval is bounded by the  values of the objects included in this leaf node. Let $n$ be the leftmost leaf node; assume that $n$ contains $x$  objects from $D$.
Then, we have that $n.I^-=S[1].o$ and $n.I^+=S[x].o$, 
where $S$ is the ordered object set resulting from $D$.

\item 
For an internal node, its interval is bounded by the union of the intervals of its children. That is, let $n$ be an internal node, having $k$ child nodes; then, we have $n.I^-=n.c_1.I^-$ and ${n.I^+=n.c_k.I^+}$.

\end{itemize}



Furthermore,  we present two different approaches for organizing the data in the \het.
Assume the scenario in which  a user wishes to (visually) explore and analyse 
the historic events from DBpedia \cite{AuerBKLCI07}, per decade.
In this case, user orders historic events by their date and organizes them into groups of equal ranges (i.e., decade).
In a second scenario, assume that a user wishes to analyse in the Eurostat dataset the
gross domestic product (GDP) organized into fixed groups of countries.
In this case, the user is interested in finding information like:
the range and the variance of the GDP values over the top-10 countries with the highest GDP factor.
In this scenario, the user orders countries by their  GDP and organizes them into groups of equal sizes (i.e., 10 countries per group).

In the first approach, we organize data objects into groups, 
where the object values of each group covers equal range of values.
In the second approach, we organize objects  into groups, 
where each group contains the same number of objects. 
In the following sections, we present in detail the two
approaches for organizing the data in the \het.


\subsection{A Content-based \het (\hetC)}
\label{sec:HETree-C}

In this section we introduce a version of  the \mbox{HETree}, 
named  \hetC (Content-based \het).
This \linebreak\mbox{\het} version organizes data into equally sized groups. The basic property of the \hetC  is that each leaf node
contains approximately  the same number of  objects and the content (i.e., objects) of a leaf node specifies its interval. 
For the  tree construction, the objects are first assigned to the leaves and then the intervals are defined.

An \textit{\hetC}  $(D, \ell, d)$ is an \het, with the following extra property. 
Each leaf node contains  $\lambda$ or $\lambda-1$  objects, 
where$\lambda=\ceil*{\frac{|D|}{\ell}}$.\footnote{We assume  that, the number of objects is at least as the number of leaves; i.e., $|D|\ge \ell$.} 
Particularly, the $\ell-(\lambda \cdot \ell-|D|)$ leftmost leaves contain  $\lambda$  objects, 
while the rest  leaves
 contain $\lambda-1$.\footnote{{As an alternative we can construct the \hetC, so each leaf contains $\lambda$  objects, except the rightmost leaf which will contain between $1$ and $\lambda$ objects.}}
We can equivalently define the \mbox{\hetC}  by providing the number of objects per leaf $\lambda$, instead of the number of leaves $\ell$.

\begin{figure}[t] 
 \vspace{-1mm}
\centering
\includegraphics[scale=0.45]{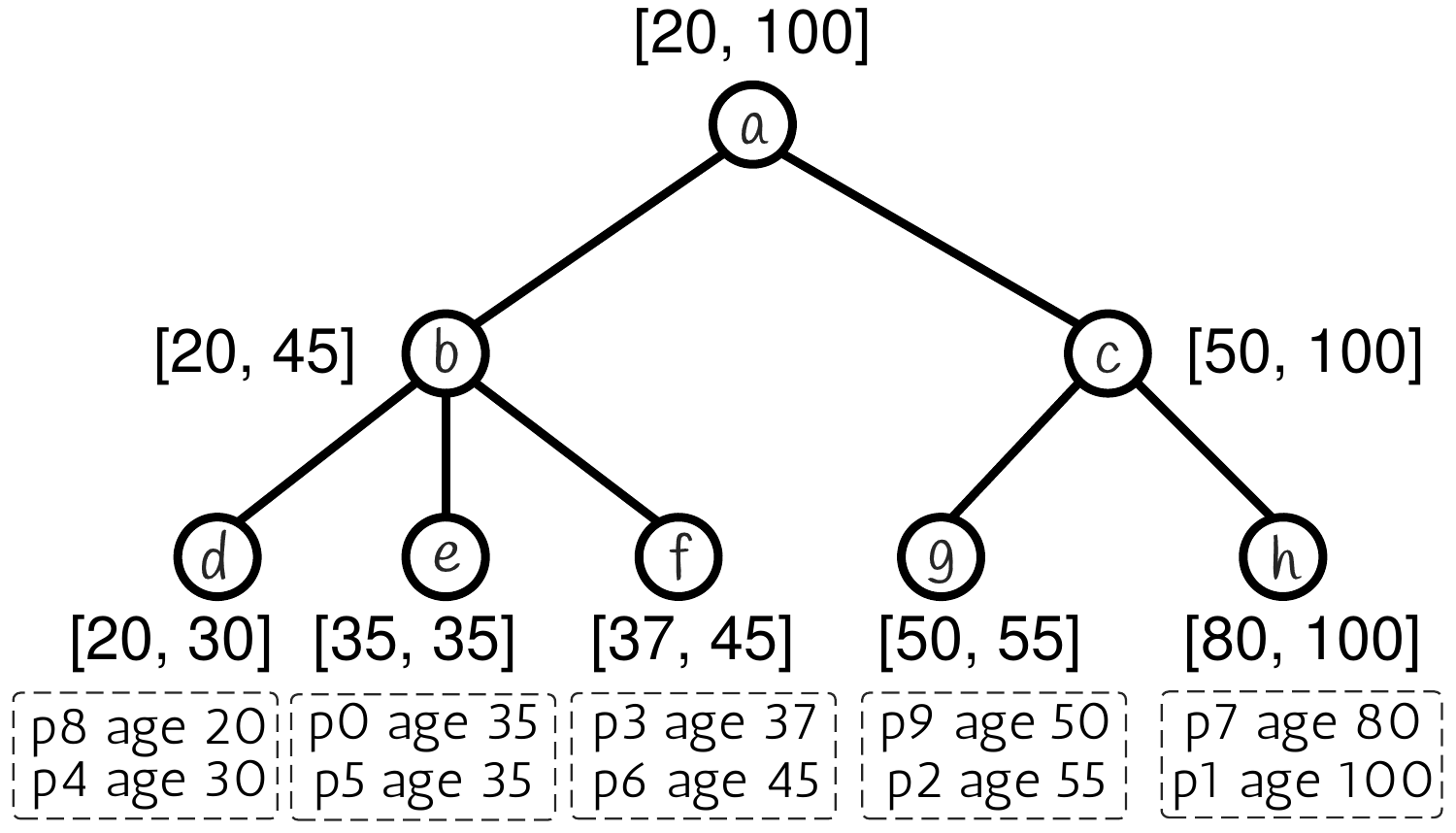}
\vspace{-1mm}
\caption{A Content-based \het (\hetC)}
\label{fig:tree-c}
\end{figure}
    
\begin{myExample}
Figure~\ref{fig:tree-c} presents an \hetC constructed by considering 
the  set of   objects $D$ from Figure~\ref{fig:data}, $\ell=5$ and $d=3$. 
As we can observe, all the leaf nodes contain equal number of objects. 
Particularly, we have that $\lambda=\ceil*{\frac{10}{5}}=2$. 
Regarding the leftmost interval, we have $d.I^-=20$ and $d.I^+=30$. 
%
\eoe
\end{myExample}

\subsubsection{The \hetC Construction}\
\label{sec:HETree-C-Constr}

We construct the \hetC in a bottom-up way. 
Algorithm~\ref{algo:HETree-C} describes the \mbox{\hetC} construction. 
Initially, the algorithm sort the object set $D$ in ascending  order, based on  objects values (\textit{line 1}). 
Then, the algorithm uses two  procedures to construct the tree nodes. 
Finally, the root node of the constructed tree is returned (\textit{line 4}).

 \begin{algorithm2e}[t]
\footnotesize
\caption{createHETree-C/R ($D$, $\ell$, $d$)}
\label{algo:HETree-C}
\KwIn{$D$:   set of objects; $\ell$: number of leaf nodes;  \linebreak
$d$: tree degree}
\KwOut{$r$:   root node of the HETree tree}
\vspace{2mm}

$S\gets$	sort $D$ based on objects values\;
$L\gets	\mathsf{constrLeaves\text{-}C/R}(S, \ell)$\;
$r \gets	 \mathsf{constrtInterlNodes}(L, d)$\;
\Return $r$\;
\end{algorithm2e}

The $\mathsf{constrLeaves\text{-}C}$ procedure (Procedure~1) construct $\ell$ leaf nodes (\textit{\mbox{lines~4--16}}).
For the first $k$ leaves, $\lambda$ objects are inserted, while for the rest leaves,  $\lambda-1$ objects  are inserted (\textit{lines 6--9}).
Finally, the set of created  leaf nodes is returned (\textit{line 17}).

\begin{procedure}[!t]
\footnotesize
\SetAlgoProcName{\small Procedure 1:}{}
\caption{{constrLeaves-C} ($S$, $\ell$)}
\label{proc:leaves-C}
\KwIn{$S$: ordered set of objects; $\ell$: number of leaf nodes}
\KwOut{$L$: ordered set of leaf nodes}

\vspace{1mm}
$\lambda \gets \ceil*{\frac{|S|}{\ell}}$\;
$k \gets \ell-(\lambda \cdot  \ell-|S|)$\;
$beg \gets	  1$ \;
\For{$i\gets 1$ \KwTo $\ell$}{
	create an empty leaf node $n$\;
	\eIf{$i\leq k$}{
		  	$num \gets	  \lambda$ \;
   }{ 			
		  	$num \gets	  \lambda-1$ \; 	 
	}
	$end  \gets	beg + num$ \;
	\For{$t \gets beg$ \KwTo $end$}{
			$n.data \gets S[t]$ 
	}
	$n.I^- \gets S[beg].o$\;
	$n.I^+ \gets S[end].o$\;
	$L[i] \gets n$ \;
	$beg \gets end +1$ \;
}
\Return $L$\;
\end{procedure}



The  $\mathsf{constrtInterNodes}$ procedure (Procedure~2) builds the internal nodes in a recursive manner. 
For the nodes $H$, their parents nodes $P$ are created (\textit{lines 4-16});
then, the procedure  calls itself using as input the parent nodes $P$ (\textit{line 21}). 
The recursion terminates when the number of created parent nodes is equal to one (\textit{line 17});  i.e., the root of the tree is created.

\stitle{Computational Analysis.}
The computational cost for the \hetC construction   (Algorithm~\ref{algo:HETree-C})
 is the sum of three parts.  
The first is sorting the input data, which can be done in the worst case in $O(|D| log |D|)$, employing a linearithmic sorting algorithm (e.g., merge-sort).
The second part is the $\mathsf{constrLeaves\text{-}C}$ procedure, which 
requires $O(|D|)$ for scanning all data objects.
The third part is the  $\mathsf{constrtInterNodes}$ procedure, which requires
$d\cdot(\ceil*{\frac{\ell}{d}} +\ceil*{\frac{\ell}{d^2}} +\ceil*{\frac{\ell}{d^3}} + \ldots+1)$, 
with the sum being the number of internal nodes in the tree. 
Note that the maximum number of internal nodes in a $d$-ary tree corresponds to the number of internal nodes in a perfect $d$-ary tree of the same height. 
Also, note the number of  internal nodes of a perfect $d$-ary tree of height $h$ is $\frac{d^h-1}{d-1}$.
In our case, the height of our tree is $h=\ceil*{log_d\ell}$. 
Hence, the maximum number of internal nodes is 
$\frac{d^{\ceil*{log_d\ell}}-1}{d-1}\leq \frac{d\cdot\ell-1}{d-1}$.
Therefore,  the $\mathsf{constrtInterNodes}$ procedure, in worst case requires
$O(\frac{d^2 \cdot \ell-d}{d-1})$.
Therefore, the overall computational cost for the \hetC construction in the worst case is%

$O(|D|   log |D| + |D| +\frac{d^2 \cdot \ell-d}{d-1})
=O(|D| log |D|+ \frac{d^2 \cdot \ell - d}{d-1})$.\footnote{{In the complexity computations presented through the paper,  terms that are dominated by others (i.e., having lower growth rate) are omitted.}}

\begin{procedure}[t!]
\footnotesize
\SetAlgoProcName{\small Procedure 2:}{}
\caption{constrtInterNodes($H$, $d$)}
\label{proc:parents}
\KwIn{$H$: ordered set of nodes; $d$: tree degree}
\KwOut{$r$:   root node for $H$}
\KwVar{$P$: ordered set of  $H$'s parent nodes }
\vspace{1mm}

$p_{num} \gets    \ceil*{\frac{|H|}{d}}$
 \Comment*[r]{\mycomment{{number of parents nodes}}}
 
${t \gets    d-(p_n  \cdot  d-|H|)}$\hspace{-0.5mm}
 \Comment*[r]{\mycomment{{last parent's number of children}}}

$c_{beg} \gets	  1$  \Comment*[r]{\mycomment{{first child node}}}

\For{$p \gets 1$ \KwTo $p_{num}$}{
	create an empty internal node $n$\;
		\eIf{$p = p_{num}$}{
		  	$c_{num} \gets	  t$  \Comment*[r]{\mycomment{{number of children}}}
   }{ 			
		  	$c_{num} \gets	  d$ \;
	}
		$c_{end} \gets	c_{beg}  + c_{num}$ 
	\Comment*[r]{\mycomment{{last child node}}}
	\For{$j \gets c_{beg}$ \KwTo $c_{end}$}{
			$n.c[j] \gets H[j]$ \;
	}	
	$n.I^- \gets H[c_{beg}].I^-$\;
	$n.I^+ \gets H[c_{end}].I^+$\;
		$P[p] \gets n$ \;
	$c_{beg}  \gets c_{end} +1$ \;
}
	\eIf{$p_{num} =1$}{
		$r\gets P$\;
		\Return{r}\;
	}{
		\Return{$\mathsf{constrtInterlNodes}(P, d)$}\;
	}
\end{procedure}

\subsection{A Range-based \het (\hetR)}
\label{sec:HETree-R}

The second version of  the \het is called \hetR (Range-based \het). 
{\hetR} organizes data into equally ranged groups. The basic property of the \hetR  is that each leaf node covers an equal range of values. 
Therefore, in  \hetR, the data space defined by the objects values is equally divided over the leaves. 
As opposed to \hetC, in \hetR the interval of a leaf  specifies its content.
Therefore, for the \hetR construction, the intervals of all leaves are first defined  and then objects are inserted.

An \textit{\hetR}  $(D, \ell, d)$ is an \het, with the following extra property. 
The interval of each leaf node has the same length; i.e., covers equal range of values. 
Formally, let $S$ be the sorted RDF set resulting from $D$, for each leaf node  its interval has length $\rho$, 
where
$\rho=\frac{|S[1].o - S[|S|].o|}{\ell}$.\footnote{We assume here that, there is at least one object in $D$ with different   value than the rest objects.}
Therefore, for a leaf node $n$, we have that $|n.I^- -n.I^+|=\rho$.
For example, for the leftmost leaf, its interval is $[S[1].o, S[1].o+\rho)$.
The \hetR is equivalently defined by providing the interval length $\rho$,  instead of the number of leaves $\ell$.

\begin{figure}[!h] 
 \vspace{-1mm}
\centering
\includegraphics[scale=0.49]{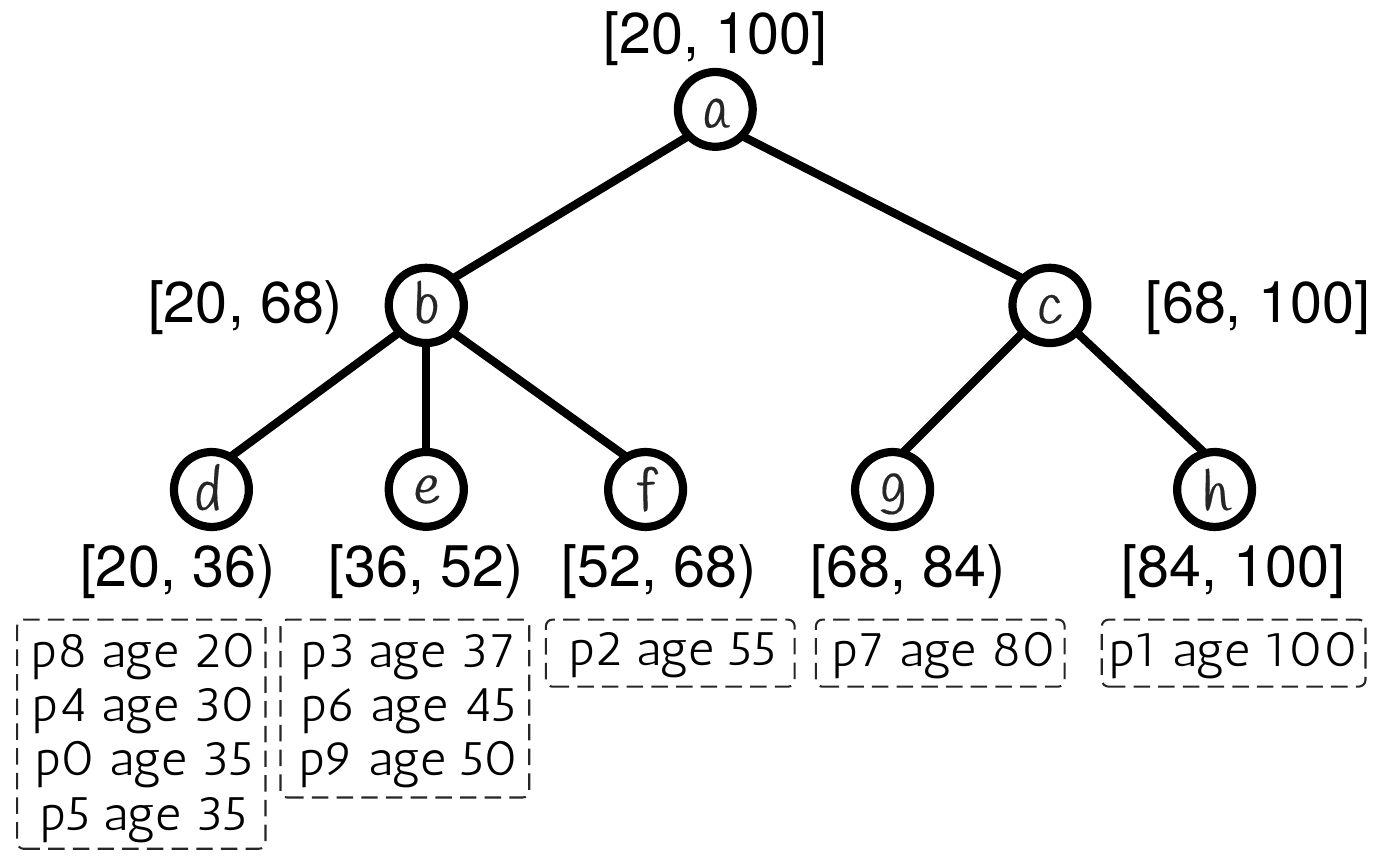}
 \vspace{-1mm}
\caption{A Range-based \het (\hetR)}
\label{fig:tree-r}
\end{figure}

\begin{myExample}
Figure~\ref{fig:tree-r} presents an \hetR tree constructed by considering 
the set of objects $D$ (Figrue~\ref{fig:data}), $\ell=5$ and $d=3$. 
As we can observe from Figure~\ref{fig:tree-r},  each leaf node covers equal range of values. 
Particularly, we have that the interval of each leaf must have length $\rho=\frac{|20 - 100|}{5}=16$.
\eoe
\end{myExample}

\subsubsection{The \hetR Construction}
This section studies the construction of the  \hetR structure.
The \hetR is also constructed in a bottom-up fashion. 

Similarly with the \hetC version, 
Algorithm~\ref{algo:HETree-C} is used  for the \hetR construction. 
The only difference   is the 
 $\mathsf{constrLeaves\text{-}R}$ procedure (\textit{line 2}), which creates the leaf nodes of the \hetR and is presented in Procedure~3.
 
The procedure 
The procedure constructs $\ell$ leaf nodes (\textit{\mbox{lines 2--9}}) and assigns same intervals to all of them (\textit{lines 4--8}),  it traverses all objects in $S$ (\textit{lines 10--12}) and places them to  the appropriate leaf node (\textit{\mbox{line 12}}). 
Finally, 
returns the set of created leaves  (\textit{line 13}).

%
%
%
%
%

\begin{procedure}[t]
\footnotesize
\SetAlgoProcName{\small Procedure 3:}{}
\caption{constrLeaves-R($S$, $\ell$)}
\label{proc:leaves-R}
\KwIn{$S$: ordered set of objects; $\ell$: number of leaf nodes}
\KwOut{$L$: ordered set of leaf nodes}

\vspace{1mm}

$\rho\gets \frac{|S[1].o - S[|S|].o|}{\ell}$\;
\For{$i\gets 1$ \KwTo $\ell$}{
	create an empty leaf node $n$\;
	\eIf{$i=1$}{
		  	$n.I^-  \gets	  S[1].o$ \;
	  }{ 			
	    $ n.I^-  \gets	  L[i-1].I^+ $ \;
	  }
  	  $n.I^+  \gets	  n.I^- + \rho$\;
	  $L[i] \gets n$\;
}
\For{$t \gets 1$ \KwTo $|S|$}{
$j \gets \floor*{\frac{S[t].o-S[1].o}{\rho}}+1$\;
$L[j].data \gets S[t]$ \;
}

\Return $L$\;
\end{procedure}

 \stitle{Computational Analysis.}
The computational cost for the \hetR construction   (Algorithm~\ref{algo:HETree-C})  for sorting the input data (\textit{line 1}) and creating the internal nodes (\textit{line 3}) is the same as in the \hetC case.
 The $\mathsf{constrLeaves\text{-}R}$ procedure (\textit{line 2})  requires 
${O(\ell+|D|)}=O(|D|)$ (since $|D|\ge \ell$). 
Using the  computational costs for the first and the third  part from Section~\ref{sec:HETree-C-Constr}, 
we have that in worst case, the overall computational cost for the \hetR construction is 
$O(|D| log |D|+ |D|+ \frac{d^2\cdot \ell-d}{d-1})=O(|D| log |D|+ \frac{d^2\cdot \ell-d}{d-1})$.


\subsection{Estimating  the \het  Parameters}
\label{sec:param}
In our working scenario, the user specifies the parameters required for the \het  construction (e.g., number of leaves $\ell$).
In this section, we describe our approach for automatically calculating the \het parameters based on the input data, when no user preferences are provided. Our goal is to derive the parameters by the input data, such that the resulting \mbox{HETree} can 
address some basic guidelines set by the visualization environment. 
In what follows, we discuss in detail the proposed approach. 
 

An important parameter in hierarchical visualizations is the minimum and maximum 
number of objects that can be effectively rendered in the most detailed level.\footnote{Similar bounds can also be defined for other tree levels.} 
In our case, the above numbers correspond to the number of objects contained in the leaf nodes.
The proper calculation of these numbers is crucial such that the resulting tree avoids overloaded  visualizations. 


Therefore, in \het construction, 
our approach  considers 
the minimum and the maximum number of objects per leaf node, denoted as 
$\lambda_{min}$ and $\lambda_{max}$, respectively.
Besides  the number of objects rendered in the lowest level, 
our approach considers perfect $m$-ary trees, such that a more "uniform" structure (i.e., all the internal nodes have exactly $m$ child nodes)  results. The following example illustrates our approach of calculating the \het parameters.

\begin{myExample}
Assume that based on an adopted visualization technique, 
the ideal number of data objects to be rendered on a specific screen is between $25$ and $50$.
Hence, we have that $\lambda_{min}=25$ and $\lambda_{max}=50$. 

Now, let's assume that we want to visualize the object set $D_1$, 
using an \hetC, where ${|D_1|=500}$.
Based on the number of objects and the $\lambda$ bounds, 
we can estimate the bounds for the number of leaves.  
Let $\ell_{min}$ and $\ell_{max}$ denote the lower and the upper bound for the number of leaves.
Therefore, we have that
$\ceil*{\dfrac{|D_1|}{\lambda_{max}}} \leq \ell \leq \ceil*{\dfrac{|D_1|}{\lambda_{min}} } \Leftrightarrow$
 $\ceil*{\dfrac{500}{50}} \leq \ell \leq \ceil*{\dfrac{500}{25}}  \Leftrightarrow$
$10 \leq \ell \leq 20$.
\vspace{4px}

Hence, our \hetC should have between  \linebreak${\ell_{min}=10}$ and ${\ell_{max}=20}$ leaf nodes.
Since, we consider perfect $m$-ary trees, from Table~\ref{tab:leafNum}
we can identify the tree characteristics that conform to the number of leaves guideline. 
The candidate setting (i.e., leaf number and degree) is indicated in \mbox{Table~\ref{tab:leafNum}}, using dark-grey colour.
Note that, the settings with $d=2$ are not examined since visualizing two groups of objects in each level 
is considered a small number under most visualization settings.
Hence,  in any case we only assume settings with $d\geq3$ and $height\geq2$. 
Therefore, an \hetC with $\ell=16$ and $d=4$ is a suitable structure for our case.

Now, let's  assume that we want to visualize the
object set $D_2$, where  $|D_2|=1000$.
Following a similar approach, we have that 
$20 \leq \ell \leq 40$.
The candidate settings are indicated in Table~\ref{tab:leafNum} using light-grey colour.
Hence, we have the following settings that satisfy the considered guideline:
$S1$:  $\ell=27$,  $d=3$;
$S2$:  $\ell=25$, $d=5$; and
$S3$:  $\ell=36$, $d=6$.

In the case where more than one setting  satisfies the considered guideline, we select the preferable one according to following set of rules.
From the candidate settings,  we prefer the setting which results in the highest tree
 ($1$st \textit{Criterion}).%
 \footnote{Depending on user preferences  and the examined task, the shortest tree may be preferable. For example, starting from the root, the user wishes to access the data objects (i.e., lowest level) by performing the smallest amount of  drill-down operations possible.}
In case that the highest tree is constructed by more than one settings, we consider 
the distance $c$, between $\ell$ and the centre of   $\ell_{min}$ and  $\ell_{max}$ ($2$nd \textit{Criterion}); 
i.e., $c=|\ell-\frac{\ell_{min}+\ell_{max}}{2}|$. 
The setting with the lowest $c$ value is selected.
Note that, based on the visualization context, different criteria and preferences may be followed.

In our example, from the candidate settings,  setting 
${S1}$ is selected, since it will construct the highest  tree (i.e., $height=3$). 
 On the other hand,    settings ${S2}$ and ${S3}$ will construct trees with lower heights (i.e., $height=2$).

 Now, assume a scenario where only  ${S2}$ and ${S3}$ are 
candidates.  In this case, since both settings result to trees with equal heights, 
 the  {\mbox{$2$nd \textit{Criterion}}} is considered. 
 Hence, for the ${S2}$   we have 
 $c_2=|25-\frac{20+40}{2}|=5$. 
 Similarly, for the ${S3}$   
  $c_3=|36-\frac{20+40}{2}|=6$. 
Therefore, between the    ${S2}$ and ${S3}$, 
the setting ${S2}$ is preferable, since $c_2<c_3$.
 
In case of \hetR, a similar approach is followed,
assuming normal distribution over the values of the objects. 
\eoe
\end{myExample}

\begin{table}[t!]
 \footnotesize
\centering
\caption{\mbox{Number of leaf nodes for perfect $m$-ary trees }}
\label{tab:leafNum}
  \begin{tabular}{ c c c c c  }
\tline
\multicolumn{1}{c}{}& \multicolumn{4}{c}{\textbf{Degree}}\\ \cmidrule[0.5pt](lr){2-5}
\multicolumn{1}{c}{\textbf{Height}}   &\textbf{3} &\textbf{4} &\textbf{5} &\textbf{6} \\
 \bline
\textbf{1}   & 3 &4 & 5 & 6\\
\textbf{2}   & 9 &\cellcolor{gray!85}16 &\cellcolor{gray!15} 25 & \cellcolor{gray!15} 36\\
\textbf{3}   & \cellcolor{gray!15}27 &64 &625 & 216\\
\textbf{4}   & 81 & 256 &3125 & 1296\\
\textbf{5}   & 243 &1024 &15625 & 7776\\
\textbf{6}  &729 & 4048 &78125 & 46656\\
\bline
\end{tabular}
 \end{table}

\subsection{Statistics Computations over \het} 
\label{sec:stat}

Data statistics is a crucial aspect in the context of hierarchical visual exploration and analysis. 
Statistical informations over  groups of objects (i.e., aggregations) offer rich insights  into  the underlying (i.e., aggregated) data.
In this way, useful information regarding different set of objects with common 
characteristics is provided. 
Additionally, this information may also guide the users through their navigation over the hierarchy. 

In this section, we present how statistics computation is performed over the nodes of the  
{\het}.
Statistics computations exploit two main aspects of the \mbox{\het} structure: 
(1) the internal nodes aggregate their child nodes; and 
(2) the tree is constructed in  bottom-up fashion. 
Statistics computation is performed during the tree construction; for the leaf nodes, we gather statistics from the objects they contain, whereas for the internal nodes we aggregate the statistics of their children.

For simplicity, here, we assume that each node contains the following extra fields, 
used for simple statistics computations, although more complex or \mbox{RDF-related} (e.g., most common subject, 
subject with the minimum value, etc.) statistics can be computed.
Assume a node $n$, as
$n.N$ we denote the \textit{number} of objects covered  by $n$;
as  $n.\mu$ and  $n.\sigma^2$ we denote the  {mean} and 
the  {variance} of the  objects' values covered  by $n$, respectively. 
Additionally, we assume the  {minimum} and the  {maximum} values, 
denoted as $n.min$ and $n.max$, respectively.
    
Statistics computations can be easily performed in the construction algorithms 
(Algorithm~\ref{algo:HETree-C})  without any modifications. 
The follow example illustrates these computations.

\begin{myExample}
In this example we assume the \linebreak   \hetC presented in Figure~\ref{fig:tree-c}. 
Figure~\ref{fig:stat} shows the \hetC  with the computed statistics in each node.
When all the leaf nodes have been constructed, 
the statistics for each leaf is computed. 
For instance,  we can see from Figure~\ref{fig:stat}, that for the rightmost  leaf $h$ we  have:
$h.N = 2$, $h.\mu = \frac{80+100}{2}=90$ and $h.\sigma^2 = \frac{1}{2} \cdot ((80-90)^2 +(100-90)^2) = 100$.
Also, we have $h.min=80$ and $h.max=100$.
Following the above process,  we compute the statistics for all leaf nodes.

\begin{figure}[h] 
  \vspace{-1mm}
\centering
\includegraphics[scale=0.575]{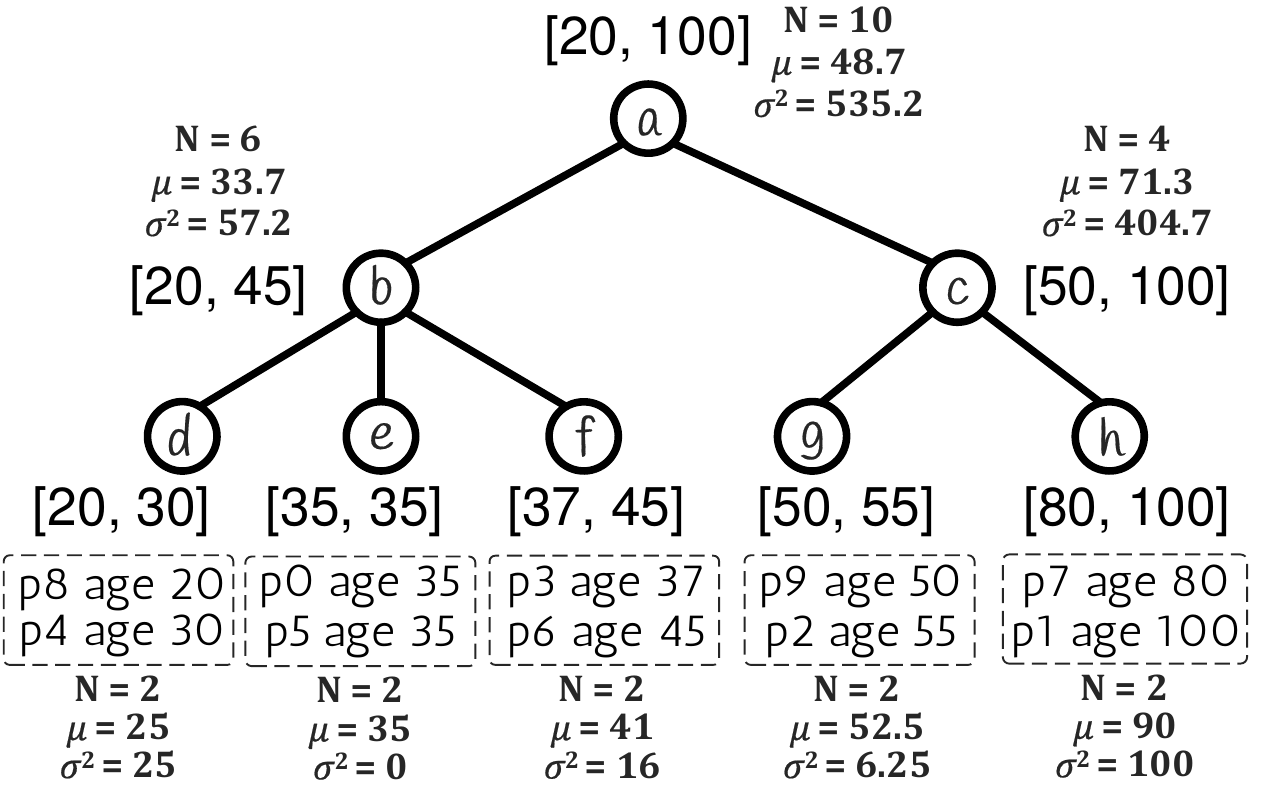}
\vspace{-1mm}
\caption{Statistics computation over \het}
\label{fig:stat}
\end{figure}

Then, for each parent node we construct, we compute its statistics 
using the computed statistics of its child nodes.
Considering the $c$ internal node, with the child nodes $g$ and $h$, we have that 
$c.min=50$ and $c.max=100$.
Also, we have that $c.N=g.N+h.N=2+2=4$.
Now we will compute the mean value by combining  the children mean values:
$c.\mu=\frac{g.N \cdot g.\mu + h.N \cdot h.\mu}{g.N+h.N}=\frac{2 \cdot 52.5+2 \cdot 90}{2+2}=71.3$.
Similarly,  for variance we have
\scalebox{0.89}{$c.\sigma^2 = \frac{g.N \cdot g.\sigma^2 + h.N\cdot h.\sigma^2 + 
g.N \cdot (g.\mu - c.\mu)^2 + h.N \cdot (h.\mu - c.\mu)^2 }{g.N+h.N} =$}\linebreak
$\frac{2 \cdot 6.25 + 2 \cdot 100 + 
2 \cdot (52.5- 71.3)^2 + 2 \cdot (90 - 71.3)^2 }{2+2} =404.7$.

The similar approach is also followed for the case of \hetR. 
\eoe
\end{myExample}

\stitle{Computational Analysis.}
Most of the well known statistics (e.g., mean, variance, skewness, etc.) can be computed  linearly w.r.t.\ the number of elements.
Therefore, the computation cost over a set of numeric values $S$
is considered  as $O(|S|)$.  
Assume a leaf node $n$ containing $k$ objects, then the cost for statistics computations for $n$ is $O(k)$. 
Also, the cost for all leaf nodes is $O(|D|)$.
Let an  internal node $n$,   then the cost for $n$ is $O(d)$; 
since the statistics in $n$ are computed by aggregating 
the statistics of the $d$ child nodes.
Considering that $\frac{d\cdot\ell-1}{d-1}$ is the maximum number  of internal nodes (Section~\ref{sec:HETree-C-Constr}), we have that in the worst case the cost for the internal nodes is $O(\frac{d^2 \cdot \ell-d}{d-1})$. 
Therefore, the overall cost for statistics computations over an \het 
is $O(|D| +\frac{d^2 \cdot \ell-d}{d-1})$.



\section{Efficient Multilevel Exploration}
\label{sec:ext}

In this section, we exploit the \het structure 
in order to  efficiently handle  different multilevel exploration scenarios. Essentially, we propose two methods for efficient hierarchical exploration over  large datasets.
The first method incrementally constructs the hierarchy via user interaction; the second one achieves dynamic adaptation of the data organization 
based on user's preferences.

\subsection{Exploration Scenarios}
\label{sec:resourceExpl}
In a typical multilevel exploration scenario, referred here as \textit{Basic
exploration scenario} (BSC), the user explores a dataset in a top-down fashion.
The user first obtains an overview of the data through the root level, 
and then drills down to more fine-grained contents for accessing the 
actual data objects  at the leaves. 
In BSC, the root of the hierarchy is the starting point of the 
exploration and, thus,    the first element to be presented (i.e., rendered).

The described scenario offers basic exploration capabilities; however it does 
not assume use cases with user-specified starting points, other than the root,
such as starting the exploration from a specific resource, or from a specific range of values.

Consider the following example, in which the user wishes to explore the \textit{DBpedia} infoboxes dataset
to find places with very large population. 
Initially, she selects the \textit{populationTotal} property and starts her exploration from the
root node, moves down the right part of the tree and ends up at the rightmost
leaf that contains the highly populated places. 
Then, she is interested in viewing the area size (i.e., \textit{areaTotal} property) for one of
the highly populated places and,  also, in exploring  places with similar area size. 
Finally, she decides to explore places based on the water area size (i.e., \textit{areaWater})
they contain. In this case, she prefers to start her exploration by considering
places that their water area size is within a given range of values.

In this example, besides BSC  one we consider two additional exploration scenarios. 
In the \textit{Resource-based exploration scenario}  ({RES}), the user specifies a resource of interest (e.g., an IRI) and a specific property;
the exploration starts from the leaf containing the specific resource and proceeds in a bottom-up fashion. 
Thus, in RES the data objects contained in the same leaf with the resource of
interest are presented first. We refer to that leaf as \textit{leaf of interest}. 

The third scenario, named \textit{Range-based exploration scenario} ({RAN}) 
enables the user to start her exploration from an arbitrary point in the hierarchy
providing a range of values; the user starts from a set of internal nodes and
she can then move up or down the hierarchy. 
The RAN scenario begins by rendering all sibling nodes that are children of the
node covering the specified range of interest; we refer to these   nodes as \textit{nodes of interest}. 


Note that, regarding the adopted rendering policy for all scenarios,  we only consider nodes belonging to the same level. 
That is, sibling nodes or data objects contained in the same leaf, are rendered.

Regarding the "navigation-related" \textit{operations}, the user can move down or up the hierarchy by performing 
a \textit{drill-down} or a \textit{roll-up} operation, respectively. 
A drill-down operation over a node $n$ enables the user to focus on $n$ and render its child nodes. 
If $n$ is a leaf node, the set of data objects  contained in $n$ are rendered. 
On the other hand, the user can perform a \textit{roll-up} operation on a set of sibling nodes $S$.
The parent node of $S$ along with the parent's sibling nodes are rendered.
Finally, the roll-up operation when applied to a set of data objects $O$ will render the leaf node
that contains $O$ along its sibling leaves, whereas a drill-down operation is not applied to a data object.


\subsection{Incremental \het Construction}
\label{sec:incrConstr}

In the Web of Data, the dataset   might be dynamically retrieved by a remote site (e.g., via a SPARQL endpoint), as a result, in all exploration scenarios, we have assumed that the \het is constructed on-the-fly at the time the user starts her exploration. In the previous DBpedia example,  the user explores three different properties; although only a small part of their hierarchy is accessed, the whole hierarchies are constructed and the statistics of all nodes are computed. Considering the recommended \het parameters for the employed properties,  this scenario requires that $29.5$K nodes will be constructed for \textit{populationTotal} property, $9.8$K nodes for the \textit{areaTotal} and $3.3$K nodes for the \textit{areaWater}, amounting to a total number of $42.6$K nodes. 
However, the construction of the hierarchies for large datasets poses a time overhead (as shown in the experimental section) and, consequently, increased response time in user exploration. 

In this section, we introduce {\ico} ({\textbf{I}ncremental \het \textbf{Co}nstruction}) method, which incrementally constructs the \het, based on user interaction. 
The proposed method goes beyond the incremental tree construction, aiming at further reducing the response time during the exploration process by "pre-constructing"  (i.e., prefetching) the parts of the tree that will be visited by the user in her next roll-up or drill-down operation. Hence, a node $n$ is not constructed when the user visits it for the first time; instead, it has been constructed in a previous exploration step, 
where the user was on  a node in which $n$ can be reached by  a roll-up or a drill-down operation. 
This way, our method offers  incremental construction of the tree, tailored to each user's exploration.
Finally,  we show that, during an exploration scenario, \ico constructs the minimum number of \het elements.

\begin{figure*}[t] 
\centering
\includegraphics[scale=0.645]{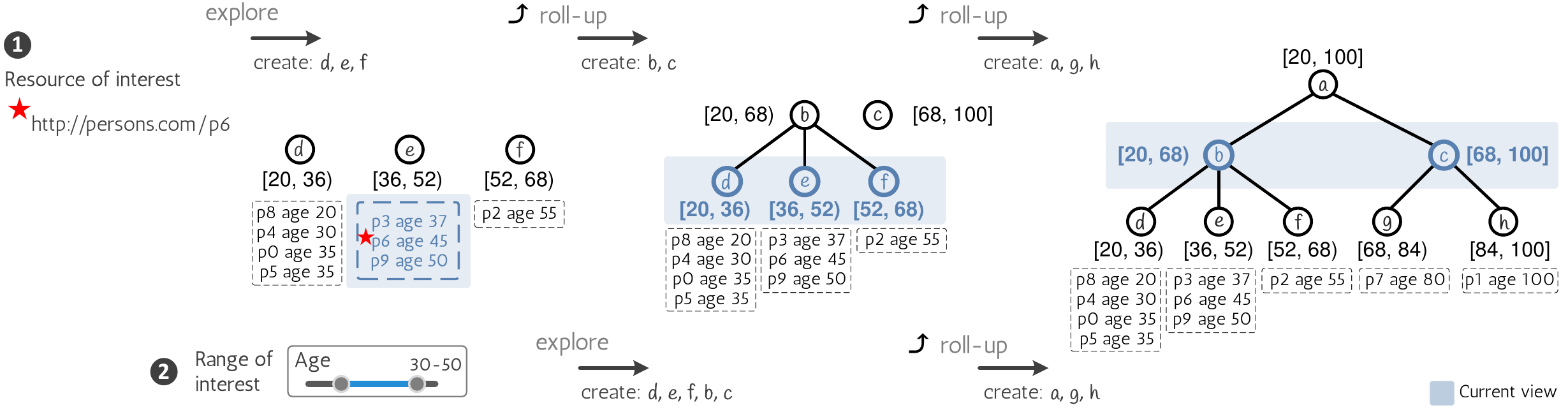}
\caption{Incremental \het construction example. {\normalsize\ding{202}} Resource-based (RES) exploration scenario;  {\normalsize\ding{203}}  Range-based (RAN) exploration scenario }
\label{fig:inter}
\end{figure*}

Employing \ico method in the DBpedia example, the \textit{populationTotal} hierarchy will only construct $76$ nodes  {(the root along its child nodes and $9$ nodes in each of the lower tree levels)} and
 the \textit{areaTotal} will construct $3$ nodes corresponding to the leaf node containing the requested resource and its siblings. Finally, the \textit{areaWater} hierarchy {initially will contain either $6$ or $15$ nodes,
 depending on whether the user's input range corresponds to a set of sibling leaf nodes, or to a set of sibling internal  nodes, respectively.}
 

\begin{myExample}
We   demonstrate the  functionality of \ico through the following example. Assume the dataset used in our running examples, describing persons and their ages.
Figure~\ref{fig:inter} presents the incremental construction of the \het presented in Figure~\ref{fig:tree-r} for the RES and RAN exploration scenarios. Blue color is used to indicate the \het elements that are presented (rendered) to the user, in each exploration stage.

In the RES scenario (upper flow in Figure~\ref{fig:inter}), the user specifies "http://persons.com/p6" as her resource of interest; all data objects contained in the same leaf (i.e., $e$) with the resource of interest are initially presented to the user. The \ico initially constructs the leaf $e$, along with its siblings, i.e., leaves $d$ and $f$. These leaves correspond to the nodes that the user can reach in a next (roll-up) step. Next, the user rolls up and the leaves $d$, $e$ and $f$ are presented to her. At the same time, parent node $b$ and its sibling $c$ are constructed. Note that all elements which are accessible to the user by moving either down (i.e., $d$, $e$, $f$ data objects), or up (i.e., $b$, $c$ nodes) are already constructed.  Finally, when the user rolls up $b$ and $c$ nodes are rendered and parent node $a$, along with the children of $c$, i.e., $g$ and $h$, are constructed.

In the RAN scenario (lower flow in Figure~\ref{fig:inter}), the user specifies [$30$, $50$] as her range of interest. The nodes covering this range (i.e., $d$, $e$) are initially presented along with their sibling $f$. Also, \ico constructs the parent node $b$ and its sibling $c$ because they are accessible by one exploration step. Then, the user performs a roll-up and \ico constructs the $a$, $g$, $h$ nodes (as described in the RES scenario above).
\eoe
\end{myExample}

In the beginning of each exploration scenario, 
\ico constructs a set of \textit{initial nodes}, which are the nodes initially
presented,  as well as the nodes potentially reached by the user's first operation (i.e., required \het elements).
The \textit{required \het elements} of an exploration step are nodes that can be reached by the user by performing one exploration operation. 
Hence, in the RES scenario, the initial nodes are the leaf of interest and its sibling leaves.
In the RAN, the initial nodes are the nodes of interest, their children, and their parent node along with its siblings. 
Finally, in the BSC scenario the initial nodes are the root node and its children. 

In what follows we describe the \textit{construction rules} adopted by \ico through the user exploration process. 
These rules provide the correspondences between the types of elements presented   in each exploration step and the elements that \ico constructs. 
Note that these rules are applied after the construction of the initial nodes, in all three exploration scenarios. The correctness of these rules is verified later in Proposition~\ref{lem:ico}. 

\vspace{3pt}
\noindent
\textbf{Rule~1:} \textit{If a set of internal sibling nodes $C$ is presented}, \ico constructs:
$(i)$ the parent node of $C$ along with the parent's siblings,
and 
$(ii)$ the children of each node in $C$.

\vspace{3pt}
\noindent
\textbf{Rule~2:}  \textit{If a set of leaf sibling  nodes $L$ is presented}, \ico does not construct anything (the required nodes have been previously {constructed}).

\vspace{3pt}
\noindent
\textbf{Rule~3:} \textit{If a set of data objects $O$ is presented}, \ico does not construct anything (the required nodes have been previously {constructed}).
\vspace{1mm}

The following proposition shows that, in all case, the required
\het elements have been constructed earlier by \ico.%
\footnote{Proofs are included in Appendix~\ref{ap:interConstr}.}

\begin{myproposition}
\label{lem:ico}
In any exploration scenario, the  \het elements  a user can reach by performing  one operation (i.e., required elements), have been previously constructed by \ico.
\end{myproposition}


Also, the following theorem shows that 
over any exploration scenario \ico constructs only the required \het elements.

\begin{mytheor}
\ico constructs the minimum
number of  \het elements in any exploration scenario. 
\end{mytheor}

\app{
\begin{proof}
As described in the proof  of Lemma~\ref{lem:ico}, 
in each case, \ico constructs only the required \het elements. 
Therefore, in  any case \ico constructs the minimum number of  required \het elements. 
\end{proof}
 }

%

\subsubsection{\ico Algorithm}

In this section, we present the incremental \het  construction algorithm.
Note that, here we include the pseudocode only  for the \hetR version, 
since the only difference with the \hetC version  is in the way that the nodes' intervals are computed and that the dataset is initially sorted. 
In the analysis of the algorithms, both versions are studied.

Here, we assume that each node $n$
contains the following extra fields.
Let a node $n$, $n.p$ denotes the parent node of $n$, and $n.h$ denotes the height of $n$ in the hierarchy.
Additionally, given a dataset $D$, $D.minv$ and $D.maxv$ denote the minimum and the maximum value for all objects in $D$, respectively.
The user preferences regarding the exploration's starting point are represented as an interval $U$.
In the RES scenario, given that the value of the explored property for the resource of interest is $o$, we have $U^- = U^+ = o$.
In the RAN scenario, given that the range of interest is $R$, we have that
$U^-=\max{(D.minv, R^-)}$ and $U^+=\min{(D.maxv, R^+)}$.
In the BSC scenario, the user does not provide any preferences regarding the starting point, so we have   $U^-=D.minv$ and $U^+=D.maxv$.
Finally, according to the definition of \het, a node $n$ \textit{encloses} a  data object (i.e., triple) $tr$ if $n.I^- \geq tr.o$ and $n.I^+ \leq tr.o$.

 

The algorithm $\mathsf{\ico\text{-}R}$ (Algorithm~\ref{algo:ico})
implements the incremental method for \hetR.
The algorithm uses two procedures to construct all required  
nodes  ({available in Appendix~\ref{app:icoapp}}). The first procedure $\mathsf{constrRollUp\text{-}R}$ (\hyperref[proc:proc4]{Procedure~4}) 
constructs the nodes which can be reached by a roll-up operation, whereas 
$\mathsf{constrDrillDown\text{-}R}$  (\hyperref[proc:proc5]{Procedure~5})
constructs the nodes which can be reached by a drill-down operation. 
Additionally, the aforementioned procedures exploit two secondary procedures ({Appendix~\ref{app:icoapp}}): 
$\mathsf{computeSiblingInterv\text{-}R}$ (\hyperref[proc:proc6]{Procedure~6}) and 
$\mathsf{constrSiblingNodes\text{-}R}$ (\hyperref[proc:proc7]{Procedure~7}), which are used for nodes' intervals computations and nodes construction.

\begin{algorithm2e}[t!]
\footnotesize
 \caption{\ico-R($D$, $\ell$, $d$, $U$,  $cur$, $H$)}
\label{algo:ico}
\KwIn{$D$:   set of objects; $\ell$: number of leaf nodes; 
$d$: tree degree;   
$U$: interval representing user's starting point;\linebreak
$cur$:   currently presented elements;\linebreak
 $H$: currently created \hetR}
\KwOut{$H$:  updated \hetR}
\KwVar{
$len$:  the length of the leaf's interval 
 }

\vspace{1mm}



\If(  \Comment*[f]{\mycomment{{first \ico call }}}){$cur = \mynull$ }{   
 $len \gets \frac{D.maxv-D.minv}{\ell}$\;   
from $U$ compute $I_0$, $h_0$
\Comment*[f]{\mycomment{{used for constructing initial nodes}}}
$cur\,, H \gets	 \mathsf{constrSiblingNodes\text{-}R}(I_0, \mynull, D, h_0)$ \;
	\lIf{\textup{{RES}} }{ \Return $H$}
}
\If{$cur[1].p = \mynull$  \Logand $D \neq \varnothing$}{   
	$H \gets	 \mathsf{constrRollUp\text{-}R}(D, d, cur, H)$	\;
	\If(  \Comment*[f]{\mycomment{{$cur$ are not leaves}}}) {$cur[1].h>0$} {   
		$H \gets	 \mathsf{constrDrillDown\text{-}R}(D, d, cur, H)$ \;
	}
}
\Return $H$\;
\end{algorithm2e}

The $\mathsf{\ico\text{-}R}$ algorithm is invoked at the beginning of the exploration scenario, in order to construct the initial nodes, as well as every time the user performs an operation.
The algorithm takes as input the dataset $D$, the tree parameters $d$ and $\ell$, 
the starting point $U$, 
the currently presented (i.e., rendered) elements $cur$, and the constructed \het $H$. $\mathsf{\ico\text{-}R}$ begins with the currently presented elements $cur$ equal to $null$ (\textit{lines~1-5}).
Based on the starting point $U$, the algorithm computes the interval $I_0$ corresponding to the sibling nodes that are first presented to the user, as well as its hierarchy height $h_0$ (\textit{line~3}).
For sake of simplicity, the details for computing $I_0$ and $h_0$ are omitted.
For example, the interval $I$ for the leaf that contains the resource of interest with object value $o$, 
is computed as 
$I^- = D.minv + len \cdot \floor*{\frac{o - D.minv}{len}}$ and 
$I^+ = \min(D.maxv, I^- + len)$.
Following a similar approach, we can easily compute $I_0$ and $h_0$.

Based on $I_0$, the algorithm constructs the sibling nodes that are first presented to the user (\textit{line~4}). 
Then, the algorithm constructs the rest initial nodes (\textit{lines~6-9}).
In the RES case, as $I_0$  we consider the interval that includes the leaf that contains the resource of interest along with its sibling leaves. 
Hence,  all the initial nodes are constructed in {line~4} and 
the algorithm  terminates (\textit{line~5}) until the next user's operation.

After the first call, in each 
\ico  execution, the algorithm initially checks if the parent node  of the currently presented elements is already constructed, 
or if all the nodes that enclose data objects%
\footnote{Note that in the \hetR version, we may have nodes that do not enclose any data objects.}
have been constructed (\textit{line~6}).
Then, procedure $\mathsf{constrRollUp\text{-}R}$ (\textit{line~7}) is used to construct the $cur$ parent node, as well as the parent's siblings. In the case that $cur$ are not leaf nodes or data objects (\textit{line~8}), procedure
$\mathsf{constrDrillDown\text{-}R}$ (\textit{line~9}) is used to construct
all $cur$ children.
Finally, the algorithm returns the updated \het (\textit{line~10}).

\subsubsection{Computational Analysis} 
\label{sec:icoanal}
Here we analyse the incremental construction for both \het versions.

\stitle{Number~of~Constructed~Nodes.}  
Regarding the number of \textit{initial nodes} constructed in each scenario: in RES scenario, at most $d$ leaf nodes are constructed; in RAN scenario, at most $2d+d^2$ nodes are constructed; finally in BSC scenario, $d+1$ are constructed.

Regarding the maximum number of nodes constructed \textit{in each operation} in RES and RAN scenarios:
(1) A \textit{roll-up operation}
constructs at most $d+d\cdot(d-1)=d^2$ nodes. 
The $d$ nodes are constructed in $\mathsf{constrRollUp}$, 
whereas the ${d\cdot(d-1)}$ in $\mathsf{constrDrillDown}$.
(2)  A \textit{drill-down operation} constructs at most $d^2$ nodes 
in $\mathsf{constrDrillDown}$.
As for the BSC scenario: 
(1) A \textit{roll-up operation} does not construct any nodes. 
(2)  A \textit{drill-down operation} constructs at most $d^2$ nodes 
in $\mathsf{constrDrillDown}$.

\stitle{Discussion.}
 The worst case for the computational cost is higher in \hetR than in \hetC, for all exploration scenarios.
Particularly, in \hetR worst case, \ico must build leaves that contain the whole dataset and the computational cost is $O(|D|log|D|)$ for all scenarios.
In \hetC, for the RES and RAN  scenarios, the cost is $O(d^2+\frac{d-1}{d}|D|)$, 
and for the BSC scenario the cost is $O(d^2+|D|)$.  A detailed computational analysis for both \hetR and  \hetC is included in Appendix~\ref{app:icocomp}.

%
%
%
%
%
%
%

  
\begin{table*}[!t]
\vspace{-1mm}
\tiny
\centering
\caption{Summary of {Adaptive \het Construction}$^\star$}
\label{tab:reconstr}
 \setlength{\tabcolsep}{3.3pt}
\begin{tabular}{lccccccccc}
\tline
\multicolumn{1}{c}{} & \multicolumn{1}{c}{}  &   \multicolumn{4}{c}{\textbf{Modify Degree}}  & \multicolumn{4}{c}{\textbf{Modify Num.\ of Leaves}}  \\ \cmidrule[0.6pt](lr){3-6} \cmidrule[0.6pt](lr){7-10}
 \multicolumn{1}{c}{} & \multicolumn{1}{c}{\textbf{Full Construction}} & $d' =d^k$ & $d' =k \cdot d$ &  $d' = \sqrt[k]d$ & \multicolumn{1}{c}{\textit{elsewhere}} & $\ell' >\ell$ & $\ell'=  \dfrac{\ell}{d^k}$ &  $\ell'=  \dfrac{\ell}{k}$  & $\ell'=\ell-k$  \\
\dlineB
\rowcolor{gray!25}
 \multicolumn{10}{l}{\textbf{\hspace{0mm}Tree Construction}} \\
\hspace{1mm}\textbf{Complexity} 
& $O(m log m\!+ \!d' e)$ &$O(m  log_{\sqrt[k]{d'}} m)$    &   $O(d'e)$ &  $O(d'^k r)$  &  $O(d'e)$   &  $O(m\!+\!d'e)$   &   $O(m)$    &  $O(m\!+\!d'e)$  & $O(m log m\!+\! d' e)$ \\
 	\rowcolor{gray!10}																				
\hspace{1mm}\#\textbf{leaves}\subscript{0} 
 & $\ell'$ &  $0$   & $0$    &    $0$&   $0$& $\ell'$    & $0$  &    0 &$\ell'$ \\
\hspace{1mm}\#\textbf{leaves}\subscript{+}
 & $0$ &    $0$&    $0$ &    $0$& $0$   &   $0$ & $\ell'$ &$\ell'$ &   0\\
  	\rowcolor{gray!10}																				
\hspace{1mm}\#\textbf{internals}\subscript{0} &
  $e$& $0$    &  $e$  &    $e-r$  &  $e$ &    $e$ &$0$   &  $e$ &$e$\\
\hspace{1mm}\#\textbf{internals}\subscript{+} 
& $0$ &   $0$ & $0$   &   $0$ &0   &    0&  0    &0 &0\\
\boldline
\rowcolor{gray!25}
 \multicolumn{10}{l}{\hspace{0mm}\textbf{Statistics Computations}} \\
\hspace{1mm}\textbf{Complexity} 
&$O(m \!+\! d'e)$ &   $O(1)$ & $O(\frac{k\ell'}{d'}\!+\!d'e)$    & $O(d'(e\!-\!r))$    & $O(d'e)$  & $O(m\!+\!d'e)$   & $O(1)$   & $O(m\!+\!d'e)$   &  $O(m \!+ \!d'e\!-\!\ell'\!-\!k)$\\
 	\rowcolor{gray!10}																				
\hspace{1mm}\#\textbf{leaves}\subscript{0} 
 & $\ell'$&    $0$&$0$    &  $0$  &  0 & $\ell'$   &  0  &  0  & \:\:\:$\ell'$ $-$ \scalebox{0.95}{$\frac{\ell'^2}{d'}$}\\
\hspace{1mm}\#\textbf{leaves}\subscript{+} 
& $0$ &    $0$ &    $0$&     $0$&   0&  0  &   0& $\ell'$&   \scalebox{0.95}{$\frac{\ell'^2}{d'}$}\\
 	\rowcolor{gray!10}																				
\hspace{1mm}\#\textbf{internals}\subscript{0} 
& $e$ &    $0$ &\: \:$e$ $-$ \scalebox{0.85}{$\ceil*{\frac{\ell'}{d'}}$}    &$e-r$    &$e$  & $e$  &   0 &  $e$ & $e$\\
\hspace{1mm}\#\textbf{internals}\subscript{+}
 & $0$ &    $0$ & \scalebox{0.85}{$\ceil*{\frac{\ell'}{d'}}$}   & 0    &0   &0    &      0& 0&0\\
\bline
\end{tabular}
\vspace{-3px}
{\begin{flushleft} \hspace{0.5cm} 
$^\star m=|D|$,     $e=\frac{d'\ell'-1}{d'-1}$ (maximum number of internal nodes), and $r=\frac{d'^k\ell'-1}{d'^k-1}$
\end{flushleft}}
\vspace{0mm}
 \end{table*}

\subsection{Adaptive \het Construction}
 \label{sec:ReConstr}
In a (visual) exploration 
scenario, users wish to modify the    organization of the data by providing user-specific preferences for the whole hierarchy or   part of it. The user can select a specific subtree  and alter 
the number of groups presented in each level (i.e., the tree degree) 
or the size of the groups (i.e., number of leaves).  
In this case, a new tree (or a part of it) pertaining to the new parameters provided by the user should be constructed {on-the-fly}.  

\begin{figure}[t] 
\centering
\includegraphics[scale=0.68]{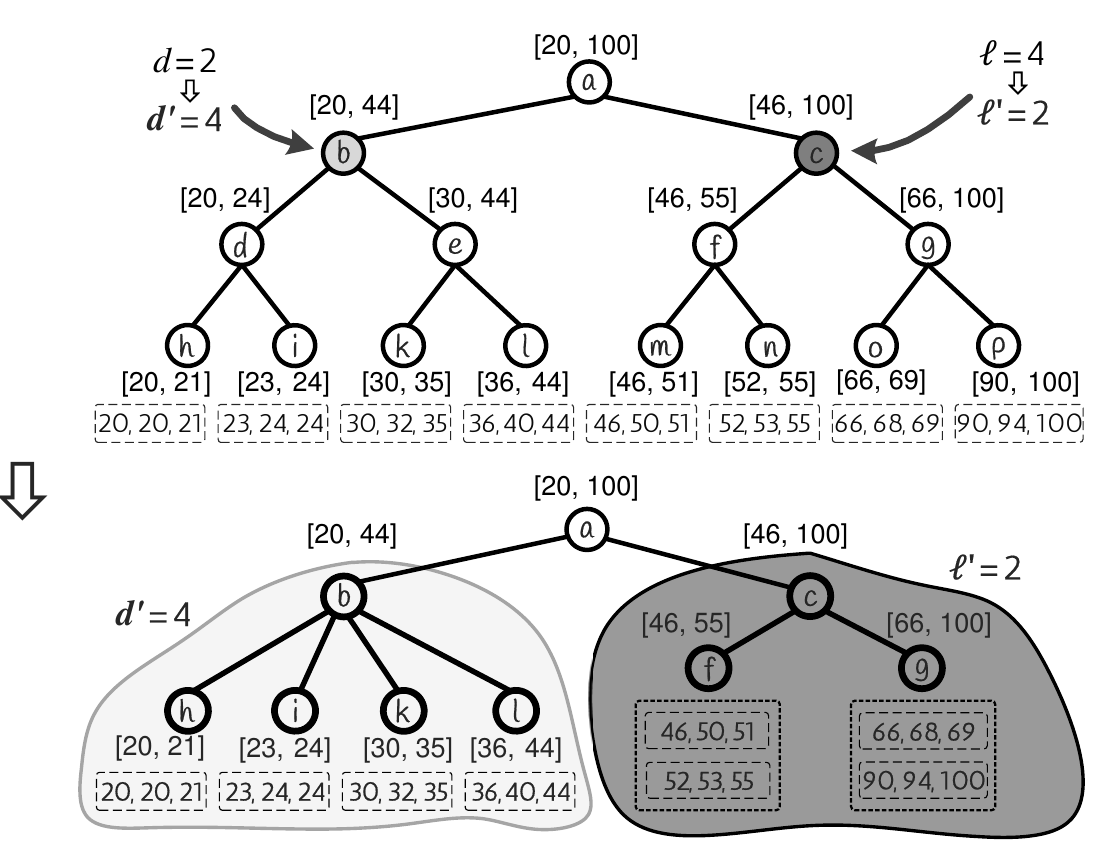}
\caption{Adaptive \het example}
\label{fig:recon_ex}
\end{figure}

For example, consider the \hetC of  Figure~\ref{fig:recon_ex} representing ages of persons.\footnote{For simplicity,  Figure~\ref{fig:recon_ex}  presents only the values of the objects.}
A user may navigate to node $b$, where she prefers to increase the number of groups presented in each level. 
Thus, she modifies the degree of $b$ from $2$ to  $4$ and the subtree is adapted to the new parameter as depicted on the bottom tree of Figure~\ref{fig:recon_ex}. 
On the other hand, the user prefers exploring the right subtree (starting from node $c$) with less details. She chooses to increase the size of the groups by reducing (from $4$ to $2$) the number of leaves for the subtree of $c$.  
In both cases, constructing the subtree from scratch based on the user-provided parameters and recomputing statistics  entails a significant time overhead, especially, when user preferences are applied to a large part of or the whole hierarchy.

In this section, we introduce \ada (\textbf{Ada}ptive \het Construction) method, which dynamically adapts an existing  \het to a new, considering a  set of user-defined parameters. 
Instead of both constructing the tree and computing the nodes' statistics  from scratch, our method reconstructs the new part(s) of the hierarchy by exploiting the existing elements (i.e., nodes, statistics)  of the tree. In this way, \ada achieves to reduce the overall construction cost and enables the on-the-fly reorganization of the visualized data.  In the example of  Figure~\ref{fig:recon_ex}, the new subtree of $b$ can be derived from the old one, just by removing the internal nodes $d$ and $e$, while the new subtree of $c$ results from merging leaves together and aggregating their statistics. 

Let $\T (D, \ell, d)$ denote the existing  \het and  $\T' (D, \ell', d')$ is the new \het corresponding to the new user preferences for the tree degree $d'$ and the number of leaves $\ell'$. Note that  $\T$ could also denote a subtree of an existing \het (in the scenario where the user modifies only a part of it). In this case, the user indicates the \textit{reconstruction root} of $\T$.  

 Then, \ada identifies the following  elements of $\T$:
(1) The elements of $\T$ that also exist in $\T'$.
For example, consider the following two cases: 
the leaf nodes of $\T'$ are internal nodes of $\T$ in  level $x$; 
the statistics of $\T'$ nodes  in  level $x$ are equal to the statistics of $\T$ nodes  in  level $y$.
(2) The elements of $\T$  that 
can be reused (as "building blocks") for constructing elements in $\T'$.
For example, consider the following two cases:  each leaf node of $\T'$ is constructed by merging $x$ leaf nodes of $\T$; 
the statistics for the node $n$ of $\T'$ can be computed by aggregating the statistics from the nodes $q$ and $w$ of $\T$.

Consequently, we consider that an element (i.e., node or node's statistics)
in $\T'$ can be:   
(1) constructed/computed from scratch%
\footnote{Note that  it is possible for a from scratch constructed node in $\T'$ to aggregate statistics from nodes in $\T$.}, 
(2) reused as is from $\T$
or (3) derived by aggregating elements from $\T$.

Table~\ref{tab:reconstr} summarizes the \ada reconstruction process. 
Particularly, the table includes: (1) the computational complexity for constructing $\T'$, denoted as $Complexity$; 
(2) the number of leaves and internal nodes of $\T'$ constructed from scratch, denoted as
\#$leaves_0$ and \#$internals_0$, respectively; and (3)  
the number of leaves and internal nodes of $\T'$ derived from nodes of $\T$, denoted as  
\#$leaves_+$ and \#$internals_+$, respectively. 
The lower part of the table presents the results for the computation of node statistics in $\T'$.
Finally, the second table column, denoted  as $Full~Construction$, presents the results of constructing $\T'$ from scratch.

The following example demonstrates the  \ada results, 
 considering a DBpedia exploration scenario. 
 
\begin{myExample}
The user   explores the \textit{populationTotal} property of the DBpedia dataset. The default system organization for this property is a hierarchy with degree $3$. The user modifies the tree parameters in order to fit better visualization results as following. 
First,  she decides to render more groups in each hierarchy level and increases the degree from $3$ to $9$ ($1$st \textit{Modification}).
Then, she observes that the results overflow the visualization area and that a smaller degree fits better; thus she re-adjusts the tree degree to a value of $6$ ($2$nd \textit{Modification}).
Finally, she navigates through the data values and decides to increase the groups' size by a factor of three (i.e., dividing by three the number of leaves) ($3$rd \textit{Modification}). Again, she corrects her decision and readjusts the final group size to twice the default size ($4$th \textit{Modification}).

Table~\ref{tab:reconstrEx} summarizes the number of nodes, constructed by a
\textit{Full Construction} and \ada in each modification, along with the required statistics computations. 
Considering the whole set of modifications, \ada constructs only the $22\%$ ($15.4$K vs.\ $70.2$K) of the nodes that are created in the case of the full construction. 
Also, \ada computes the statistics for only $8\%$ ($5.6$K vs.\ $70.2$K) of the nodes. 
\eoe
\end{myExample}

\begin{table}[!h]
\vspace{-4mm}
\tiny
\centering
\caption{Full Construction  vs.\ \ada  over DBpedia Exploration \linebreak Scenario {\tiny (cells values: Full / \ada)}}
\label{tab:reconstrEx}
 \setlength{\tabcolsep}{3.50pt}
\begin{tabular}{l cc cc }
\tline
\multicolumn{1}{c}{} &   \multicolumn{2}{c}{\textbf{Modify Degree}}  & \multicolumn{2}{c}{\textbf{Modify Num.\ of Leaves}}  \\
 \multicolumn{1}{c}{} & 1st Modification  &  2nd Modification   &3rd   Modification  & 4th   Modification \\
\dlineB \vspace{-1.0mm}\\ \rowcolor{gray!25}

 \multicolumn{5}{l}{\textbf{\hspace{-1mm}Tree Construction}} \vspace{3pt} \\ 
\textbf{\hspace{1mm}\#nodes} & 
$22.1$K  {\boldmath{$/$}} $0$  & 
$23.6$K {\boldmath{$/$}} $3.9$K  & 
$9.8$K {\boldmath{$/$}} $6.6$K & 
$14.7$K {\boldmath{$/$}} $4.9$K \vspace{5pt} \\
\vspace{-5px}\\ \rowcolor{gray!25}
 \multicolumn{5}{l}{\textbf{\hspace{-1mm} Statistics Computations}} \vspace{3pt} \\
\textbf{\hspace{1mm}\#nodes} &
$22.1$K  {\boldmath{$/$}} $0$ &  
$23.6$K {\boldmath{$/$}} $659$ & 
$9.8$K {\boldmath{$/$}}  $0$ &  
$14.7$K {\boldmath{$/$}} $4.9$K \vspace{3pt}\\
\bline
\end{tabular}
 \end{table}

\begin{figure*}[!t] 
\centering
\includegraphics[scale=0.61]{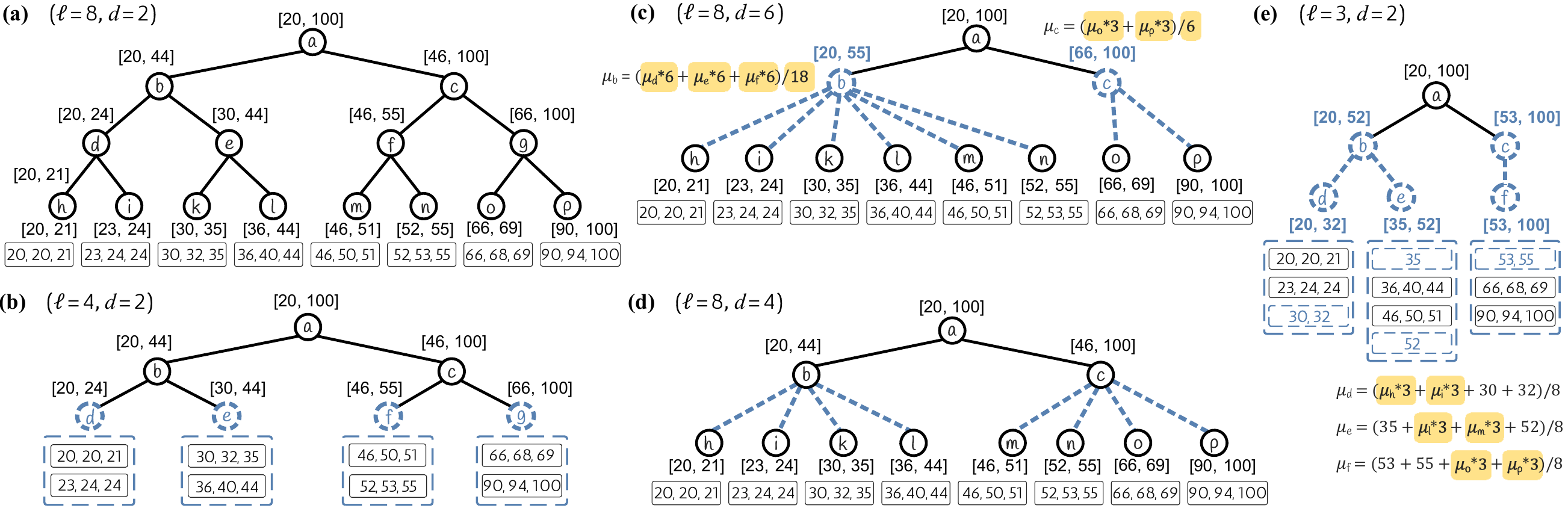}
\vspace{-4mm}
\caption{Adaptive \het construction examples}
\label{fig:reconstr}
 \end{figure*}

 In the next sections, we present in detail the  {reconstruction} process through the example trees of Figure~\ref{fig:reconstr}. 
 Figure~\ref{fig:reconstr}a presents the initial tree $\T$ that is an \hetC, 
with $\ell=8$ and $d=2$. Figures~\ref{fig:reconstr}b\textasciitilde \ref{fig:reconstr}e  present several reconstructed trees $\T'$.
Blue dashed lines are used to indicate the elements (i.e., nodes, edges)  of  $\T'$  which do not exist in $\T$.
Regarding statistics, we assume that in each node we compute the mean value. 
In each $\T'$,  we present only the mean values that are not known from $\T$. 
Also, in mean values computations,  the values that are reused from $\T$
are highlighted in yellow. All reconstruction details  and  computational analysis for each case are included in Appendix~\ref{ap:reconstr}.

\vspace{3mm}
\subsubsection{The User Modifies the Tree Degree} 
 \label{sec:reconstrDegree}
Regarding the modification of the degree parameter, we distinguish the following cases:

\vspace{2mm}
\stitle{The user increases the tree degree.}   
We have that ${d' > d}$;  based on the $d'$ value we  have the following cases:

\vspace{2mm}
\noindent 
$(1)$ $d' =d^k$, with $k \in \mathbb{N}^+$ and $k>1$:
Figure~\ref{fig:reconstr}a presents $\T$ with $d=2$
and Figure~\ref{fig:reconstr}d presents the reconstructed  $\T'$ with $d'=4$ (i.e., $k=2$).
  $\T'$ results by simply removing the nodes with height 1 (i.e., $d$, $e$, $f$, $g$) 
and connecting the nodes with height 2 (i.e., $b$, $c$) with the leaves.  

In general, $\T'$ results from $\T$ by simply removing tree levels from $\T$.
Additionally,  there is no need for computing any new statistics, since 
the statistics for all  nodes of $\T'$  remain the same as in $\T$.

\vspace{2mm}
\noindent 
$(2)$ $d' = k \cdot d$, with $k  \in \mathbb{N}^+$, $k > 1$ and $k \neq d^\nu$ where $\nu  \in \mathbb{N}^+$:
An example with $k=3$ 
is presented in  Figure~\ref{fig:reconstr}c, where we have $d'=6$.
In this case, the leaves of $\T$ (Figure~\ref{fig:reconstr}a) remain leaves in $\T'$ and all  internal nodes up to the {reconstruction root} of $\T$ are constructed from scratch.
As for the node statistics,  we can compute the mean values for $\T'$ nodes with height 1 (i.e., $\mu_b$, $\mu_c$) by aggregating already computed mean values (e.g., $\mu_d,$ $\mu_e$, etc.) from $\T$. 

In general, except for the leaves, we construct all internal nodes from scratch. For the internal nodes of height 1, we compute their statistics by aggregating the statistics of $\T$ leaves, whereas for internal nodes of height greater than 1, we compute from scratch their statistics.

\vspace{2mm}
\noindent 
$(3)$ $elsewhere$:
In any other case where the user increases the tree degree, all internal nodes in $\T'$ except for the leaves are constructed from scratch. In contrast with the previous case, the leaves' statistics from $\T$ can not be reused and, thus, for all internal nodes in $\T'$ the statistics are recomputed.

\vspace{2mm}
\stitle{The user decreases the tree degree.}  
Here we have that $d' < d$; based on the $d'$ value we have the following two cases:

\vspace{2mm}
\noindent 
$(1)$ $d' = \sqrt[k]d$, with $k \in \mathbb{N}^+$  and $k>1$:
Assume  that now Figure~\ref{fig:reconstr}d depicts $\T$, with $d=4$, 
while Figure~\ref{fig:reconstr}a presents $\T'$  with $d'=2$.
We can observe that $\T'$ contains all nodes of $\T$, 
as well as a set of extra internal nodes (i.e., $d$, $e$, $f$, $g$). 
Hence, $\T'$ results from $\T$ by constructing some new internal nodes.

\vspace{2mm}
\noindent 
$(2)$  $elsewhere$:
This case is the same as the previous case $(3)$ where the user increases the tree degree. 

\vspace{3mm}
\subsubsection{The User Modifies the Number of Leaves}
  \label{sec:reconstrLeaves}
Regarding the modification of the number of leaves parameter, we distinguish the following cases: 

\vspace{2mm}
\stitle{The user increases the number of leaves.} 
In this case we have that $\ell' > \ell$; hence, 
each leaf of $\T$ is split into several leaves in $\T'$ and the data objects  contained in a $\T$ leaf 
must be reallocated to the new leaves in $\T'$.
As a result,  all nodes (both leaves and internal nodes) in $\T'$ have different contents compared to nodes in $\T$ and must be constructed from scratch along with their statistics.

In this case, constructing $\T'$  requires \linebreak${O(|D|+\frac{d^2 \cdot \ell'-d}{d-1})}$ (by avoiding the sorting phase).  
 
\vspace{2mm}
\stitle{The user decreases the number of leaves.} 
In this case we have that $\ell' < \ell$; 
based on the $\ell'$ value we have the following three cases:

\vspace{2mm} 
\noindent 
$(1)$ $\ell'=  \dfrac{\ell}{d^k} $, with $k \in \mathbb{N}^+$:
Considering that Figure~\ref{fig:reconstr}a presents $\T$ with $\ell=8$ and $d=2$.
A  reconstruction example of this case with $k=1$,  is presented in Figure~\ref{fig:reconstr}b, 
where we have $\T'$ with $\ell'=4$. 
In  Figure~\ref{fig:reconstr}b, 
we observe that the leaves in $\T'$ result from merging $d^k$ leaves of $\T$. 
For example, the leaf $d$ of $\T'$ results from merging the leaves $h$ and $i$ of $\T$. 
Then, $\T'$ results from $\T$, by replacing the $\T$ nodes with height   $k$ (i.e., $b$, $e$, $f$, $g$), with the $\T'$ leaves. 
Finally, the nodes of $\T$ with height less than $k$ are not included in $\T'$.

Therefore, in this case, $\T'$ is constructed by merging the leaves of $\T'$ 
and removing the internal nodes of $\T'$ having height less or equal to $k$.
Also, we do not recompute the statistics of the new leaves of $\T'$ as these are derived from 
the statistics of the removed nodes with height   $k$.

%

\vspace{2mm}
\noindent 
$(2)$ $\ell'=  \dfrac{\ell}{k} $, with $k \in \mathbb{N}^+$, $k>1$ and $k \neq d^\nu$, where  $\nu \in \mathbb{N}^+$:
As in the previous case, the leaves in $\T'$ are constructed  by merging leaves
from $\T$ and their statistics are computed based on the statistics of the merged leaves. In this case, however, all internal nodes in $\T'$ have to be constructed from scratch.

\vspace{2mm}
\noindent 
 $(3)$ $\ell'=\ell-k$, with  $k \in \mathbb{N}^+$, $k>1$ and ${\ell' \neq \dfrac{\ell}{\nu}}$, where $\nu \in  \mathbb{N}^+$:
The two previous cases describe that each leaf in $\T'$  
$fully$ contains $k$  leaves from $\T$.
In this case, a leaf in $\T'$ 
may \textit{partially} contains  leaves from $\T$. 
A leaf in $\T'$ fully contains a leaf from $\T$ when the $\T'$ leaf contains all data objects belonging to the $\T$ leaf.
Otherwise, a leaf in $\T'$ partially contain a leaf from $\T$ when the $\T'$ leaf  contains a subset of the data objects from the $\T$ leaf.

An example of this case is shown in Figure~\ref{fig:reconstr}e that depicts a reconstructed $\T'$ resulting from the $\T$ presented in Figure~\ref{fig:reconstr}a. 
The $d$ leaf of $\T'$  fully contains   leaves $h$, $i$ of $\T$ 
and partially   leaf $k$ for which   value $35$  belongs to a different leaf (i.e., $e$). 

Due to this partial containment, we have to construct all leaves and internal nodes from scratch and recalculate their statistics. 
Still, the statistics of the fully contained  leaves of $\T$ can be reused, by aggregating them with the individual values of the data objects included in the leaves.
For example, as we can see in  Figure~\ref{fig:reconstr}e, the mean value $\mu_d$ of the  leaf $d$
is computed by aggregating the mean values $\mu_h$  and $\mu_i$ corresponding
to the fully contained  leaves $h$ and $i$, with the individual values $30$, $32$ of the partially contained leaf $k$.


%




\begin{figure*}[!thb] 
\centering
\includegraphics[scale=0.39]{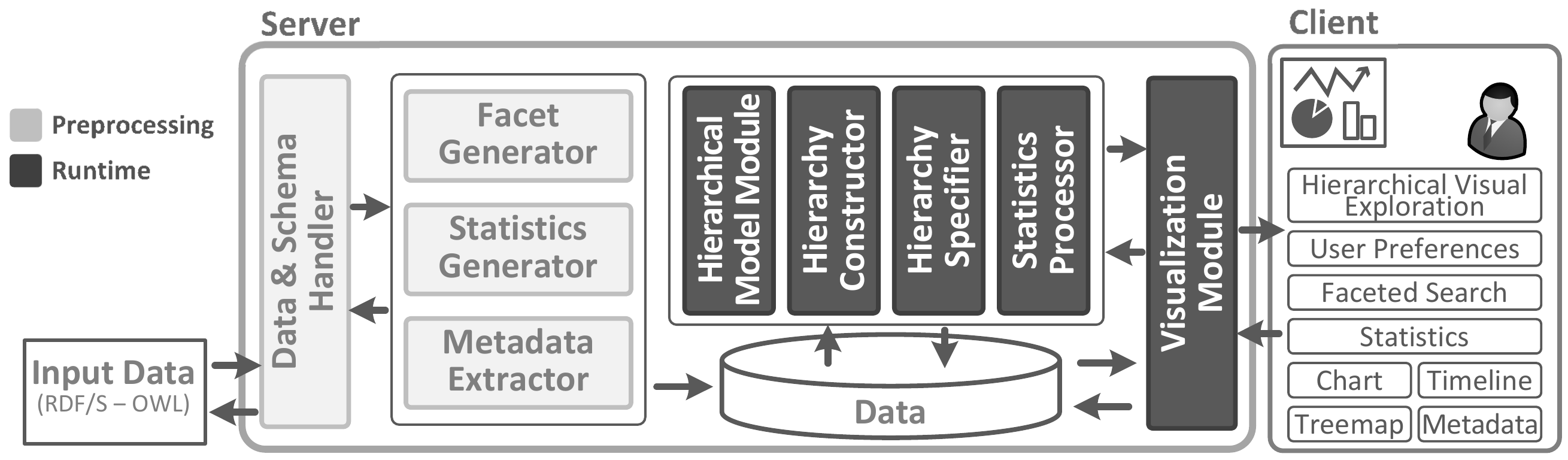}
\caption{System architecture}
\label{fig:arch}
\end{figure*}

\section {The SynopsViz Tool }
\label{sec:system}

Based on the proposed hierarchical model, we have developed a web-based prototype called \textit{SynopsViz}%
\footnote{\href{http://synopsviz.imis.athena-innovation.gr}{{synopsviz.imis.athena-innovation.gr}}}. 
The key features of SynopsViz  are summarized as follows:
(1) It supports the aforementioned \textit{hierarchical model} 
for RDF data visualization, browsing and analysis. 
(2) It offers \textit{automatic} on-the-fly  hierarchy construction, 
as well as \textit{user-defined} hierarchy construction based on users' preferences.
(3) Provides  \textit{faceted}    browsing and filtering  over  classes and properties.
(4) Integrates \textit{statistics with visualization};
visualizations have been enriched with useful statistics and data information. 
(5) Offers several visualization techniques  (e.g., timeline, chart, treemap).
(6) Provides a large number of dataset's \textit{statistics} regarding the:
\textit{data-level} (e.g., number of sameAs triples),  
\textit{schema-level} (e.g., most common classes/properties), 
and \textit{structure level} (e.g., entities with the larger in-degree).
(7) Provides numerous  \textit{metadata}  related to the dataset: licensing, provenance, linking, availability, undesirability, etc. The latter  can be considered useful for assessing   data quality \cite{ZRMP+13}.

In the rest of this section,
Section~\ref{sec:arch}  describes the system architecture, 
Section~\ref{sec:demo} demonstrates the basic functionality of the SynopsViz. 
Finally, Section~\ref{sec:impl} provides technical information about the implementation.

\subsection{System Architecture}
\label{sec:arch}

The architecture of SynopsViz  is presented in Figure~\ref{fig:arch}.
Our scenario involves three main parts: the Client UI, the  SynopsViz, and the Input data. 
The \textit{Client} part, corresponds to the system's front-end offering several functionalities to the end-users. 
For example, hierarchical visual exploration, facet search, etc. (see Section~\ref{sec:demo} for more details).
SynopsViz  consumes RDF data as \textit{Input data}; optionally, OWL-RDF/S
vocabularies/ontologies describing the input data can be loaded.
Next,  we describe the basic components of the SynopsViz. 

In the preprocessing phase,  the \textit{Data and Schema Handler}
parses the input data and inferes schema information (e.g., properties domain(s)/range(s), 
class/ property  hierarchy, type of instances, type of properties, etc.).
\textit{Facet Generator} generates class and property facets over input data. 
\textit{Statistics Generator} computes several statistics regarding the
schema, instances and graph structure of the input dataset.
\textit{Metadata Extractor} collects dataset metadata.
Note that  the model construction does not require any preprocessing, it is  performed online, according to user interaction.

During runtime the following components are involved.
\textit{Hierarchy Specifier} is responsible for managing the
configuration parameters of our hierarchy model, e.g., the number of
hierarchy levels, the number of nodes per level, and providing this
information to the Hierarchy Constructor. 
\textit{Hierarchy Constructor} implements our tree structure. 
Based on the selected facets, and the hierarchy configuration,  
 it determines the hierarchy of groups and the contained triples. 
\textit{Statistics Processor} computes statistics about the groups included in the hierarchy. 
\textit{Visualization Module} allows the interaction between the user and the back-end, 
allowing several operations (e.g., navigation, filtering, hierarchy specification) over the visualized data.  Finally, the \textit{Hierarchical Model Module}
maintains the in-memory tree structure for our model and communicates with the  {Hierarchy Constructor}  for the model construction, the  {Hierarchy Specifier} for the model customization, the  {Statistics Processor} for the statistics computations, and   the  {Visualization Module} for the visual representation of the model. 

 \begin{figure*}[!th]
 \centering
\includegraphics[width=6in]{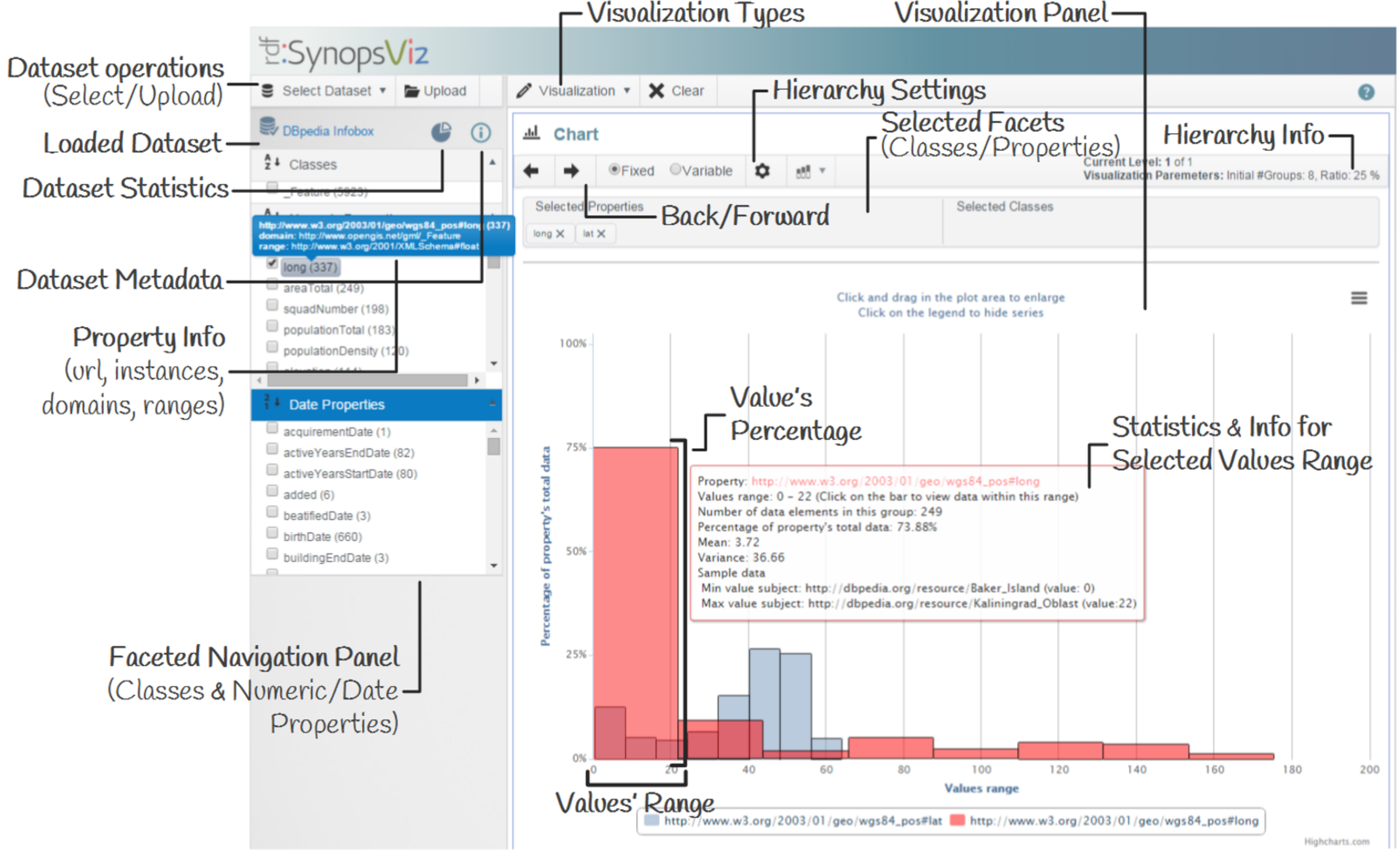}
\caption{Web user interface}
\label{fig:UI}
\end{figure*}  

\subsection{SynopsViz In-Use}
\label{sec:demo}

 In this section we outline the basic functionality of SynopsViz prototype. 
Figure~\ref{fig:UI} presents the web user interface of the main 
window.
SynopsViz  UI consists of the following main panels: 
\textit{Facets panel}: presents and manages facets on classes and properties; 
\textit{Input data control panel}: enables the user to import and manage input datasets; 
\textit{Visualization panel}: is the main area where interactive charts and statistics are presented;
\textit{Configuration  panel}: handles visualization settings.

Initially, users are able to  select a dataset from a number of offered real-word LD datasets 
(e.g., DBpedia,  Eurostat) or upload their own. 
Then, for the selected dataset, the users are able to examine several of the \textit{dataset's  metadata}, 
 and explore several \textit{datasets's statistics}.

Using the \textit{facets panel}, users are able to navigate and \textit{filter} data based on classes, 
numeric and date properties.
In addition, through facets panel several information about the classes and properties 
(e.g.,  number of instances, domain(s), range(s), IRI, etc.) are provided to the users through the UI.

\begin{figure*}[!t]
{\subfloat[Groups of numeric RDF  data (Area chart)] {\includegraphics[trim = 0mm 0mm 51mm 20mm, clip,width=3.05in]{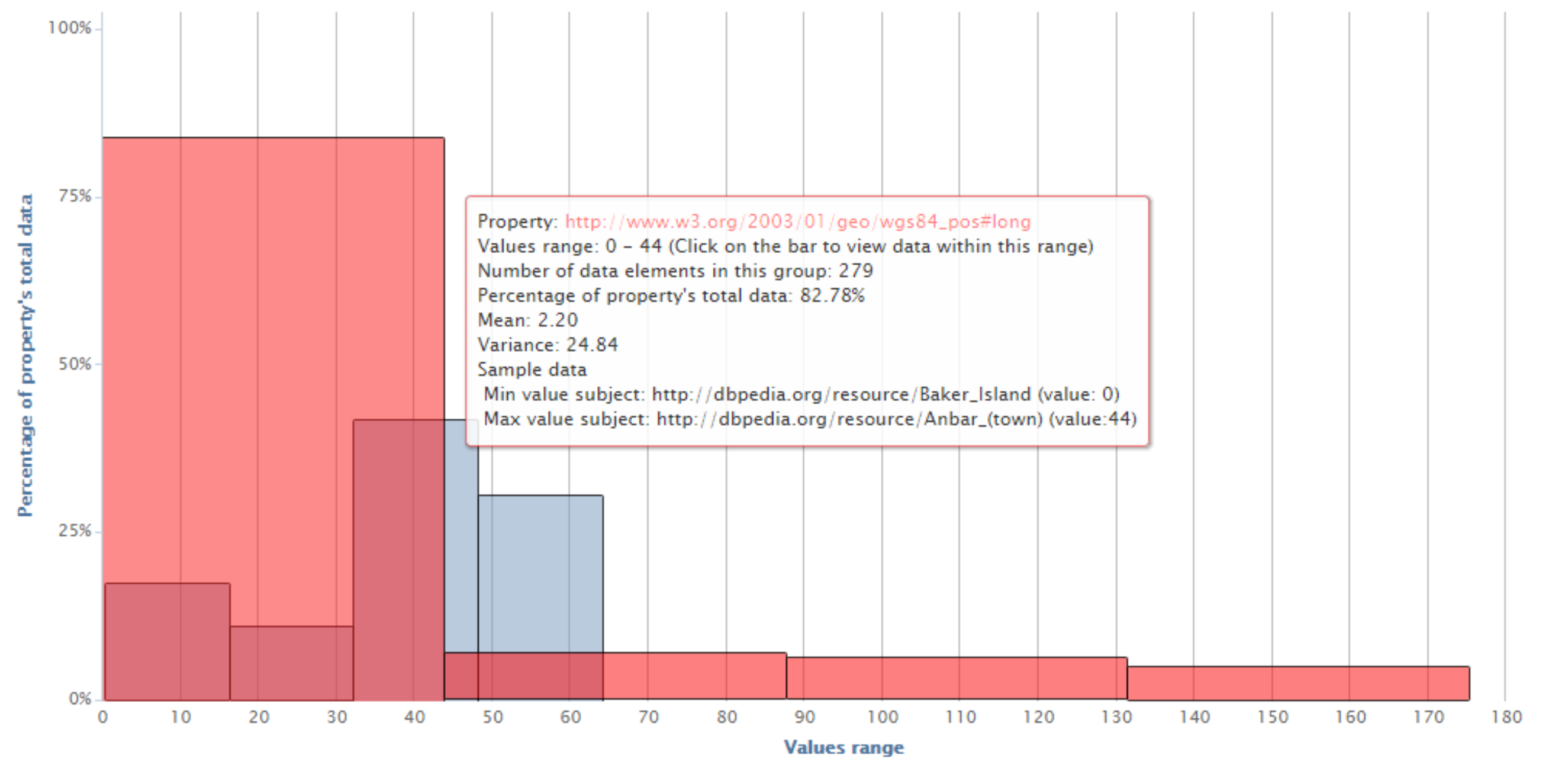}\label{fig:raw1}}}
\hspace{5mm}
{\subfloat[Numeric RDF data  (Column chart)] {\includegraphics[trim = 0mm 0mm 25mm 0mm, clip, width=3.05in]{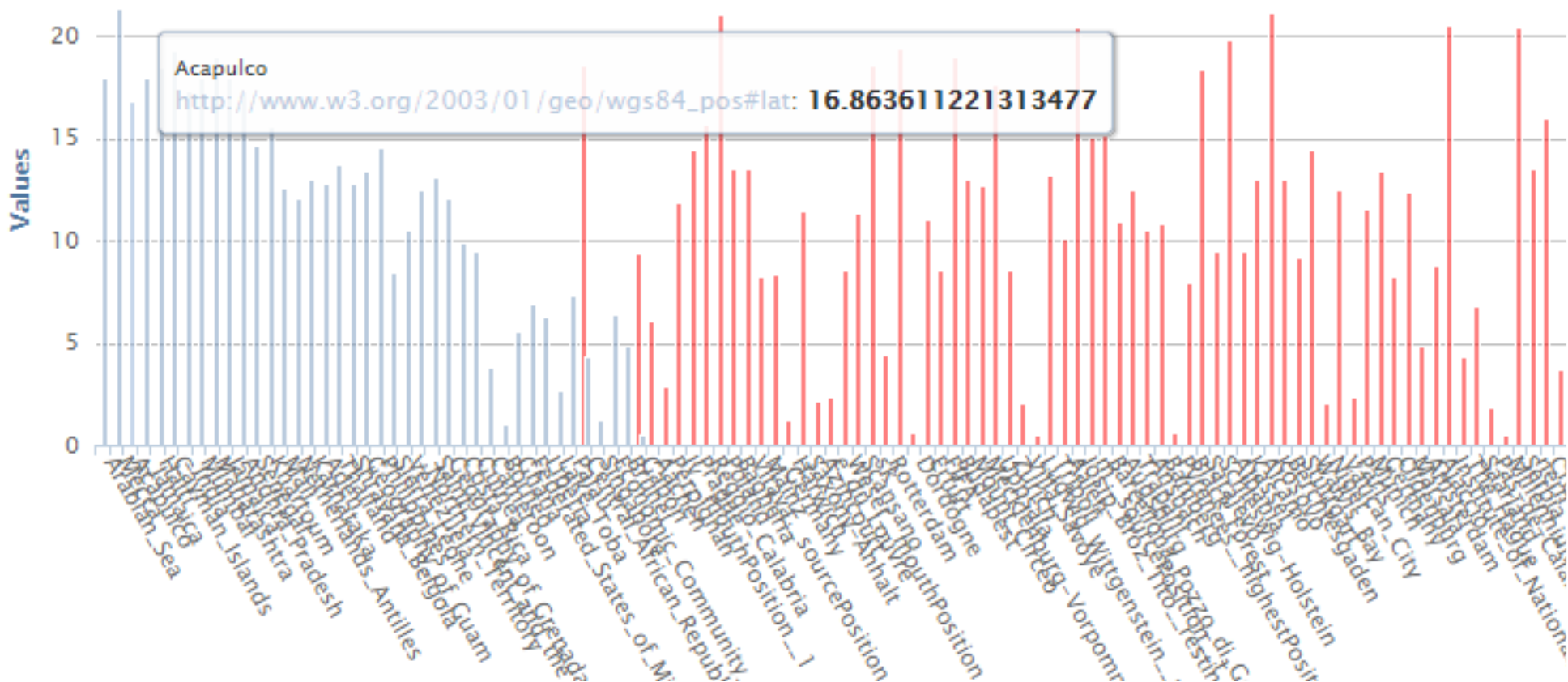}\label{fig:raw2}}}
\\
{
\subfloat[Class hierarchy (Treemap chart)]
{\includegraphics[width=3.2in]{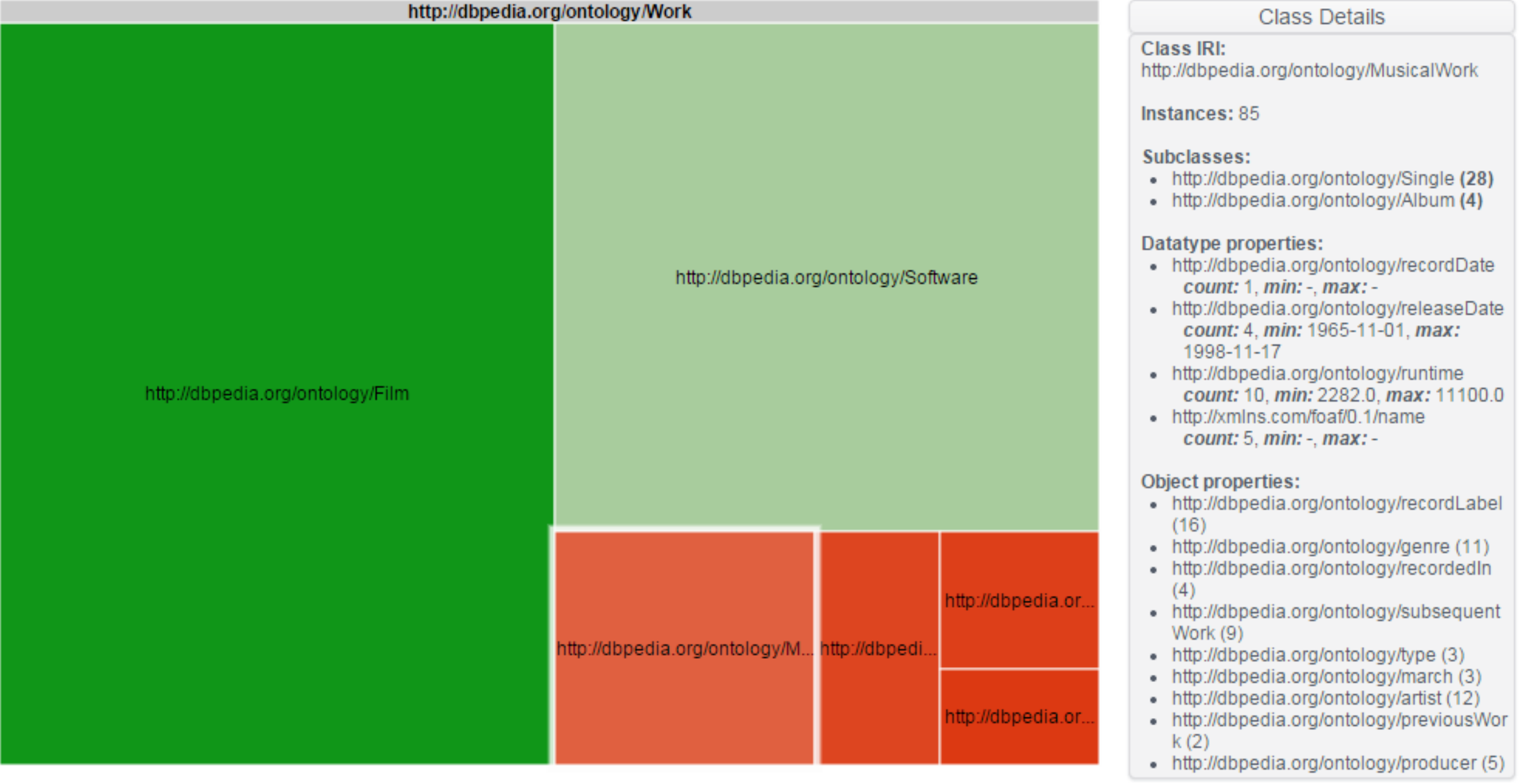} \label{fig:treemap}}
}
\hspace{15mm}
{\subfloat[Class hierarchy (Pie chart)]
{\includegraphics[width=2.0in]{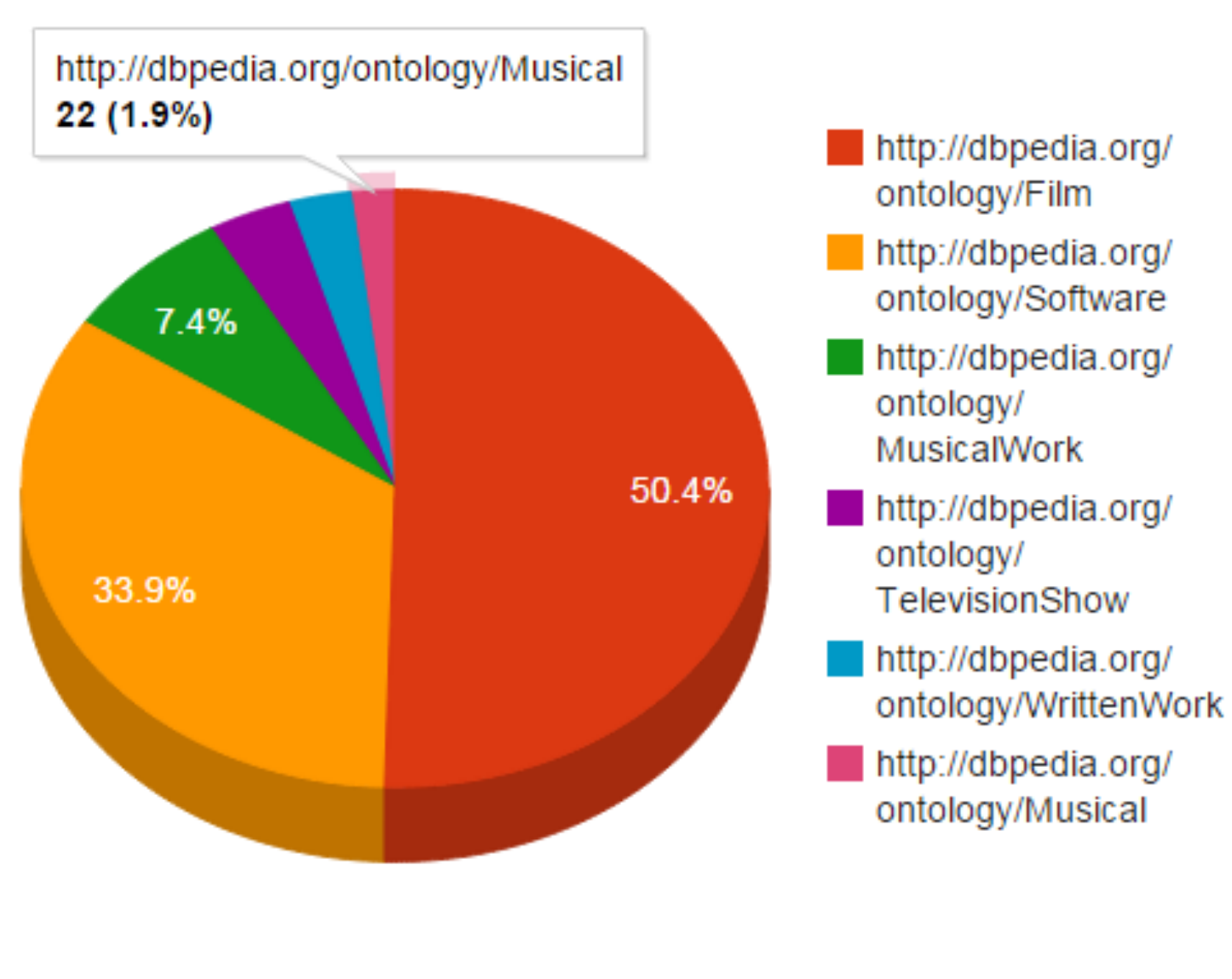}
\label{fig:pie}}
}
\caption{Numeric data  \& class hierarchy visualization examples }
\label{fig:rawViz}
\end{figure*}

Users are able to visually explore data by considering  properties' values. 
Particularly, \textit{area charts} and \textit{timeline-based area charts} are used to 
visualize the resources considering  the user's selected properties. 
Classes' facets can also be used to \textit{filter} the visualized data.
Initially, the top level of the hierarchy is
presented providing an \textit{overview} of the data, organized into top-level groups;
the user can interactively \textit{drill-down} (i.e., zoom-in) and \textit{roll-up} (i.e., zoom-out) over the group of interest, 
up to the actual values of the input data (i.e., LD resources). 
At the same time, statistical information concerning the hierarchy groups as well as their 
contents (e.g., mean value, variance, sample data, range) is presented through the UI (Figure~\ref{fig:raw1}).
Regarding the most detailed level (i.e., LD resources), several visualization types are offered; i.e., area, column, line, spline and areaspline (Figure~\ref{fig:raw2}).


In addition, users are able to visually explore data, through class hierarchy. 
Selecting one or more classes, users can interactively navigate over the class hierarchy using treemaps (Figure~\ref{fig:treemap}) or pie charts (Figure~\ref{fig:pie}). 
Properties' facets can also be used to {filter} the visualized data.
In SynopsViz  the treemap visualization has been enriched with schema and statistical information.
For each class, schema metadata (e.g., number of instances,
 subclasses, datatype/object  properties) and statistical information
  (e.g., the cardinality of each property,  min, max value for datatype properties) are provided.
  
Finally, users can interactively modify the hierarchy specifications. 
Particularly, they are able to increase or decrease the level of abstraction/detail presented, by modifying  both the number of hierarchy levels, and number of nodes per level.

A video presenting the basic functionality of our prototype is available at 
\href{http://youtu.be/n2ctdH5PKA0}{\myFontC{youtu.be/n2ctdH5PKA0}}.
Also, a demonstration of SynopsViz tool is presented in \cite{bsp14}.

\subsection{Implementation}
\label{sec:impl}

SynopsViz  is implemented on top of several open source tools and libraries. 
The back-end of our system is developed in Java, 
Jena framework
is used for RDF data handing and Jena TDB is used for disk-based RDF storing.
The front-end prototype, is developed using HTML and Javascript. 
Regarding visualization libraries, we use
Highcharts,
 for
the area, column, line, spline, areaspline and timeline-based charts and 
Google Charts
for treemap and pie charts.


%
%
%


 \section{Experimental Analysis}
\label{sec:eval}

In this section we present the evaluation   of our approach. 
In Section~\ref{sec:evaldataSetting}, we present the dataset   and the experimental setting. Then, in Section~\ref{sec:effEval} we present the performance results and in Section~\ref{sec:userEval}  the user evaluation we performed.

\subsection{Experimental Setting}
\label{sec:evaldataSetting}
 In our evaluation, we use the well known \textit{DBpedia} 2014
LD dataset.
Particularly, we use the \textit{Mapping-based Properties (cleaned)} dataset%
\footnote{\href{http://downloads.dbpedia.org/2014/en/mappingbased\_properties\_cleaned\_en.nt.bz2}{downloads.dbpedia.org/2014/en/mappingbased\_properties\_ \newline cleaned\_en.nt.bz2}}
 which contains high-quality data, extracted from Wikipedia Infoboxes. 
This  dataset contains $33.1$M   triples and includes a large number of numeric and temporal properties of varying sizes. 
The largest numeric property in this dataset has $534$K triples, whereas the largest temporal property has  $762$K.

Regarding the methods used in our evaluation, 
we consider our \het hierarchical approaches, as well as 
a simple non-hierarchical  visualization approach, referred as \textit{FLAT}.
FLAT is considered as a competitive method against our hierarchical approaches. 
It provides single-level visualizations, rendering only the actual data objects; 
i.e., it is the same as the visualization provided by SynopsViz at the most detailed level.
In more detail, the FLAT approach corresponds to a  column chart in which the resources 
are sorted in ascending  order based on their object values, the horizontal axis contains
the resources' names (i.e., triples' subjects),  and the vertical axis  corresponds to objects' values. 
By hovering over a resource, a tooltip appears including the resource's name and object value. 

Regarding the \het approaches,  the tree parameters (i.e., number of leaves, degree and height) 
are automatically computed following the approach described in Section~\ref{sec:param}. 
In our experiments, the lower and the upper bound 
for the objects rendered at the most detailed level have been set to $\lambda_{min}=10$ and $\lambda_{max}=50$, respectively.
Considering the visualizations provided by the default Highcharts settings, these numbers are reasonable for our screen size and resolution.


Finally, our backend system is hosted on a server with a quad-core CPU at 2GHz
and 8GB of RAM running Windows Server 2008. 
As client, we used a laptop with i5 CPU at 2.5GHz with 4G RAM, 
running Windows 7, Firefox 38.0.1 and  ADSL2+ internet connection.
Additionally, in the user evaluation, the client is employed with a 24" (1920$\times$1200) screen.

\subsection{Performance Evaluation}
\label{sec:effEval}

In this section, we study the performance of the proposed model, as well as the behaviour of our tool, in terms of construction and response time, respectively. 
 Section~\ref{sec:effEvalSetup} describes the setting of our performance evaluation, 
and Section~\ref{sec:effevalres} presents the evaluation results.

 \subsubsection{Setup}
\label{sec:effEvalSetup}

In order to study the performance, a number of numeric and temporal properties from 
the employed dataset are visualized using the two hierarchical approaches
(i.e., \mbox{HETree-C/R}), as well as the FLAT approach. 
We select one set from each type of properties; each set contains $15$ properties with varying sizes, starting from small properties having $50$-$100$ triples up to the largest properties. 

In our experiment, for each of the three approaches, we measure the tool response time.  
Additionally, for the two hierarchical approaches we also measure the time required for the \het construction.

Note that in hierarchical approaches  through user interaction, the server sends to the browser
only the data required for rendering the current visualization level (although the whole tree is constructed at the backend). 
Hence, when a user requests to  generate a visualization we have the following workflow. 
Initially, our system constructs the tree. Then, the data regarding the 
top-level groups (i.e., root node children) are sent to the browser which renders the result. 
Afterwards, based on user interactions (i.e., drill-down, roll-up), the server retrieves 
the required data from the tree and sends it to the browser.
Thus,  the tree is constructed the first time a visualization is requested for the given input dataset; 
for any further user navigation over the hierarchy, the response time does not include the 
construction time. 
Therefore, in our experiments, in the hierarchical approaches, as response time we measure the time required 
by our tool to provide the first response (i.e., render the top-level groups), 
which corresponds to the slower response in our visual exploration scenario. Thus, we consider the following measures in our experiments: 

$Construction$ $Time$: the time required to build the \het structure.
This time includes (1) the time for sorting the triples; 
(2) the time for building the tree; and (3) the time for the statistics computations.
 
$Response$ $Time$:  the time required to render the charts,
starting from the time the client sends the request. 
This time includes
(1) the time required by the server to compute and build the response. In the hierarchical approaches, this time corresponds to the \textit{Construction Time},
plus the time required by the server to build the JSON object sent to the client. 
In the FLAT approach, it corresponds to the time spent in sorting the triples plus the time for 
the JSON construction;
(2) the time spent in the client-sever communication; and 
(3) the time required by the visualization library to render the charts on the browser.

\begin{table*}[!t]
\centering
\caption{Performance Results for Numeric \& Temporal Properties}
\label{tab:effresults}
\setlength{\tabcolsep}{1.9pt}
\scriptsize
\begin{tabular}{lccccccccc}
\tline
 \multicolumn{1}{c }{}& \multicolumn{4}{c}{\textbf{Tree Characteristics}}  &   \multicolumn{2}{c}{\textbf{\hetC}} &\multicolumn{2}{c}{\textbf{\hetR}} &\textbf{FLAT}\\ \cmidrule[0.6pt](lr){2-5} \cmidrule[0.6pt](lr){6-7}  \cmidrule[0.6pt](lr){8-9} \cmidrule[0.6pt](lr){10-10}%
 \multicolumn{1}{l}{\textbf{Property (\#Triples)} }& \textbf{\#Leaves} &	\textbf{Degree}  &\textbf{Height} &	\multicolumn{1}{c}{\textbf{\#Nodes }} & 
	 \specialcell{\textbf{Construction} \\ \textbf{Time} (msec)} &  
 \multicolumn{1}{c}{\specialcell{\textbf{Response} \\ \textbf{Time} (msec)}} & \specialcell{\textbf{Construction} \\ \textbf{Time} (msec)} &  \specialcell{\textbf{Response} \\ \textbf{Time} (msec)}&\specialcell{\textbf{Response} \\ \textbf{Time} (msec)} \\
\dlineB
\rowcolor{gray!25}
 \multicolumn{10}{l}{\textbf{Numeric Properties}} \\
\hspace{1mm}rankingWins $(50)$		&	9	&	3	&	2	&	13	&			5	&	324	&	1	&	323	&	 415 	\\
\rowcolor{gray!10}																						
\hspace{1mm}distanceToBelfast $(104)$		&	9	&	3	&	2	&	13	&			7	&	337	&	4	&	329	&	 419 	\\
\hspace{1mm}waistSize $(241)$		&	16	&	4	&	2	&	21	&			10	&	346	&	9	&	336	&	 440 	\\
\rowcolor{gray!10}																						
\hspace{1mm}fileSize $(492)$		&	27	&	3	&	3	&	40	&			18	&	347	&	16	&	345	&	 575 	\\
\hspace{1mm}hsvCoordinateValue $(995)$		&	81	&	3	&	4	&	121	&			74	&	403	&	50	&	383	&	 980 	\\
\rowcolor{gray!10}																						
\hspace{1mm}lineLength  {$(1,923)$}		&	81	&	3	&	4	&	121	&			77	&	409	&	55	&	391	&	 1,463 	\\
\hspace{1mm}powerOutput $(5,453)$		&	243	&	3	&	5	&	364	&			234	&	560	&	217	&	540	&	 2,583 	\\
\rowcolor{gray!10}																						
\hspace{1mm}width $(11,049)$		&	729	&	3	&	6	&	1,093	&			506	&	830	&	467	&	799	&	 6,135 	\\
\hspace{1mm}numberOfPages  $(21,743)$		&	729	&	3	&	6	&	1,093	&			2,888	&	3,219	&	2,403	&	2,722	&	 12,669 	\\
\rowcolor{gray!10}																						
\hspace{1mm}inseeCode $(36,780)$		&	2,187	&	3	&	7	&	3,280	&			4,632	&	4,962	&	4,105	&	4,436	&	 19,119 	\\
\hspace{1mm}areaWater $(40,564)$		&	2,187	&	3	&	7	&	3,280	&			4,945	&	5,134	&	5,274	&	5,457	&	 29,538 	\\
\rowcolor{gray!10}																						
\hspace{1mm}populationDensity $(52,572)$		&	2,187	&	3	&	7	&	3,280	&			6,803	&	7,127	&	6,080	&	6,404	&	 44,262 	\\
\hspace{1mm}areaTotal $(140,408)$		&	6,561	&	3	&	8	&	9,841	&			16,158	&	16,482	&	13,298	&	13,627	&	 219,018 	\\
\rowcolor{gray!10}																						
\hspace{1mm}populationTotal $(304,522)$		&	19,683	&	3	&	9	&	29,524	&			31,141	&	31,473	&	25,866	&	26,196	&	 1,523,675 	\\
\hspace{1mm}lat $(533,900)$		&	19,683	&	3	&	9	&	29,524	&			73,528	&	73,862	&	71,784	&	72,106	&		\textbf{---}
\vspace{2pt}																																						
\\
\boldline
\rowcolor{gray!25}
 \multicolumn{10}{l}{\textbf{Temporal Properties}} \\
 	\hspace{1mm}retired $(155)$		&	9	&	3	&	2	&	13	&	8	&	330	&	4	&	327	&	425	 \\ 
 	\rowcolor{gray!10}																				
	\hspace{1mm}endDate $(341)$		&	27	&	3	&	3	&	40	&	17	&	339	&	16	&	339	&	 468 	 \\ 
	\hspace{1mm}lastAirDate $(704)$		&	64	&	4	&	3	&	85	&	34	&	359	&	30	&	359	&	 853 	 \\ 
\rowcolor{gray!10}																					
	\hspace{1mm}buildingStartDate $(1,415)$		&	81	&	3	&	4	&	121	&	73	&	406	&	53	&	384	&	 1,103 	 \\ 
	\hspace{1mm}latestReleaseDate $(2,925)$		&	243	&	3	&	5	&	364	&	162	&	496	&	146	&	480	&	 1,804 	 \\ 
\rowcolor{gray!10}																					
	\hspace{1mm}orderDate $(3,788)$		&	243	&	3	&	5	&	364	&	210	&	542	&	195	&	523	&	2,011	 \\ 
	\hspace{1mm}decommissioningDate $(7,082)$		&	243	&	3	&	5	&	364	&	405	&	735	&	383	&	717	&	 3,423 	 \\ 
\rowcolor{gray!10}																					
	\hspace{1mm}shipLaunch $(15,938)$		&	729	&	3	&	6	&	1,093	&	1,772	&	2,094	&	1,595	&	1,919	&	 6,935 	 \\ 
	\hspace{1mm}completionDate $(17,017)$		&	729	&	3	&	6	&	1,093	&	1,987	&	2,311	&	1,793	&	2,121	&	7,814	 \\ 
\rowcolor{gray!10}																					
	\hspace{1mm}foundingDate $(19,694)$		&	729	&	3	&	6	&	1,093	&	2,745	&	3,069	&	2,583	&	2,905	&	 8,699 	 \\ 
	\hspace{1mm}added $(44,227)$		&	2,187	&	3	&	7	&	3,280	&	5,912	&	5,943	&	6,244	&	6,265	&	33,846	 \\ 
\rowcolor{gray!10}																					
	\hspace{1mm}activeYearsStartDate $(98,160)$		&	6,561	&	3	&	8	&	9,841	&	10,368	&	10,702	&	8,952	&	9,282	&	 107,587 	 \\ 
	\hspace{1mm}releaseDate $(169,156)$		&	6,561	&	3	&	8	&	9,841	&	19,122	&	19,451	&	16,526	&	16,856	&	 950,545 	 \\ 
\rowcolor{gray!10}																					
	\hspace{1mm}deathDate $(321,883)$		&	19,683	&	3	&	9	&	29,524	&	32,990	&	33,313	&	27,936	&	28,271	&	\textbf{---}	 \\ 
	\hspace{1mm}birthDate $(761,830)$		&	59,049	&	3	&	10	&	88,573	&	85,797	&	86,120	&	83,982	&	84,314	&		 \textbf{---} \\ 
\bline
\end{tabular}
 \end{table*}

\subsubsection{Results} 
\label{sec:effevalres}

Table~\ref{tab:effresults}   presents the evaluation 
results regarding the numeric (upper half) and the temporal properties (lower half). 
The properties are sorted in ascending order of the number of triples. 
For each property, the table contains the number of triples, the characteristics of the constructed \het structures
(i.e., number of leaves, degree,  height, and number of nodes),  as well as the 
construction and the response time for each approach.
The presented time measurements are the average values from $50$ executions.

Regarding the comparison between the \het and   FLAT, 
the FLAT approach can not provide results for properties having  more than $305$K triples, indicated in the last rows for both numeric and temporal properties 
with "--" in the FLAT response time. 
For the rest properties, we can observe that the \het approaches clearly outperform   FLAT in all cases,
even in the smallest property (i.e., \textit{rankingWin}, $50$ triples). 
As the size of properties increases, the difference between the \het approaches and   FLAT increases, as well. 
In more detail, for large properties having more than $53$K triples 
(i.e., the numeric properties larger  than the \textit{populationDensity} \mbox{-$12$th row-,} 
and the temporal properties larger than the \textit{added} -$11$th row-),
the \het approaches outperform the FLAT by  one order of magnitude.

Regarding the time required for the construction of the \het structure, 
  from Table~\ref{tab:effresults} we can observe the following:
The performance of both HETtree structures is very close for most of the examined properties, 
with the \hetR performing slightly better than the \hetC (especially in the relatively small numeric properties). 
Furthermore, we can observe that the response 
time follows a similar trend as  the construction time.
This is expected since the communication cost, as well as the times required for constructing and rendering the JSON object
are almost the same for all cases. 
 
Regarding the comparison between the construction and the response time 
in the \het approaches, from Table~\ref{tab:effresults} we can observe the following. 
For properties having up to $5.5$K triples 
(i.e., the numeric properties smaller  than the \textit{width} \mbox{-$8$th row-,} 
and the temporal properties smaller than the \textit{decommissioningDate} \mbox{-$7$th row-}),
the response time is dominated
by the communication cost, and the time required for the JSON construction and rendering.
For properties with only a small number of triples
(i.e., \textit{waistSize}, $241$ triples),   
only   $1.5$\% of the response time is  spent on constructing the \het.
Moreover, for a property with a larger number of triples (i.e., \textit{buildingStartData}, $1.415$ triples), $18$\% of the time is spent on constructing the \het.
Finally, for the largest property for which the time spent in 
communication cost, JSON construction and rendering  is larger than 
the construction time   (i.e., \textit{powerOutput}, $5.453$ triples), $42$\% of the time is spent on constructing the \het.

 \begin{figure*}[]
 \centering
 {  
\subfloat[All Properties (50 to  762K triples)]{\hspace{-0.8cm}\includegraphics[width=2.3in]{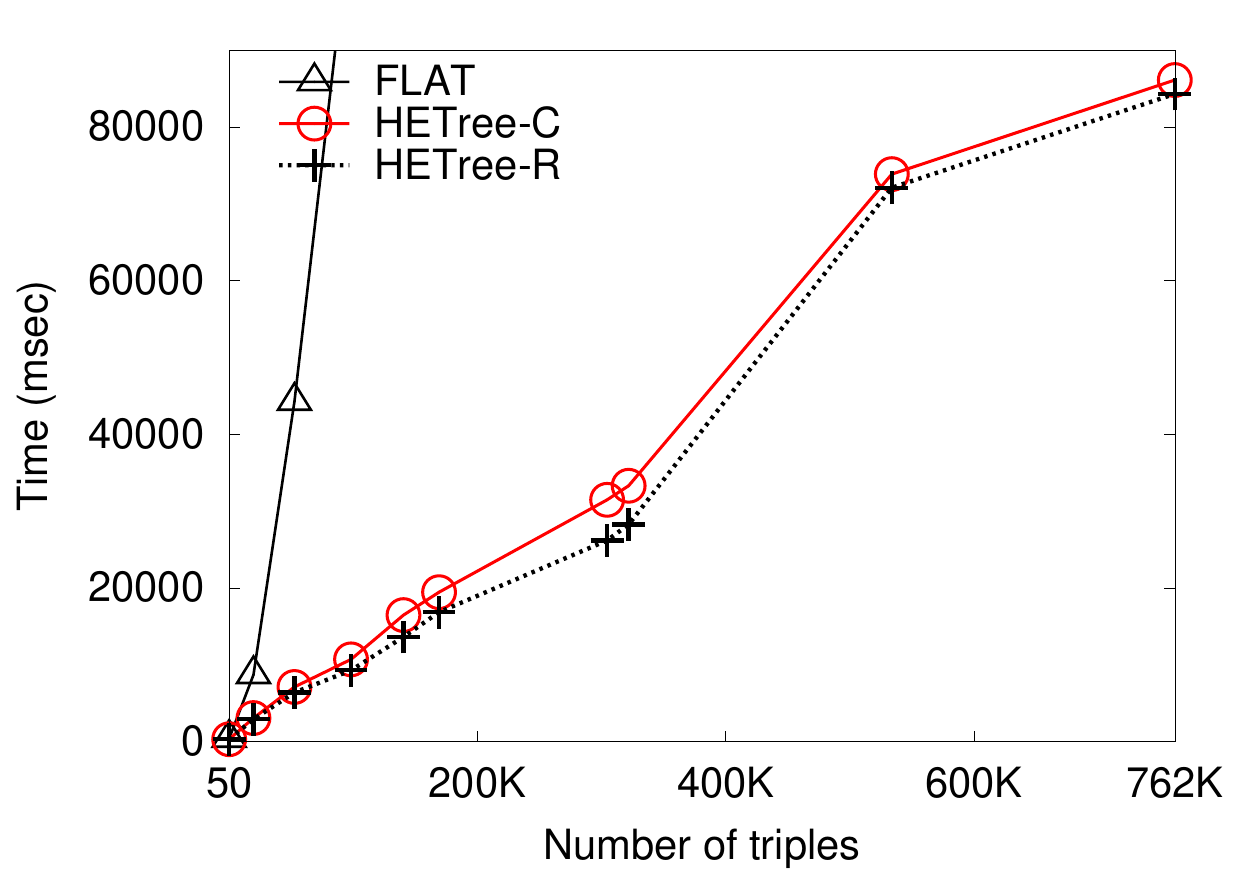}}}
\hspace{1cm}
\subfloat[Small Properties (50 to  20K triples)]{\hspace{-0.5cm}\includegraphics[width=2.3in]{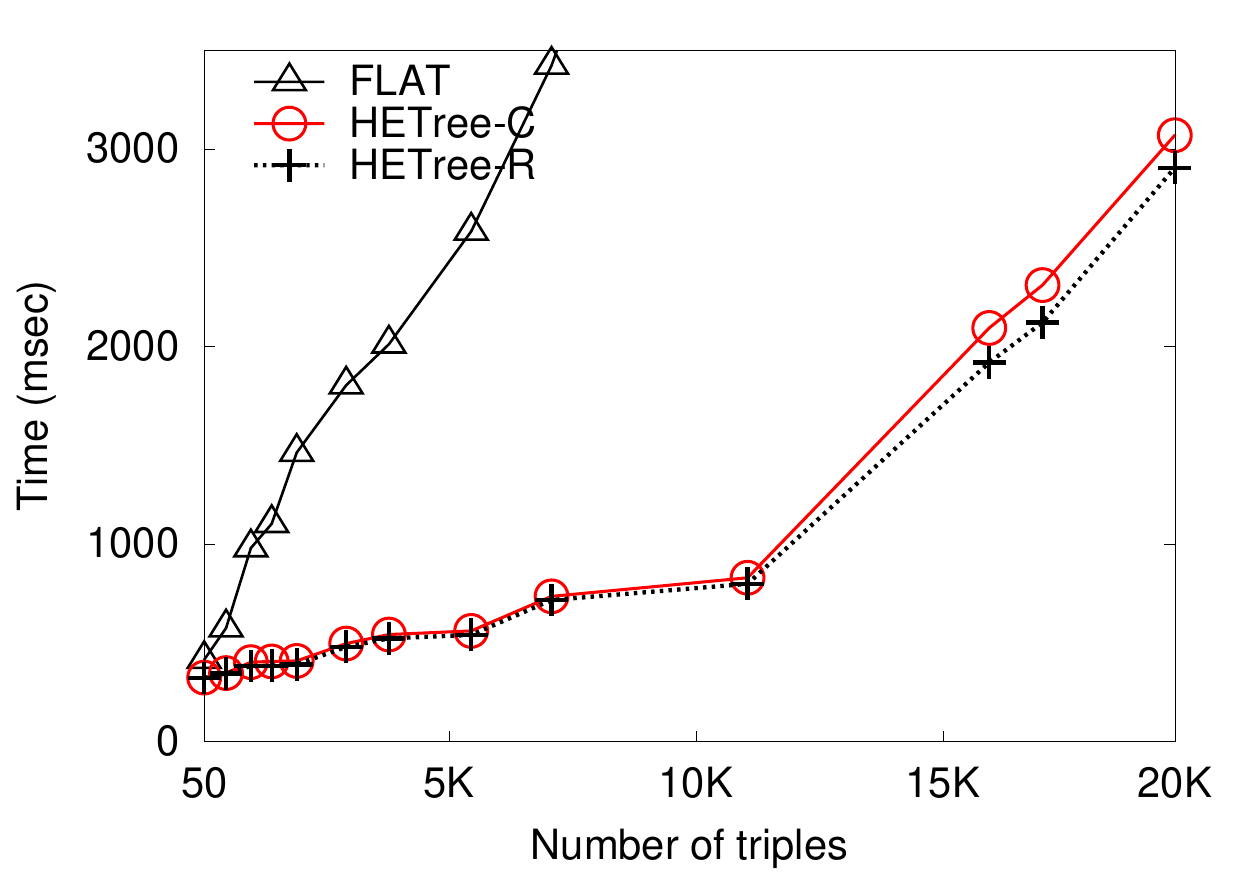}}
\vspace{-1mm}
\caption{Response   Time  w.r.t.\ the number of triples }
\label{fig:resp-time-tr}
\end{figure*}

Figure~\ref{fig:resp-time-tr} summarizes the results from Table~\ref{tab:effresults}, 
presenting the response time for all approaches w.r.t.\ the number of triples.
Particularly, Figure~\ref{fig:resp-time-tr}a includes all properties sizes (i.e., $50$ to $762$K).
Further, in order to have a precise observation over small property sizes (Small properties) in which the difference between the FLAT and the HETree approaches is smaller, 
we report properties with less than $20$K triples separately in 
Figure~\ref{fig:resp-time-tr}b.
Once again, we observe that  \hetR performs slightly better than the \hetC.
Additionally, from Figure~\ref{fig:resp-time-tr}b we can indicate that for up to $10$K triples
the performance of the \het approaches is almost the same. 
We can also observe the significant difference between the FLAT and the \het approaches. 


However our method clearly outperforms the non-hierarchical method, 
as we can observe from the above results, the construction of the whole hierarchy 
can not provide an efficient solution for datasets containing more than $10$K objects. 
As discussed in Section~\ref{sec:incrConstr}, for efficient exploration over large datasets an incremental hierarchy construction is required. 
In   the incremental exploration scenario, the number of hierarchy nodes that have to be processed and constructed is significantly fewer compared to the non-incremental. 

For example, adopting an non-incremental construction in \textit{populationTotal}   ($305$K triples), $29.6$K nodes are to be initially constructed (along with their statistics). 
On the other hand, with the incremental approach (as analysed in Section~\ref{sec:incrConstr})  
at the beginning of each exploration scenario, 
only the \textit{initial nodes} are constructed. 
Initial nodes are the nodes  
initially presented, as well as the nodes potentially
reached by the user's first operation.

In the RES scenario, the initial nodes are the
leaf of interest  ($1$ node) and its sibling leaves (at most $d-1$ nodes). 
In the RAN, the
initial nodes are the nodes of interest (at most $d$ nodes), 
their children (at most $d^2$ nodes),
and their parent node along with its siblings (at most $d$ nodes). 
Finally,
in the BSC scenario the initial nodes are the root node ($1$ node)
and its children (at most $d$ nodes).
Overall, at most
$d$, $2d+d^2$, and $d+1$ nodes are initially constructed in the RES, RAN, and BSC scenarios respectively.
Therefore, in \textit{populationTotal}   case  where $d=3$ at most $3$, $15$  and $4$ nodes are initially constructed  in the RES, RAN, and BSC scenarios respectively.




\subsection{User Study}
\label{sec:userEval}

In this section we present the user evaluation of our tool, where we have employed   three approaches: the two hierarchical and the FLAT.
Section~\ref{sec:userTasks} describes the user tasks, Section~\ref{sec:userProc} outlines the evaluation procedure and setup,  Section~\ref{sec:userRes} summarizes the evaluation results, and Section~\ref{sec:userDis}  discusses issues related to the evaluation  process.

\subsubsection{Tasks}
\label{sec:userTasks}

In this section we describe the different types of tasks that are used in the user evaluation process. 


\stitle{Type 1 [Find resources with specific value]:}
This type of tasks requests the resources having value $v$ (as object). 
For this task type, we define   task T1 by selecting a value $v$ that corresponds to $5$ resources. 
Given this task, the participants are asked to provide the number of resources that pertain to this value. 
In order to solve this task, the participants first have  to find   a resource with   value $v$
and then check which of the nearby resources also have the same value.

%

\stitle{Type 2 [Find resources in a range of values]:}
This type of tasks requests the  resources having value greater than $v_{min}$ and less than $v_{max}$.
We define two tasks of this type, by selecting different combinations of $v_{min}$ and $v_{max}$ values, 
such that tasks which consider different numbers of resources are defined.
We define two tasks, the first task considers a relative small number of resources while the second a larger. In our experiments we select 
$10$ as a small number of resources, while as a large number we select $50$.
Particularly, in the first task, named T2.1, we specify 
the values $v_{min}$ and $v_{max}$ such that a relatively small set of (approximately $10$) resources are included, whereas the second task, T2.2, considers a relatively larger set of (approximately $50$)  resources.
Given these tasks, the participants are asked to provide the number of  resources included in the given range. 
This task can be  solved by first finding a resource with a value included in the given range, 
and then explore  the nearby resources in order to identify the resources in the given range.

\stitle{Type 3 [Compare distributions]:}
This type of tasks requests from the participant to identify whether more resources appear above or below a given value $v$.
For this type, we define  task T3, by selecting the value $v$ near to the median. 
Given this task, the participants are asked to provide the number
of resources appearing either above or below the value $v$.
The answer for this tasks requires from the participants to indicate the value $v$ and 
determine the number or resources appearing either before or after this value.

\begin{table*}[!t]
\hspace{-6mm}
\scriptsize
%
\parbox{.45\linewidth}{
 \caption{Average Task Completion Time (sec)}
\label{tab:userTimes}
\setlength{\tabcolsep}{2.2pt}
\begin{tabular}{l cccc cccc }
\tline
 \multicolumn{1}{c }{}& \multicolumn{4}{c}{\textbf{Small Property}}  &   \multicolumn{4}{c}{\textbf{Large Property}} \\ 
  \cmidrule[0.6pt](lr){2-5} \cmidrule[0.6pt](lr){6-9}
   \multicolumn{1}{l}{ }&  {FLAT} &	 {\hetC}  & {\hetR}    & 	$p$    &   {FLAT} &	 {\hetC}  & {\hetR}&$p$ \\
    \dlineB 
\textbf{T1} &		$54$	&	$29$	&	$28$ &   \mystar\mystar &	$85$ &		$52$	&	$47$	  & \mystar\mystar\\
\textbf{T2.1}&		$63$	&	$57$	&	$64$	& 	\mydiamond 	& $74$	 & 	$60$ &  $69$ & \mystar  \\
\textbf{T2.2}&		$120$	&	$69$	 &	$74$	& \mystar\mystar &	 $128$&$72$	 	& $77$& 			\mystar\mystar\\
\textbf{T3}&			$262$	&		$41$	& $40$&  \mystar\mystar &	{---}	&	 $64$	&   $62$&  ---			\\																			
			\bline
\end{tabular}
 \vspace{-4px}
{\begin{flushleft} 
\hspace{0.3cm} \tiny
 \mystar\mystar~$(p<0.01)$   \hspace{2mm} \mystar~$(p<0.05)$ \hspace{2mm}  \mydiamond~$(p>0.05)$
\end{flushleft}}
}
\hspace{10mm}
\parbox{.45\linewidth}{
 \caption{Error Rate (\%)}
\label{tab:userError}
\setlength{\tabcolsep}{2.2pt}
\begin{tabular}{l cccc cccc }
\tline
 \multicolumn{1}{c }{}& \multicolumn{4}{c}{\textbf{Small Property}}  &   \multicolumn{4}{c}{\textbf{Large Property}} \\ 
  \cmidrule[0.6pt](lr){2-5} \cmidrule[0.6pt](lr){6-9}
   \multicolumn{1}{l}{ }&  {FLAT} &	 {\hetC}  & {\hetR}    & 	$p$    &   {FLAT} &	 {\hetC}  & {\hetR}&$p$ \\
   \dlineB
\textbf{T1}  &  $0$ &$0$ &$0$ & \mydiamond   &$0$ &$0$&$0$ 	& \mydiamond   	\\
 \textbf{T2.1}& $0$ &$0$ &$0$ & \mydiamond   &$0$ &$0$&$0$ 	&  \mydiamond   
		\\
\textbf{T2.2}&	 $20$ 		&$0$ &$0$  & \mydiamond   &$20$ &$0$&$10$ & \mydiamond 	\\
\textbf{T3}&		$70$ &$0$ &$0$			& \mystar\mystar & ---  & $0$ & $0$ & --- \\																					
			\bline
\end{tabular}
 \vspace{-4px}
{\begin{flushleft} 
\hspace{0.3cm} \tiny
 \mystar\mystar~$(p<0.01)$   \hspace{2mm} \mystar~$(p<0.05)$ \hspace{2mm}   \mydiamond~$(p>0.05)$
\end{flushleft}}
}
 \end{table*}

\subsubsection{Setup}
\label{sec:userProc}

In order   to study the effect of the property size in the selected tasks,
we have selected two properties of different sizes from the employed dataset (Section~\ref{sec:evaldataSetting}).
The \textit{hsvCoordinateHue} numeric property containing  $970$ triples, is referred to as $Small$, and the  \textit{maximumElevation}  numeric property, containing $37.936$ triples, is referred to as $Large$. 
The first one corresponds to a  hierarchy of height $4$ and degree $3$, 
and the latter corresponds to a hierarchy of height $7$ and degree $3$.
We should note here that through the user evaluation, the hierarchy parameters were fixed for all the tasks, and the participants were not allowed to modify them, such that the setting has been the same for everyone.

%


In our evaluation, $10$ participants took part.
The participants were computer science graduate students and researchers. 
At the beginning of the evaluation, each participant has introduced to the 
system by an instructor who provided a brief tutorial over the 
features required for the tasks. 
After the instructions, the participants familiarized themselves with the system. 
Note that we have integrated in the SynopsViz the FLAT approach along with the \het approaches. 
 
During the evaluation, each participant performed the previously described four tasks, 
using all approaches (i.e., \mbox{HETree-C/R} and FLAT), over both the small and large properties. 
In order to reduce the  learning effects and fatigue we  defined three groups. 
In the first group, the participants start their tasks with the \hetC approach, 
in the second with \hetR,  and in the third with FLAT. 
Finally, the property (i.e., small, large) first used in each task was counterbalanced among the participants and the tasks. 
The entire evaluation did not exceed   75 minutes.

Furthermore, for each task (e.g., T2.1, T.3), three task instances were specified by slightly modifying the task parameters. 
As a result, given a task, a participant has to solve a different instance of this task, in each approach. 

For example, in task T2.1,  for the \hetR, the selected $v$ corresponds to a solution of 11 resources, 
in \hetC, to 9 resources, whereas for FLAT $v$ corresponded to a solution of 8 resources.
The task instance assigned to each approach varied among the participants.


During the evaluation the instructor measured the time required for each 
participant to complete a task,  as well as the number of incorrect answers. 
Table~\ref{tab:userTimes} presents the average time required for the participants to complete each task. 
The table contains the measurements for all approaches, and for both properties. 
Although we acknowledge that the number of participants in our evaluation is small, we have computed the statistical significance of the results.
Essentially, for each property, the $p$-value of each task is presented in the last column. 
The $p$-value is computed using  one-way repeated measures ANOVA. 
 
In addition, the results regarding the number of tasks that 
were not correctly answered 
 are presented in Table~\ref{tab:userError}. 
 Particularly, the table presents the percentage of incorrect answers for each task and property, referred to as \textit{error rate}.
Additionally, for each task and property, the table includes the $p$-value.
Here, the $p$-value has been computed using  \textit{Fisher's exact test}. 


\subsubsection{Results}
\label{sec:userRes}
 
 
 \stitle{Task T1.}
Regarding the first task, as we can observe from Table~\ref{tab:userTimes},
the \het approaches outperform FLAT, in both property sizes.
Note that the time results on T1  are statistically significant ($p<0.01$).

As expected, all approaches require more time for the Large property compared to the Small one. This overhead in FLAT is caused by   the larger number of resources
that the participants have to scroll over and examine, until they indicate the requested resource's value.
On the other hand, in \het, the overhead is caused by the larger 
number of levels that the Large property hierarchy has. 
Hence, the participants have to perform more drill-down operations and examine more groups of objects, 
until they reach the LD resources. 

We can also observe that in this task, the \hetR performs slightly better than the \hetC in both property sizes. 
This is due to the fact that, in \hetR structure, resources having the same value are always contained in the same leaf. As a result, the participants had to inspect only one leaf. 
On the other hand, in \hetC this does not always hold, hence the participants could have explored more than one leaf.

Finally, as we can observe from Table~\ref{tab:userError}, in all cases 
only correct answers have been provided.
However, none of those results are statistically significant ($p>0.05$).

 \stitle{Task T2.1.}
In the next task, where the participants had to indicate a small set of resources in a range of values, 
the FLAT performance is very close to the \het, especially in the Small property 
(Table~\ref{tab:userTimes}). 
In addition,  we can observe that the \hetC approach performs slightly better than the \hetR.
Finally, regarding the statistical significance of the results, in Small property we have 
$p>0.05$, while in Large we have $p<0.005$.

The poor performance of the \het approaches in this task 
can be explained by the
small set of resources requested and the \het parameters adopted in the user evaluation.
In this setting, the resources contained in the task solution are distributed over more than one leaves. 
Hence, the participants had to perform several roll-up and  drill-down operations in order to find all the resources. 
On the other hand, in FLAT, once the participants had indicated one of the requested resources, it was very easy for them to find out the rest of the solution's resources. 
To sum up, in FLAT,  most of the time is spent on identifying the first of the resources, while in \het the first resource is identified very quickly. 
Regarding the  difference in   performance between the \het approaches we have the following.
In \hetC due to the fixed number of objects in each leaf, the participants had to visit at most one or two leaves in  order to solve this task. 
On the other hand, in \hetR, the number of objects in each leaf is varied, so   most times the participants had to inspect more than two leaves in order to solve the task.  
Finally, also in this case only correct answers were given (Table~\ref{tab:userError}).

 \stitle{Task T2.2.}
In  this task  the participants had to indicate a larger set (compared to the previous task) of resources given a range of values. 
\het approaches noticeably outperform the FLAT approach with  statistical significance ($p<0.01$), while similar results are observed in both properties.

In the FLAT approach a considerable time was spent to identify and navigate over a large number of resources. 
On the other hand,  due to the large number of resources involved in the task's solution, there are  groups in the hierarchy  that explicitly contain resources of solutions (i.e., they do not contain  resources not included in the solution).
As a result, the participants in \het could easily indicate and compute the whole solution 
by combining the information related to the groups (i.e., number of enclosed resources) and individual resources.
Due to the same reasons stated in the previous task (i.e., T2.1), similarly in  T2.2 the \hetC performs slightly better than the \hetR. 
Finally, we can observe  from   Table~\ref{tab:userError} (but without statistical significance), 
that it was more difficult for participants to solve correctly this task with FLAT than with \het. 

 \stitle{Task T3.}
In the last task,   participants were requested to find which of the two ranges contained more resources.  
As   expected,   Table~\ref{tab:userTimes} shows that the \het approaches clearly outperform the FLAT approach with statistical significance in the Small property. This is due to the fact that the participants in FLAT   had to overview and  navigate over almost half of the dataset. 
As a result, apart from the long time required for this process, it was also very difficult to find the 
correct solution. This is also verified by Table~\ref{tab:userError}
on a statistically significant level. 
On the other hand, in the \het approaches, the participants could easily find out the answer by considering the resources enclosed by several groups.

Regarding the Large property, as it is expected, it was impossible for participants to solve this task with FLAT, since this required to parse over and count about 19K resources. 
As a result, none of the participants completed this task using FLAT (indicated with "--" in Table~\ref{tab:userTimes}), 
considering the 5 minute time limit used in this task. 


\subsubsection{Discussion} 
\label{sec:userDis}

The user evaluation showed that the hierarchical 
approaches can be efficient (i.e., require short time in solving tasks) and effective (i.e., have lower error rate) in several cases. 
In more detail,  the \het approaches  performed very well 
on indicating specific values over a dataset, 
and given the appropriate parameter setting
are marginally affected by the dataset size. 
Also note that due to the "vertical-based" exploration, 
the position (e.g., towards the end) of the requested value in the dataset does not affect the efficiency of the approach.
Furthermore, it is shown that the hierarchical approaches can efficiently and effectively handle visual exploration tasks that involve 
large numbers of objects.
 
At the end of the evaluation, the participants gave us valuable feedbacks on possible improvements of our tool. 
Most of the participants  criticized several aspects in the interface,
since our tool  is an early prototype. 
Also, several participants mentioned difficulties in obtaining their 
"position" 
(e.g.,  which is the currently visualized range of values,
or the previously visualized range of values) during the exploration.  
Finally, some participants mentioned that some hierarchies contained more levels than needed. 
As previously mentioned, the adopted parameters are not well suited for the evaluation, 
since hierarchies with a degree  larger than 3 (and as result less levels) are required.


 Finally, additional tasks for demonstrating the capabilities of our model can be considered.
However, most of these tasks were not selected in this evaluation, because it was not possible for the participants to perform them with the FLAT approach. An indicative set includes:
(1) Find  the number of resources (and/or statistics) in the 1st and 3rd quartile;
(2) Find statistics (e.g., mean value, variance) for the top-10 or 50 resources; 
(3) Find the decade (i.e., temporal data) in which   most events take place.

\begin{table*}[!t]
\scriptsize
\centering
\caption{Visualization Systems Overview}
\label{tab:related}
\setlength{\tabcolsep}{3.3pt}
\begin{tabular}{l c c c c c c c c c c }
\tline\vspace{-6pt}
\\
\textbf{System} &
\textbf{WoD} &
\textbf{Hierarchical} &
\textbf{Data Types}$^\star$ &
\textbf{Vis.\ Types}$^{\star\star}$ &
\textbf{Statistics}& 
\textbf{Recomm.\ } &
\textbf{Incr.\ } &
\textbf{Preferences} &
\textbf{Domain} &
\textbf{App.\ Type} \vspace{1pt}\\
\dlineB
\textbf{Rhizomer} \cite{BGG12}& \yes	& \no	& \textsf{N, T, S, H, G}	& C, M, T, TL & \no	&	\yes & \no &\no	&	generic & 	Web\\
 	\rowcolor{gray!10}																				
\textbf{Payola} \cite{KHN13}	 &	\yes &	\no	&	\textsf{N, T, S, H, G} & C, CI, G, M,  T, TL, TR  & \no		& 	\no 	& \no			 & \no &	generic &Web	\\
\textbf{LDVM} \cite{BrunettiAGKN13} 	&	 	\yes & \no		& \textsf{S, H, G} & B, M, T, TR 	&	 \no	&	\yes &	 \no		 & \no & generic	&	Web\\
 	\rowcolor{gray!10}																				
\textbf{Vis Wizard} \cite{TschinkelVMS14}&	\yes 	&\no		&	\textsf{N, T, S}&	B, C, M, P, PC, SG &	 \no	&	\yes & \no		 &\yes	&	generic &		Web\\
\textbf{LDVizWiz} \cite{EURECOM+4380} 	&	\yes 	&	\no	&\textsf{S, H, G} 	&M, P, TR	& \no		&	 \yes & 	\no & \no & generic &		Web\\
 	 	\rowcolor{gray!10}																				
 	\textbf{{LinkDaViz}} \cite{ThellmannGOS15} & \yes & \no &  \textsf{N, T, S} & B, C, S, M, P & \no & \yes & \no & \yes & generic &		Web\\
 \textbf{{VizBoard}} \cite{VoigtSGK12} & \yes & \no & \textsf{N, H} & C,S, T & \no &\yes &\no & \yes & generic &		Web\\	
 	 
 	\rowcolor{gray!10}																				
\textbf{SemLens} \cite{HeimLTE11}&		\yes & 	\no	&	\textsf{N} &	S&	 \no	& 	\no 	& \no		 & \yes	&	generic &		Web\\
\textbf{LODeX} \cite{BenedettiPB14}&		\yes &\no		& \textsf{G}	& G, M, P 	&\yes	& 		\no &	\no & \no &	generic	&	Web\\
 	\rowcolor{gray!10}																				
\textbf{LODWheel} \cite{SDN11}&	\yes 	&\no		& \textsf{N, S, G} &C, G, M, P 	&	 \no	&	\no 	&	\no		 & \no &	generic	&	Web\\
\textbf{RelFinder} \cite{HeimLS10}&	 	\yes &	\no	&	\textsf{G}&	G &	 \no	&	\no 	&\no		 & 	\yes &	generic	&	Web\\
 	\rowcolor{gray!10}																				
\textbf{Fenfire} \cite{HastrupCB08}&		\yes &	\no	&\textsf{G}	&	G& \no		&		\no & \no	& \no &	generic	&	Desktop \\
\textbf{Lodlive} \cite{CamardaMA12}&		\yes &\no		& \textsf{G}	&G	&	 \no	&		\no &\no & \yes &	generic	&Web	\\
	\rowcolor{gray!10}							

\textbf{IsaViz} \cite{pietriga03} & \yes & \no		&  \textsf{G}	& G	&	 \no	&		\no &	\no & \yes &	generic	& Desktop	\\											

\textbf{graphVizdb} \cite{BikakisLKG16,Bikakis15}&		\yes &\no		& \textsf{G}	&G	&	 \no	&		\no &	\no		& \yes &	generic	&Web	\\
 	\rowcolor{gray!10}							

\textbf{ViCoMap} \cite{RistoskiP15} &		\yes &	\no	& \textsf{N, T, S}	&M	&	\yes &\no  &	\no		& \no &	generic	&	Web\\

\textbf{EDT} \cite{MansmannS07}&	\no &\no	&  \textsf{N, T, H}	& C, CM, T, SP 	&	\yes &		\no &	\no		& \no &	OLAP 	&	Desktop\\
 	\rowcolor{gray!10}																				
\textbf{Polaris} \cite{StolteTH02} & 	\no 	&	\no &\textsf{N, T, S, H}	& C, M, S	&\yes	&		\no &\no				& \yes & OLAP &	Desktop\\

\textbf{{XmdvTool}} \cite{Ward94}  & \no &  \yes & \textsf{N} & DS, PC, S, ST & \no & \no & \no		& \no & generic & Desktop \\ 

 	\rowcolor{gray!10}																				
\textbf{GrouseFlocks} \cite{ArchambaultMA08}&		\no &	\yes&		\textsf{G}&	G& \no		&	\no 	& \yes  &\yes	&	generic	& Desktop\\
\textbf{GMine} \cite{RodriguesTPTTF13}&		\no &	\yes &	\textsf{G}&	G&	 \no	&	\no 	&	\no		& \no &	generic	&	Desktop\\

 	\rowcolor{gray!10}																				
\textbf{Gephi} \cite{BastianHJ09}&		\no &	\yes &	 \textsf{G}& G	&\yes	&	 	\no &	\no			& \yes &	generic	&Desktop	\\
\textbf{CGV} \cite{TominskiAS09}&		\no &	 \yes &\textsf{G}	& G	&	 \no	&\no	&	 \yes & \yes 	&	generic	& Desktop	\\
 	\rowcolor{gray!10}																				

\textbf{SynopsViz} & \yes 	&\yes	& \textsf{N, T, H}	&	 C, P, T, TL $^\$$&	\yes &\yes 	& 	\yes & \yes 	&	generic & Web	\\

\bline
\end{tabular}
\vspace{-2px}
{\begin{flushleft} 
\hspace{0.8cm} 
 $^\star$
\textsf{N}: Numeric, 
\textsf{T}: Temporal,
\textsf{S}: Spatial,
\textsf{H}: Hierarchical (tree),
\textsf{G}: Graph (network)
\end{flushleft}}
\vspace{-14px}
{\begin{flushleft} 
\hspace{0.8cm} 
$^{\star\star}$  
B: \textit{bubble chart},  
C: \textit{chart},  
CI: \textit{circles}, 
CM:  \textit{colormap}, 
DS: \textit{dimensional stacking},
G: \textit{graph}, 
M: \textit{map}, 
P: \textit{pie}, 
PC: \textit{parallel coordinates}, 

\hspace{1.8cm} \quad \,\,
S: \textit{scatter}, 
SG: \textit{streamgraph}, 
SP: \textit{solarplot}, 
ST: \textit{star glyphs},
T: \textit{treemap}, 
TL: \textit{timeline}, 
TR: \textit{tree}
\end{flushleft}}
\vspace{-16.5px}
{\begin{flushleft} 
\hspace{0.8cm} 
$^\$$ The \het model is not restricted to these visualization types. 
\end{flushleft}}
 \end{table*}
 
 \section{Related Work} 
 \label{sec:related}

This section reviews works related to our approach on  visualization and exploration in the Web of Data (WoD). Section~\ref{sec:LDV} presents systems and techniques for  WoD visualization and exploration, Section~\ref{sec:statRelated} discusses techniques on WoD statistical analysis, Section~\ref{sec:hRelated} present hierarchical data visualization techniques, and finally, Section~\ref{sec:dsp} discusses works on data structures \&  processing related to our \het data structure.

In Table~\ref{tab:related} we provide an overview and compare several visualization systems that offer similar features to our  {SynopsViz}. 
The \textit{WoD} column indicates systems	 that target the Semantic Web and Linked Data area 
(i.e., RDF, RDF/S, OWL).
The \textit{Hierarchical} column indicates systems that provide 
hierarchical visualization of non-hierarchical data. 
The \textit{Statistics} column captures the provision of statistics about the visualized data. 
The \textit{Recomm}.\ column indicates systems, which offer recommendation mechanisms for visualization settings (e.g., appropriate visualization type, visualization  parameters, etc.).
The \textit{Incr}.\ column indicate systems that provide incremental visualizations. 
Finally, the \textit{Preferences} column captures the ability of the users to 
apply data (e.g., aggregate) or visual (e.g., increase abstraction) operations.

\subsection{Exploration \& Visualization in the Web of Data }
\label{sec:LDV}
A large number of works studying issues related to WoD
visual exploration and analysis have been proposed in the literature
\cite{DR11,bs16,MarieG14a,AlahmariTMW12}.
In what follows, we classify these works into the following categories: 
(1) Generic visualization systems, 
(2) Domain, vocabulary \& device-specific visualization systems, and 
(3) Graph-based visualization systems.
\subsubsection{Generic Visualization Systems} 
 In the context of {WoD visual exploration}, 
there is a large number of  {generic visualization frameworks},
that offer a wide range of visualization types and operations. 
Next, we outline the best known systems in this category.

\textit{Rhizomer} \cite{BGG12} provides WoD exploration based on a overview, zoom and filter workflow.  
Rhizomer  offers various types of visualizations such as maps, timelines, treemaps and charts.
\textit{VizBoard} \cite{VoigtSGK12,vpm13} is an information
visualization workbench for WoD build on top of a mashup
platform. VizBoard presents datasets in a dashboard-like, composite,
and interactive visualization. Additionally, the system
provides visualization recommendations.
\textit{Payola} \cite{KHN13}  is a generic framework for WoD visualization and analysis. 
The framework offers a  variety of domain-specific (e.g., public procurement) analysis
plugins (i.e., analyzers), as well as several visualization techniques (e.g., graphs, tables, etc.).
In addition, Payola offers collaborative features for users to create and share analyzers. 
In Payola the visualizations can be customized according to ontologies used in the resulting data.

The  \textit{Linked Data Visualization Model} (LDVM) \cite{BrunettiAGKN13}
 provides an abstract visualization process for WoD datasets. 
 LDVM enables  the connection of different datasets with various kinds of visualizations in a dynamic way. 
The visualization process follows a four stage workflow:  Source data, 
Analytical abstraction, Visualization abstraction, and View. 
A prototype based on LDVM considers several visualization techniques, e.g., circle, sunburst, treemap, etc. 
Finally, the LDVM has been adopted in several use cases  \cite{KlimekHN15}.
\textit{Vis Wizard} \cite{TschinkelVMS14} is a Web-based visualization system, which exploits 
data semantics to simplify the process of setting up visualizations. 
Vis Wizard is able to analyse 
multiple datasets using brushing and linking methods.
Similarly, 
\textit{Linked Data Visualization Wizard} {(LDVizWiz)} \cite{EURECOM+4380} provides a 
semi-automatic way for the production of possible visualization for WoD datasets. 
In a same context, \textit{LinkDaViz} \cite{ThellmannGOS15} finds the suitable visualizations
for a give part of a dataset. The framework uses heuristic
data analysis and a visualization model in order to facilitate
automatic binding between data and visualization options.

\textit{Balloon Synopsis} \cite{SchlegelWSSGK14}  provides a WoD visualizer based on 
HTML and JavaScript.  It adopts a node-centric visualization approach in a tile design.
Additionally, it supports automatic  information enhancement of the local RDF
data  by accessing either remote SPARQL endpoints or performing 
federated queries over endpoints using the Balloon Fusion service. 
Balloon Synopsis offers customizable filters,  namely    ontology templates,
for the users to handle and transform (e.g., filter, merge) input data. 
\textit{SemLens} \cite{HeimLTE11} is a visual system that combines scatter plots and semantic lenses, offering 
visual discovery of correlations and patterns in data. 
Objects are arranged in  a scatter plot and are analysed using user-defined semantic lenses. 
\textit{LODeX} \cite{BenedettiPB14}
is a system that generates a representative summary of a WoD source. 
The system takes as input a  SPARQL endpoint and generates a visual (graph-based) summary of the WoD source, 
accompanied  by statistical and structural information of the source. 
 \textit{LODWheel} \cite{SDN11}
is a Web-based visualizing system which combines JavaScript libraries 
(e.g., MooWheel, JQPlot) in order to visualize RDF data in charts and graphs.
 \textit{Hide the stack} \cite{DRP11}  
proposes an approach for visualizing WoD  for mainstream end-users.
Underlying Semantic Web technologies (e.g., RDF, SPARQL) are utilized, but are "hidden"  from the end-users. 
Particularly, a template-based visualization approach is adopted, where the information for each resource is presented based on its rdf:type.

\subsubsection{Domain, Vocabulary \& Device-specific Visualization Systems}
In this section, we present systems that target visualization needs for specific types of data and domains, RDF vocabularies or  {devices}. 

Several systems 
focus on visualizing and exploring geo-spatial data.
\textit{Map4rdf} \cite{Map4rdf} is a faceted browsing tool that enables
RDF datasets to be visualized on an OSM or Google Map. 
\textit{Facete}  \cite{StadlerMA14}  is an exploration and visualization system
for SPARQL accessible data, offering faceted  filtering functionalities. 
\textit{SexTant} \cite{BeretaNKKK13} and \textit{Spacetime}  \cite{Valsecchi14} focus on visualizing and  exploring  time-evolving geo-spatial data.  
The \textit{LinkedGeoData Browser}  \cite{StadlerLHA12} is a faceted
browser and editor which is developed in the context of 
LinkedGeoData  project.
Finally, in the same context \textit{DBpedia Atlas} \cite{ValsecchiABTM15} offers exploration over the DBpedia dataset by exploiting the dataset's spatial data. 
Furthermore,  in the context of linked university data,   \textit{VISUalization
Playground} (VISU) \cite{AKSH13} is an interactive tool for specifying and creating
visualizations using the contents of linked university data cloud. 
Particularly, VISU offers a  novel SPARQL interface for creating
data visualizations.
Query results from selected SPARQL endpoints are visualized with Google Charts.
 
A variety of systems target multidimensional WoD modelled with the Data Cube vocabulary. 
\textit{CubeViz} \cite{ErmilovMLA13,SMB+12}  is a faceted browser  for exploring statistical data.
The system provides data visualizations using different types of charts (i.e., line, bar, column, area and pie).
The \textit{Payola Data Cube Vocabulary} \cite{HelmichKN14}
adopts the LDVM stages \cite{BrunettiAGKN13}  in order to visualize RDF data described  by the Data Cube vocabulary. 
The same types of charts as in CubeViz are provided in this system. 
%
The \textit{OpenCube  Toolkit} \cite{Kalampokis14} offers several systems related to statistical WoD.
For example, \textit{OpenCube Browser} explores RDF data cubes by
presenting a two-dimensional table. 
Additionally,  the \textit{OpenCube Map View} offers
interactive map-based visualizations of RDF data cubes based on their geo-spatial dimension.
The \textit{Linked Data Cubes Explorer} (LDCE) \cite{KampgenH14}
allows users to explore and analyse statistical datasets.
Finally, \cite{petrou2014e} offers several map and chart visualizations of 
demographic, social and statistical linked cube data.\footnote{\href{www.linked-statistics.gr}{www.linked-statistics.gr}}


%
Regarding device-specific systems, \textit{DBpedia Mobile} \cite{BeckerB09}    is a location-aware mobile application
for exploring and visualizing DBpedia resources. 
\textit{Who's Who} \cite{CanoDH11}  is an  application  for exploring and visualizing information
focusing on several issues that appear in the mobile environment. 
For example, the application considers the
usability and data processing challenges related to 
the small display size and   limited resources of the mobile devices.

\subsubsection{Graph-based Visualization Systems}

A large number of systems visualize WoD
datasets adopting a \textit{graph-based} (a.k.a., node-link) approach.
 \textit{RelFinder} \cite{HeimLS10} is a Web-based tool that offers  interactive discovery and visualization 
of relationships (i.e., connections) between selected WoD resources. 
\textit{Fenfire} \cite{HastrupCB08} and    \textit{Lodlive} \cite{CamardaMA12}
are exploratory systems that allow users to browse WoD using interactive graphs.
Starting from a given URI, the user can explore WoD by following the links.
\textit{IsaViz} \cite{pietriga03} allows users to zoom and navigate over the RDF graph, 
and also it offers several "edit" operations (e.g., delete/add/rename nodes and edges).
In the same context, \textit{graphVizdb} \cite{BikakisLKG16,Bikakis15}  is built on top of spatial and
database techniques offering interactive visualization over very large (RDF) graphs. 
A different approach has been adopted in \cite{SundaraAKDWCS10},
where  sampling techniques have been exploited. 
 Finally, \textit{ZoomRDF} \cite{ZhangWTY10} employs a space-optimized visualization algorithm
 in order to increase the number of resources which are  displayed.

\subsubsection{Discussion}
 
In contrast to the aforementioned approaches, our work does
not focus solely on proposing techniques for  WoD visualization.
Instead, we introduce a generic model for organizing, exploring and analysing
numeric and temporal data in a multilevel fashion.
The underlying  model is not bound to any specific type of visualization (e.g., chart); 
rather it can be adopted by several "flat" techniques and offer multilevel visualizations over non-hierarchical data.
Also, we present a prototype system that employs
the introduced hierarchical model and offers efficient multilevel visual
exploration over WoD datasets, using charts and timelines.

\subsection{Statistical Analysis in the Web of Data}
\label{sec:statRelated}

A second area related to the analysis features of the proposed model deals with WoD statistical analysis. 
\textit{RDFStats} \cite{LW09} calculates statistical information about RDF datasets. 
\textit{LODstats} \cite{ADML12} is  an extensible framework, offering scalable statistical analysis of WoD datasets.
\textit{RapidMiner LOD Extension} \cite{Ristoski14,Paulheim12}
is an extension of the data mining platform RapidMiner\footnote{\href{https://rapidminer.com}{rapidminer.com}}, offering  
sophisticated data analysis  operations over WoD.
\textit{SparqlR}\footnote{\href{http://cran.r-project.org/web/packages/SPARQL/index.html}
{cran.r-project.org/web/packages/SPARQL/index.html}}
is a package of the R\footnote{\href{http://www.r-project.org}{www.r-project.org}}
statistical analysis platform. SparqlR executes SPARQL queries over SPARQL endpoints
and provides statistical analysis and visualization over   SPARQL results.
Finally, \textit{ViCoMap} \cite{RistoskiP15} combines WoD
statistical analysis and visualization, in a Web-based tool, which offers 
correlation analysis and data visualization on maps.

\subsubsection{Discussion}
In comparison with these systems, our work does not focus on 
new techniques for WoD statistics computation and analysis. 
We are primarily interested on enhancing the visualization and user exploration 
functionality by providing statistical properties of the visualized datasets and objects,
making use of existing computation techniques. 
Also, we demonstrate how in the proposed structure, 
computations can be  efficiently performed on-the-fly and enrich our hierarchical model. 
The presence of statistics provides quantifiable overviews of the underlying WoD resources at each exploration step. 
This is particularly important in several tasks when you have to explore a large number of 
either numeric or temporal data objects. 
Users can examine next levels' characteristics at a glance, this way 
are not enforced to drill down in lower hierarchy levels. 
Finally, the statistics over the different hierarchy levels enables
analysis over different granularity levels.


%
%

\subsection{Hierarchical Visual Exploration}
\label{sec:hRelated}

The wider area of data and information visualization has provided a variety of 
approaches for hierarchical analysis and presentation.

\textit{Treemaps} \cite{Shneiderman92} visualize tree structures using a
space-filling layout algorithm based on recursive subdivision of space. 
Rectangles are used to represent tree nodes, the size of each node is  
proportional to the cumulative size of its descendant nodes. 
Finally, a large number of treemaps  variations have been proposed 
(e.g., Cushion Treemaps, Squarified Treemaps, Ordered Treemaps, etc.).


Moreover, hierarchical visualization techniques have been extensively employed to
visualize very large graphs using the node-link paradigm. 
In these techniques the graph is recursively decomposed into 
smaller sub-graphs that form a hierarchy of abstraction layers.
In most cases, the hierarchy is constructed by exploiting clustering and partitioning methods \cite{AbelloHK06,Auber04,BastianHJ09,RodriguesTPTTF13,TominskiAS09,LBW15}.
In other works, the hierarchy is defined with hub-based \cite{LinCTWKC13}
and density-based \cite{ZinsmaierBDS12} techniques. 
 \textit{GrouseFlocks} \cite{ArchambaultMA08} supports ad-hoc hierarchies which are manually defined by the users.
Finally, there also some edge bundling techniques which  join
graph edges to bundles. The edges are often aggregated  
based on clustering techniques \cite{Gansner2011,Ersoy2011,Phan2005},
a mesh \cite{Lambert2010,Cui2008} or explicitly by a hierarchy \cite{Holten2006}.



In the context of data warehousing and \textit{online analytical processing} (OLAP), 
several approaches provide hierarchical visual exploration, by exploiting the predefined hierarchies in the dimension space. 
\cite{MansmannS07} proposes a class of OLAP-aware hierarchical visual layouts; 
similarly, \cite{TechapichetvanichD05} uses OLAP-based hierarchical stacked bars.
\textit{Polaris} \cite{StolteTH02} offers visual exploratory analysis of data 
warehouses with rich hierarchical structure.

Several hierarchical techniques have been proposed in the context of \textit{ontology visualization  and exploration} \cite{FuN14,DudasZS14,HaagLNE14,LanzenbergerSR09}.
\textit{CropCircles} \cite{WangP06a} adopts a hierarchical geometric containment approach,
representing the class hierarchy as a set of concentric circles.
\textit{Knoocks} \cite{KriglsteinM08} combines containment-based 
and node-link approaches. 
In this work, ontologies are visualized as 
nested blocks where each block is depicted as a
rectangle containing a sub-branch shown as tree map.
A different approach is followed by  \textit{OntoTrix} \cite{BachPL13}
which combine graphs with adjacency matrices.

Finally, in  the context of hierarchical navigation, 
\cite{KashyapHPT11} organizes query results using the MeSH concept hierarchy. 
In \cite{ChakrabartiCH04} a hierarchical structure is dynamically constructed to categorize numeric and categorical query results. Similarly, \cite{ChenL07} constructs personalized hierarchies by considering diverse users preferences.

\subsubsection{Discussion}
In contrast to above approaches that target graph-based or hierarchically-organized data, our work focuses on handling arbitrary numeric and temporal data, 
with out requiring it to be described by an hierarchical schema.
As an example of  hierarchically-organized data, 
consider class hierarchies or multidimensional data organized in multilevel hierarchical dimensions (e.g., in OLAP context, temporal data is hierarchically organized based on years, months, etc.).
In contrast to aforementioned approaches, our work dynamically constructs the hierarchies from raw numeric and temporal data.
Thus the proposed model can be combined with "flat" visualization techniques (e.g., chart, timeline), in order to provide multilevel visualizations over non-hierarchical data. 
In that sense, our approach can be considered more flexible compared to the techniques that rely on predefined hierarchies, as it can enable exploratory functionality on dynamically retrieved datasets, by (incrementally) constructing hierarchies on-the-fly, and allowing users to modify these hierarchies.



\subsection{Data Structures \& Data Processing}
\label{sec:dsp}
 In this section we present the data structures and the data (pre-)processing techniques 
which are the most relevant to our approach. 

\textit{R-Tree} \cite{Guttman84} is disk-based multi-dimensional  
indexing structure, 
which has been widely used in order to efficiently handle spatial queries. 
R-Tree adopts the notion of minimum bounding rectangles (MBRs) 
in order to hierarchical organize multi-dimensional  objects.  

\textit{Data discretization} \cite{GarciaLSLH13,DoughertyKS95}
 is a process where continuous attributes are transformed into discrete. 
 A large number of methods (e.g., supervised, unsupervised, univariate, multivariate)
  for data discretization have been proposed. 
\textit{Binning} is a simple unsupervised discretization method in which a predefined number of bins is created. 
Widely known binning methods are the \textit{equal-width}  and  \textit{equal-frequency}.  
In  equal-width approach,  the range of an attribute is  divided into intervals 
that have equal width and each interval represents a bin. 
In  equal-frequency approach, an equal number of  values are placed in each bin.

By recursively applying discretization techniques,  a hierarchical
discretization of attribute's values can be produced (a.k.a.\ \textit{concept}/\textit{generalization hierarchies}).
In \cite{ShenC08}  a dynamic programming algorithm for generating numeric concept hierarchies is proposed. 
The algorithm attempts to maximize both the similarity between the objects 
stored in the same hierarchy's  node, as well as the dissimilarity between the objects stored in different nodes.
The generated hierarchy is a balanced tree where different nodes may have different number of children. Similarly, 
\cite{HanF94} constructs hierarchies based on data distribution. 
Essentially, both the leaf and the interval nodes are created in such a way that an even distribution is achieved. 
The hierarchy construction considers also a threshold specifying the
maximum number of  distinct values enclosed by nodes in each hierarchy level. 
Finally, binary concept hierarchies (with degree equal to two) are  
generated in \cite{ChuC94}. 
Starting from the whole dataset, 
it performs a recursive binary partitioning over the dataset's values; 
the   recursion is terminated when the number of
distinct values in the resultant partitions is less than a pre-specified threshold.

Using the data objects from our running example (Figure~\ref{fig:data}),
Figure~\ref{fig:hiera} shows the hierarchies generated from the aforementioned approaches.
Figure~\ref{fig:hiera}(a) presents the hierarchy resulting from \cite{ChuC94} and Figure~\ref{fig:hiera}(b) depicts the result using the method from \cite{HanF94}.
The parameters in each method are set, so that the resulting hierarchies   
are as much as possible similar to our hierarchies (Figures~\ref{fig:tree-c}~\&~\ref{fig:tree-r}) .
Hence, the threshold in (a) is set to 3, and in (b) is set to 2.


\subsubsection{Discussion}
The basic concepts of \het structure can be considered similar
to a simplified version  of a static 1D \mbox{R-Tree}.
However, in order to provide efficient query processing in disk-based environment, 
R-Tree considers a large number of I/O-related issues
 (e.g., space coverage, nodes overlaps,  fill guarantees, etc.).
On the other hand, we introduce  a lightweight, main memory structure that 
 efficiently constructed on-the-fly. 
Also, the proposed structure aims at organizing the data in a practical manner for a
(visual) exploration scenario, rather than for disk-based indexing and querying efficiency.

\begin{figure}[t]
 \centering
\includegraphics[scale=0.89]{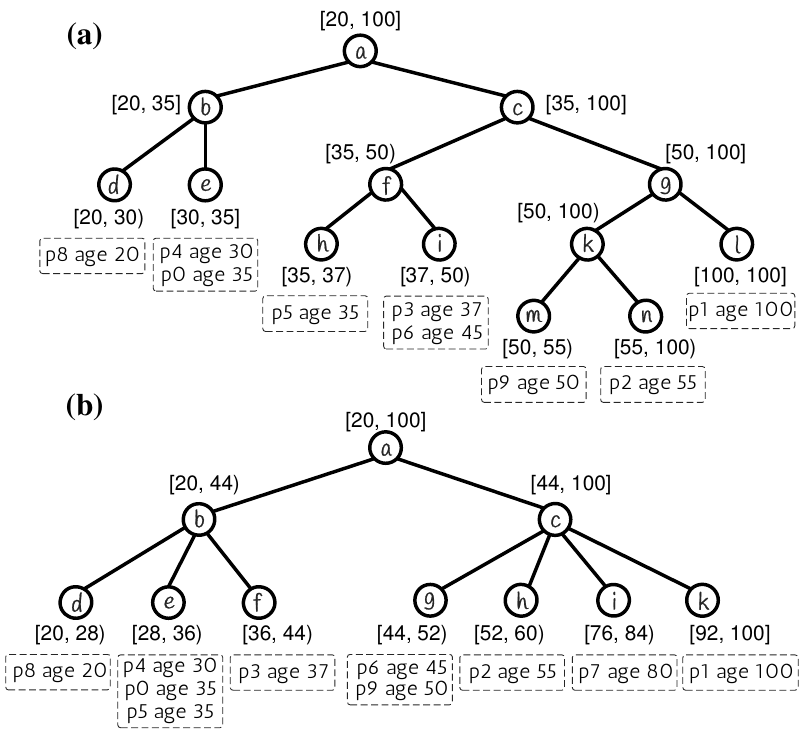}
 \caption{{Hierarchies generated from different approaches.
 a) based on \cite{ChuC94} 
 b) based on \cite{HanF94} 
 }}
\label{fig:hiera}
\end{figure}

Compared to  discretization techniques, our tree model exhibits several similarities, 
namely, the \hetC version can be considered
as a hierarchical version of the equal-frequency binning, and the \hetR of the equal-width binning. 
However, the goal of data organization in \het is to enable visualization 
and hierarchical exploration capabilities over dynamically retrieved non-hierarchical data. 
Hence, compared to the binning methods we can consider the following basic differences.
First, in contrast with binning methods that require from the user to specify some parameters (e.g., the number/size of the bins,  the number of distinct values in each bin, etc); our approach is able to automatically estimate the hierarchy parameters and adjust the visualization results by considering the visualization environment characteristics.
Second, in hierarchical approaches the user is not always allowed to specify the hierarchy characteristics (e.g., degree). 
For example, the hierarchies in  \cite{ChuC94}  have always degree equal to two (Figure~\ref{fig:hiera}(a)), while in \cite{HanF94} 
the nodes have varying degrees (Figure~\ref{fig:hiera}(b)). 
On the other hand, in our approach the hierarchy characteristics can be specified precisely. 
In addition, when not specific hierarchy characteristics are requested, 
our approach generates perfect trees (Section~\ref{sec:param}), 
offering a "uniform" hierarchy structure. 
Third, the computational complexity in some of the hierarchical approaches (e.g., \cite{ShenC08}) is prohibitive (i.e., at least cubic) for using them in practise;
 especially  in settings  where the hierarchies have to constructed on-the-fly. 
Fourth, the proposed tree structure is exploited in order to allow efficient statistics computations over different groups of data; 
then, the statistics are used in order to enhance the overall  exploration functionality.
Finally, the construction of the model is tailored to the user interaction and preferences; our model offers
incremental construction  considering the user interaction, 
as well as efficiently adaptation to the users preferences.

 \section{Conclusions}
 \label{sec:concl}

In this paper we have presented  {\het},  a generic model that
 combines personalized multilevel exploration with online analysis of numeric and temporal data.
Our model is built on top of a  lightweight tree-based structure,
 which can be efficiently constructed on-the-fly for a  given set of data.
We have presented two variations for constructing our model: the  \hetC structure organizes input data into fixed-size groups, whereas the  \hetR structure organizes input data into fixed-range groups. In that way, the users can customize the exploration experience, allowing them to organize data into different ways,  by parameterizing the number of groups, the range and cardinality of their contents, the number of hierarchy levels, and so on. We have also provided a way for efficiently computing statistics over the tree, as well as a method  for automatically deriving from the input dataset the best-fit parameters for the construction of the model. Regarding the performance of multilevel exploration over large datasets,
 our model offers incremental \het construction and prefetching, as well as efficient \het adaptation based on user preferences.
Based on the introduced model, a Web-based  prototype system, called 
SynopsViz, has been developed. 
Finally, the efficiency and the effectiveness of the presented approach are demonstrated via a thorough performance evaluation and an empirical user study.

Some insights for future work include 
the support of  sophisticated methods for data organization 
in order to effectively handle skewed data distributions and outliers. 
Particularly, we are currently working on hybrid \het  versions,  
that integrate concepts from both \hetC and \hetR version. 
For example, a hybrid \hetC considers a threshold regarding the maximum range of a group; 
similarly, a threshold regarding the maximum number of objects in a group is considered in hybrid  \hetR version.
Regarding the SynopsViz tool, we are planning to redesign and extend the graphical user interface, so our tool to be able to use data resulting from SPARQL endpoints, as well as to offer more sophisticated filtering techniques (e.g., SPARQL-enabled browsing over the data). Finally, we are interested in including more visual techniques and libraries.

\vspace{3mm}
\stitle{Acknowledgements.}
The authors would like to thank the editors and the three reviewers for their hard work in reviewing our article, 
their  comments helped us to significant improve our work. 
Further, we  thank  Giorgos Giannopoulos
and Marios Meimaris 
 for many helpful comments on earlier versions of this article.
 This work was partially supported by the
  EU/Greece funded KRIPIS: MEDA Project
 and the EU project "SlideWiki" (688095).

\bibliographystyle{abbrv}
\bibliography{biblio}
\appendix
 
 


\section{Incremental HETree Construction}
\label{ap:interConstr}

\begin{myremark}
\label{re:sibling}
Each time \ico constructs a node (either as part of initial nodes or due to a construction rule), it also constructs all of its sibling nodes.
\end{myremark}

\begin{proof}[\textbf{Proof of Proposition 1}]

Considering the different cases of currently presented HETree elements
and the available exploration operations,  we have the following. 

\vspace{2mm}
$(1)$ \textit{A set of (internal or leaf) sibling nodes $S$ are presented and the user performs a roll-up action.}
Here, the roll-up action will render the parent node of $S$ along with parent's sibling nodes.
In the case that  $S$ are the nodes of interest (RAN scenario), 
the rendered nodes have been constructed in the beginning of the exploration (as part of RAN initial nodes). 
Otherwise, the presented nodes have been previously constructed due to construction $Rule~1(i)$
 (see Section~\ref{sec:incrConstr}). 

\vspace{2mm}
$(2)$ \textit{A set of internal sibling nodes $C$ are presented and the user 
performs a drill-down action over a node $c \in C$.}
In this case, the drill-down will render $c$ child nodes.
If $C$ are the nodes of interest (RAN scenario), 
then the child nodes of $c$ have been constructed at the beginning of the exploration (as part of RAN initial nodes). 
Else, if $C$ is the root node (BSC scenario), 
then again the child nodes of $c$ have been constructed at the beginning of the exploration (as part of BSC initial nodes). 
Otherwise, the children  of $c$, have been constructed before due to construction  $Rule~1(ii)$  (see Section~\ref{sec:incrConstr}). 

\vspace{2mm}
$(3)$ \textit{A set of leaf sibling nodes $L$ are presented and the user 
performs a drill-down action over a leaf $l \in L$.}
In this case the drill-down action will render data objects contained in $l$.
Since a leaf is constructed together with its data objects, 
all data objects here have been previously constructed along with $l$.

\vspace{2mm}
$(4)$ \textit{A set of data objects $O$ are presented and the user performs a roll-up action.}
Here, the roll-up action will render  the leaf that contains $O$ along with the leaf's siblings.
In RAN and BSC exploration scenarios, data objects are reachable only via a drill-down action over the leaf over the leaf that  are contained, whereas in the RES scenario, the data objects, 
contained in the  leaf  of interest, are the first elements that are presented to the user.

In the general case, since $O$ are reached only via a drill-down, their parent leaf has already been constructed. 
Based on Remark~\ref{re:sibling}, all  sibling nodes of this leaf have also been constructed. In the case of the RES scenario, 
where $O$ includes the resource of interest, the  leaf that contains $O$ along with leaf's siblings have been constructed at the beginning of the exploration. 

Thus, it is shown that, in all cases, 
the HETree elements that a user can reach by performing 
one operation, have been previously constructed by \ico.
This concludes the proof of   Proposition~1.
\end{proof}

\vspace{1mm}
\begin{proof}[\textbf{Proof of Theorem 1}]
We will show that, during an exploration scenario, in any exploration step, \ico constructs only the required HETree elements. 
Considering an exploration scenario, \ico constructs nodes only either as initial nodes, or via construction rules. 
The initial nodes are constructed once, at the beginning of the exploration process; based on 
the definition of the initial nodes, these nodes are the required HETree elements for the first user operation. 

During the exploration process,   \ico constructs nodes only via the construction rules.
Essentially, from construction rules,  only the $Rule~1$ construct new nodes. 
Considering the part of the tree rendered when $Rule~1$  (Section~\ref{sec:incrConstr}) is applied, 
it is apparent that the nodes constructed by $Rule~1$ are  only the required HETree elements.

Therefore, it is apparent  that in any exploration step, \ico constructs only the required HETree elements. 
By considering all the steps comprising a user exploration scenario, the overall number of  elements constructed is the minimum. 
This concludes the proof of Theorem~1.

\end{proof}
\vspace{-0mm}

%
%
%
%

\section{\ico Algorithm}
\label{app:icoapp}

The $\mathsf{constrRollUp\text{-}R}$ (\hyperref[proc:proc4]{Procedure~4}) initially
constructs the $cur$ parent node $par$ (\textit{lines~1-7}). 
Next, it  computes the interval $I_{ppar}$ corresponding to $par$ parent node interval (\textit{lines~10-11}). 
Using $I_{ppar}$, it computes the intervals for each of $par$ sibling nodes (\textit{line~13}).
Finally, the computed sibling nodes' intervals $I_{spar}$ are used for the nodes construction (\textit{line~15}).

\begin{procedure}[]
\label{proc:proc4}
\footnotesize
\SetAlgoProcName{\small Procedure 4:}{}
\caption{{constrRollUp}-R($D$, $d$, $cur$,   $H$)}
 \KwIn{$D$:   set of objects; 
$d$: tree degree;  $cur$:   currently presented elements; 
$H$: currently created \hetR}
\KwOut{$H$:  updated \hetR}
 
\vspace{0.5mm}

\Comment{\mycomment{{Computed in $\mathsf{\ico\text{-}R}$:  \hspace{1mm} $len$:   the length of the leaf's interval} }}

\vspace{1mm}

	create an empty  node $par$ \Comment*[r]{\mycomment{$cur$ parent node}}
	$par.h \gets cur[1].h+1$\;
	$par.I^- \gets cur[1].I^- $ \;   
	$par.I^+ \gets cur[|cur|].I^+ $\;

\For( \Comment*[f]{\mycomment{create parent-child relations}}){$i\gets 1$ \KwTo $|cur|$}{
	$par.c[i] \gets cur[i]$\;
	$cur[i].p \gets par$ \;
}

insert $par$ into $H$\;

$l_{p} \gets par.I^+ - par.I^- $
 \Comment*[r]{\mycomment{{$par$ interval length}}}
$I_{ppar}^-  \gets D.minv + d \cdot l_{p} \cdot \floor*{\frac{par.I^- - D.minv}{d  \cdot  l_{p}}}$\;
$I_{ppar}^+ \gets \min(D.maxv, I_{ppar}^-  + d \cdot l_{p} )$\;
\Comment*[r]{\mycomment{compute interval for $par$ parent, $I_{ppar}$}}


$l_{sp} \gets (len \cdot d ^{cur[1].h})$
 \Comment*[r]{\mycomment{{interval length for a $par$ sibling node}}}

$I_{spar} \gets \mathsf{computeSiblingInterv\text{-}R}(I_{ppar}^-, I_{ppar}^+, l_{sp}, d)$\;
   \Comment*[r]{\mycomment{{compute intervals for all $par$ sibling nodes}}}


remove $par.I$ from $I_{spar}$ 
\scalebox{0.92}{\hspace{0.8mm}\Comment*[r]{\mycomment{{remove $par$ interval, $par$ already constructed}}}}

$S \gets \mathsf{constrSiblingNodes\text{-}R}(I_{spar}, \mynull,  D, cur[1].h+1)$ 
 
insert $S$ into $H$\;

\Return $H$\;
\end{procedure}

In the $\mathsf{constrDrillDown\text{-}R}$ (\hyperref[proc:proc5]{Procedure~5}), 
for each node in $cur$, its children are constructed as follows (\textit{line~2}). 
First, the procedure computes the intervals $I_{ch}$ of each child and then it constructs all children (\textit{line~5}).  Finally, the child relations for the parent node $cur[i]$ are constructed (\textit{line~6-7}).

\begin{procedure}[h!]
\label{proc:proc5}
\footnotesize
\SetAlgoProcName{\small Procedure 5:}{}
\caption{{constrDrillDown}-R($D$, $d$, $cur$, $H$)}
 \KwIn{$D$:   set of objects; 
$d$: tree degree;  $cur$:   currently presented elements; 
 $H$: currently created \hetR}
\KwOut{$H$:  updated \hetR}
 
\vspace{0.5mm}

\Comment{\mycomment{{Computed in $\mathsf{\ico\text{-}R}$:  \hspace{1mm} $len$:   the length of the leaf's interval} }}

\vspace{1mm}

$l_c=len \cdot d^{cur[1].h-1}$   \Comment*[r]{\mycomment{{length of the children's intervals}}}

\For{$i\gets 1$ \KwTo $|cur|$}{
\lIf(\hspace{-0.1mm}{\scalebox{0.95}{\Comment*[f]{\mycomment{{nodes previously constructed}}}}}){$cur[i].c[0] =\mynull$}{\mycontinue}
${I_{ch} \gets \mathsf{computeSiblingInterv\text{-}R}(cur[i].I^-, cur[i].I^+, l_c, d)}$
\Comment*[r]{\mycomment{{compute intervals for $cur[i]$ children}}}
\scalebox{0.93}{${S \gets \mathsf{constrSiblingNodes\text{-}R}(I_{ch}, cur[i], cur[i].data, cur[1].h-1)}$}\; 
\vspace*{-8pt}
\For{$k \gets 1$ \KwTo $|S|$}{
$cur[i].c[k] \gets S[k]$\;
}
insert $S$ into $H$\;
}
 \Return $H$\;
\end{procedure}

 \begin{procedure}[]
\label{proc:proc6}
\footnotesize
\SetAlgoProcName{{  \small Procedure 6:}}{}
 \caption{computeSiblingInterv-R($low$, $up$, $len$, $n$)}
 \KwIn{
$low$:  intervals' lower bound;  
$up$:  intervals' upper bound;  
$len$:  intervals'  length; 
$n$:   number of siblings
}
\KwOut{$I$:  an ordered set with at most $n$ equal length intervals} 
\vspace{1mm}
$I_t^-\,, I_t^+ \gets low$\;
\For{$i\gets 1$ \KwTo $n$}{
	$I_t^- \gets I_t^+$ \;
	$I_t^+\gets  \min(up\,,  len +I_t^-)$ \;
		append $I_t$ to $I$\;
		\lIf{$I_t^+ =up$}{ \mybreak }
}
\Return $I$\;
\end{procedure}


 \begin{procedure}[!h]
\label{proc:proc7}
\footnotesize
\SetAlgoProcName{\small  Procedure 7:}{}
 \caption{constrSiblingNodes-R($I$,   $p$,  $A$, $h$ )}
\KwIn{
$I$:  an ordered  set with equal length intervals
$p$:   nodes' parent node;  
$A$: available  data objects; 
$h$: nodes' height}
\KwOut{$S$:  a set of  \hetR sibling nodes} 
\vspace{1mm}
$l = I[1]^+ - I[1]^-$ \Comment*[r]{\mycomment{{intervals' length}}}
$T[ \,\,] \gets \varnothing$\; 
\ForEach(\Comment*[f]{\mycomment{{indicate enclosed data for each node}}})
{$tr \in A$}{
$j \gets	\floor*{ \frac{tr.o - I[1]^-}{l} }+1$\;
\If{$j\geq 0 $ \Logand $ j \leq |I|$}{
 			  	insert object $tr$ into $T[j]$\;
			  	remove object $tr$  from $A$\;
	}
}
\For(\Comment*[f]{\mycomment{{construct  nodes}}}){$i\gets 1$ \KwTo $|I|$}{
	\lIf{$T[i] = \varnothing$}{\mycontinue}
create a new node $n$\;
$n.I^- \gets I[i]^-$\;
$n.I^+ \gets I[i]^+$\;
$n.p \gets p$\;
$n.c \gets \mynull$\;
$n.data \gets T[i]$\;
$n.h \gets h$\; 
\If( \Comment*[f]{\mycomment{{node is a leaf}}}){$h=0$}{	
	 sort $n.data$ based on objects values\;
}
append $n$ to $S$
}
\Return $S$\;
\end{procedure}

\section{Incremental \het Construction Analysis}
\label{app:icocomp}
In this section, we analyse in details the worst case of \ico algorithm, i.e., when the construction cost is maximized.  

\subsection{The \hetR Version}
The worst case in \hetR occurs when  
the whole dataset $D$ is contained in a set
of sibling leaf nodes $L$,  where $|L|\leq d$.

Considering the above setting, in  the RES scenario, 
the cost is maximized when $\mathsf{\ico\text{-}R}$  constructs $L$ (as initial nodes).
In this case,  the  cost is \linebreak $O(|D|+|D|log|D|)=O(|D|log|D|)$.

In a RAN scenario, 
the cost is maximized when the  parent node $p$ of $L$ along with $p$'s sibling nodes are considered as nodes of interest. 
First, let's note that in this case $p$ has no sibling nodes, 
since all the sibling nodes are empty (i.e., they do not enclose data). 
Hence,  the $p$  has to be constructed 
 in $\mathsf{\ico\text{-}R}$ as initial nodes,
 as well as the $L$  in $\mathsf{constrDrillDown\text{-}R}$, and 
 the parent of $p$ in $\mathsf{constrRollUp\text{-}R}$.
The $p$ construction  in $\mathsf{\ico\text{-}R}$   requires $O(|D|)$.
Also, the $L$ construction in  $\mathsf{constrDrillDown\text{-}R}$ requires 
   $O(d+|D|+|D|log|D|+d)$.
Finally, the construction of the parent of $p$  in  $\mathsf{constrRollUp\text{-}R}$ requires 
$O(1)$. 
Therefore, in RAN the overall cost in the worst case is \linebreak
$O(|D|+d+|D|+|D|log|D|+d)=O(|D|log|D|)$.

Finally, in BSC  scenario, the cost is maximized when the 
$L$  have to be constructed by \linebreak  $\mathsf{constrDrillDown\text{-}R}$, 
which requires  \linebreak$O(d+|D|+|D|log|D|+d)=O(|D|log|D|)$.

\subsection{The \hetC Version}
First  let's note that in \hetC version, 
the dataset is sorted at the  beginning of the exploration and the leaves contain equal number of data objects. 
As a result, during a node construction, the data objects, enclosed by it, can be directly identified by computing its position over the dataset and without the need of scanning the dataset or the enclosed data values.
However, in \ico we assume that the node's statistics are computed each time the node is constructed. 
Hence, in each node construction, we scan the data objects that are enclosed by this node.
 In RES scenario, the worst case occurs, when 
the user rolls up for the first time to the nodes at level $2$ (i.e., two levels below the root). 
In this case, \ico has to construct the $d$ nodes at level $1$, as well 
the children for the $d-1$ nodes in level $2$. 
Note that  the construction of the parent of the nodes in level $2$ 
does not require to process any data objects or construct 
children, since these nodes are already constructed.
Now regarding the construction of the rest $d-1$ nodes at level $1$,
\ico will process at most the $\frac{d-1}{d}$ of all data objects.%
\footnote{This number can be easily computed by considering the number of leafs enclosed by these nodes.}
Thus, the  cost for $\mathsf{constrRollUp\text{-}C}$ is $O(d+\frac{d-1}{d}|D|+d-1)$.
Finally, for constructing the child nodes for the $d-1$ nodes in level $2$, 
we are required to process at most the $\frac{d-1}{d^2}$ of all data objects.
Hence, the cost for 
$\mathsf{constrDrillDown\text{-}C}$ is $O(d^2+\frac{d-1}{d^2}|D|+d^2)$.
Therefore, in RES the cost in worst case is 
${O(d+\frac{d-1}{d}|D|+d-1+d^2+\frac{d-1}{d^2}|D|+d^2)}$ $=O(d^2+\frac{d-1}{d}|D|)$.

In RAN scenario, the worst case occurs, when 
the user starts from any set of sibling nodes at level $2$.
Hence, the cost is maximized at the beginning of the exploration.
In this case, \ico has to construct the $d$ initial nodes at level $2$, 
the $d$ nodes at level $1$, and 
the children  for all the $d$ nodes in level $2$. 
First the $d$ initial nodes at level $2$ are constructed by $\mathsf{\ico\text{-}R}$, 
which can be done in $O(d+|D|+d)$. 
Then, the $d$ nodes at level $1$ are constructed by  $\mathsf{constrRollUp\text{-}C}$. 
Similarly as in RES scenario, this can be done in  $O(d+\frac{d-1}{d}|D|+d-1)$.
Finally, the construction of the child nodes for all the $d$ nodes in level $2$ 
requires to process  $\frac{|D|}{d} $ data objects.
Hence, the cost for 
$\mathsf{constrDrillDown\text{-}C}$ is $O(d^2+\frac{|D|}{d}+d^2)$.
Therefore, in RAN the cost, in the worst case, is 
 $O(|D|+d+d+\frac{d-1}{d}|D|+d-1+d^2+\frac{|D|}{d}+d^2)=O(d^2+\frac{d-1}{d}|D|)$.

Finally, in BSC  scenario, 
the worst case occurs, when  the user visits for the first time any of the node at level $1$. 
In this case, \ico has to construct the children   for the $d$ nodes in level $1$.
Hence, 
$\mathsf{constrDrillDown\text{-}C}$ has to process $|D|$ data objects in order to construct the $d^2$ child nodes. 
Therefore, in BSC the cost in worst case is
$O(d^2+|D|+d^2)=O(d^2+|D|)$.


\section{Adaptive \het Construction}
\label{ap:reconstr}

\subsection{Preliminaries}

In order to perform traversal over the levels of the HETree (i.e., level-order traversal), 
we use an array $\H$ of pointers to the ordered set of nodes at each level, with $\H[0]$ referring to the set of leaf nodes 
and $\H[k]$ referring to the set of nodes at height $k$. Moreover, we consider the following simple procedures that are used for the \ada implementation:



\textbf{--} $\mathsf{mergeLeaves}(L, m)$, where $L$ is an 
ordered set of leaf nodes and $m \in \mathbb{N}^+$, with $m > 1$. 
This procedure returns an ordered set of $\ceil*{\frac{L}{m}}$ new leaf nodes, i.e., each new leaf merges $m$ leaf nodes from $L$.
The procedure traverses $L$, constructs a new leaf for every $m$ nodes in $L$ and appends the data items from the $m$ nodes to the new leaf. 
This procedure requires $O(|L|)$.

\textbf{--} $\mathsf{replaceNode}(n_1, n_2)$, replaces the node $n_1$ with the node $n_2$;
it removes $n_1$, and updates the parent of $n_1$ to refer to $n_2$. 
This procedure requires constant time, hence $O(1)$. 

\textbf{--}  $\mathsf{createEdges}(P, C, d)$, where $P$, $C$ are ordered sets of nodes 
and $d$ is the  tree degree.  It creates the edges (i.e., parent-child relations) from the parent nodes $P$ to the child nodes $C$, with degree $d$.
The procedure traverses over  $P$ and connects each node $P[i]$ with the nodes from $C[(i-1)d+1]$ to $C[(i-1)d+d]$.
This procedure requires $O(|C|)$. 

\subsection{The User Modifies the Tree Degree}

\subsubsection{The user increases the tree degree}  

 \vspace{5mm}
 \noindent
 $(1)$ $d' =d^k$, with $k \in \mathbb{N}^+$ and $k>1$

\sstitle{$Tree~Construction.$}
For the $\T'$ construction, we perform a reverse level-order traversal over   $\T$, using the $\H$ vector.
Starting from the leaves ($\H[0]$), we skip (i.e., remove) $k-1$
levels of nodes. Then, for the nodes of the above level ($\H[k]$), 
we create  child relations with the (non-skipped) nodes in the level below. 
The above process continues until we reach the root node of $\T$.

Hence, in this case all nodes in $\T'$ are obtained "directly" from  $\T$. 
Particularly,   $\T'$ is constructed using the root node of $\T$,
 as well  the $\T$  nodes from 
$\H[j \cdot  k], j \in \mathbb{N}^0$. 

The $\T'$ construction requires the execution of \linebreak $\mathsf{createEdges}$ procedure, $j$ times. 
For computing   $j$, we have that $j \cdot k \leq |\H| \Leftrightarrow j \cdot k \leq log_d\ell$. 
Considering that $d'=d^k$, we have that $k=log_dd'$. 
Hence,  $j \cdot log_dd' \leq log_d\ell \Leftrightarrow j \leq log_d(\ell-d')$. 
So, considering that worst case complexity for $\mathsf{createEdges}$ is $O(\ell)$, 
we have that the overall complexity is $O(\ell \cdot log_d(\ell-d'))$.
Since we have that $\ell \leq |D|$,
then in worst case the   $\T'$  can be constructed in $O(|D|  log_d(|D|)) = O(|D|  log_{\sqrt[k]{d'}} (|D|))$.

\sstitle{$Statistics~Computations.$}
In this case  there is no need for computing any new statistics.

 \vspace{5mm}
\noindent $(2)$ $d' = k \cdot d$, with $k  \in \mathbb{N}^+$, $k > 1$ and $k \neq d^\nu$ where $\nu  \in \mathbb{N}^+$

\sstitle{$Tree~Construction.$}
As in $\T'$ the leaves remain the same as in $\T$, 
we only use the  $\mathsf{constrInterNodes}$ (Procedure~2) to build the rest of the tree. 
Therefore, in the worst case, the complexity for constructing the $\T'$ is $O(\frac{d'^2  \cdot \ell -d'}{d'-1})$.

\sstitle{$Statistics~Computations.$}
The statistics for $\T'$ nodes of height 1 can be computed by
aggregating statistics from $\T$. 
Particularity,  in $\T'$ the statistics computations for each internal node of height 1,  
require $O(k)$ instead of $O(d')$, where $k = \frac{d'}{d}$.
Hence, considering that there are  $\ceil*{\frac{\ell}{d'}}$  internal nodes
of height 1 in $\T'$, the cost for their statistics is   
$O(k \cdot \ceil*{\frac{\ell}{d'}})=O(\frac{k\cdot \ell}{d'} +k) $.

Regarding the cost of recomputing them from scratch, consider that there are $\frac{\ell-1}{d'-1}$ internal nodes%
\footnote{Take into account that  the maximum number of internal nodes  (considering all levels)
is  $\frac{d^{\ceil*{log_d\ell}}-1}{d-1}$.}
with heights greater than 1;
the statistics computations  for these nodes require
$O( \frac{d'\cdot \ell-d'}{d'-1})$.
Therefore, the overall cost for statistics computations is  
$O(\frac{k\cdot \ell}{d'} +k+ \frac{d'\cdot \ell-d'}{d'-1})$.

 \vspace{5mm}
\noindent
$(3)$ ${elsewhere}$

\sstitle{$Tree~Construction.$}
Similar to the previous case, the $\T'$ construction requires $O(\frac{d'^2 \cdot \ell-d'}{d'-1})$.

\sstitle{$Statistics~Computations.$}
In this case, the statistics should be computed from scratch for all  internal nodes in $\T'$.
Therefore, the complexity is  $O( \frac{d'^2 \cdot \ell-d'}{d'-1})$.

 \vspace{5mm}

\subsubsection{The user decreases the tree degree}  

 
 \vspace{5mm}
\noindent 
$(1)$ $d' = \sqrt[k]d$, with $k \in \mathbb{N}^+$  and $k>1$
 
\sstitle{$Tree~Construction.$}
For the $\T'$ construction we perform a reverse  level-order traversal over   
$\T$ using the $\H$ vector and starting from the nodes having  height of 1. 
In each level, for each node $n$ we call the  $\mathsf{constrInterNodes}$
(Procedure~2) using as input the $d$ child nodes of $n$ and the new degree $d'$.
Note that, in this reconstruction case, the $\mathsf{constrInterNodes}$ does require to construct the root node; 
the root node here is always corresponding to the node $n$.
Hence, the complexity of $\mathsf{constrInterNodes}$ for one call is $O(d)$.
Considering that, we perform a  procedure call for all the internal nodes, 
as well as that the maximum number of internal nodes  is $\frac{d \cdot\ell-1}{d-1}$, 
we have that,  in the worst case the $\T'$ can be constructed in  
$O(\frac{d^2 \cdot \ell-d}{d-1})=O(\frac{d'^{2k} \cdot \ell -d'^k}{d'^k-1})$.
 
Regarding the number of internal nodes that we have to construct from scratch. 
Since $\T'$ has all the nodes of $\T$, for $\T'$ we have 
to construct from scratch $\frac{d' \cdot \ell-1}{d'-1} - \frac{d \cdot \ell-1}{d-1}$ new internal nodes, 
where the first part corresponds to the number of internal nodes of $\T'$,
and the second part corresponds to $\T$.
Considering that, $d' = \sqrt[k]d$, we have to build 
$\frac{d' \cdot \ell-1}{d'-1} - \frac{d'^k \cdot \ell-1}{d'^k-1}$ 
 internal nodes.
  
\sstitle{$Statistics~Computations.$}
Statistics should be computed only for the new internal nodes of $\T'$. 
Hence,  the cost here is $O(d' \cdot (\frac{d' \cdot \ell-1}{d'-1} - \frac{d'^k \cdot \ell-1}{d'^k-1}))$


 \vspace{5mm}

\subsection{The User Modifies the Number of Leaves}
  \label{app:reconstrLeaves}

\subsubsection{The user decreases the number of leaves} 

 \vspace{5mm}
 \noindent 
$(1)$ $\ell'=  \dfrac{\ell}{d^k} $, with $k \in \mathbb{N}^+$

\sstitle{$Tree~Construction.$}
In this case,
each  leaf in $\T'$ results by merging $d^k$ leaves from $\T$.
Hence, $\T'$ leaves are constructed by calling   $\mathsf{mergeLeaves}(\ell, d^k)$.
So, considering the $\mathsf{mergeLeaves}$ complexity, 
in worst case the new leaves construction  requires $O(|D|)$. 
Then, each  leaf of $\T'$  replace an internal nodes of $\T$ having height of $k$. 
Therefore, in worst case ($k=1$), we  call  $\ceil*{\frac{\ell}{d}}$ times the $\mathsf{replaceNode}$ procedure, which requires $O(\ceil*{\frac{\ell}{d}})$.
Therefore,    the overall cost for constructing $\T'$ in worst case is  
$O(|D| + \ceil*{\frac{\ell}{d}})= O(|D|)$.
Note that in this, as well as in the following case, we assume that $\T$ is a perfect tree.
In case where $\T$ is not perfect,  we can use as $\T$ the perfect tree that initially proposed by our system.


  \vspace{5mm}
\noindent 
$(2)$ $\ell'=  \dfrac{\ell}{k} $, with $k \in \mathbb{N}^+$, $k>1$ and $k \neq d^\nu$, where  $\nu \in \mathbb{N}^+$
 
\sstitle{$Tree~Construction.$}
In this case, each  leaf in $\T'$  results by merging $k$ leaves from $\T$.
Hence, the $\T'$ leaves are constructed by calling the 
$\mathsf{mergeLeaves}(\ell, k)$, which in worst case requires $O(|D|)$. 
Then, the rest of the tree is constructed from   scratch using the  $\mathsf{constrInterNodes}$.
Therefore, the overall cost for $\T'$ construction is 
 $O(|D|+\frac{d^2 \cdot \ell' -d}{d-1})$.


\sstitle{$Statistics~Computations.$}
The statistics for all internal nodes have to be computed from scratch. 
Regarding the leaves, the statistics for each leaf in $\T'$ are computed 
by aggregating the statistics of the $\T$ leaves it includes. 
Essentially, for computing the statistics in each leaf in an \hetC,
we have to process $k$ values instead of ${\frac{|D|}{\ell'}}$.
However, in the worst case (i.e., $\ell=|D|$), we have that $k=\frac{|D|}{\ell'}$. 
Therefore, in the worst case (for both HETree versions) the complexity 
is the same as computing statistics from scratch.


 \vspace{5mm}
\noindent
$(3)$ $\ell'=\ell-k$, with  $k \in \mathbb{N}^+$, $k>1$ and $\ell' \neq \dfrac{\ell}{\nu}$, where $\nu \in  \mathbb{N}^+$
 
\sstitle{$Tree~Construction.$}
In order to construct $\T'$ we have to construct all  nodes 
from scratch, which in the worst case requires $O(|D| log |D|+ \frac{d^2\cdot \ell'-d}{d-1})$.

\sstitle{$Statistics~Computations.$}
The statistics of the $\T$ leaves that are fully contained in $\T'$ can be used for calculating the statistics of the new leaves.
The worst case is when the number of leaves that are fully contained in $\T'$ is minimized.
For \hetC (resp.\ \hetR), this occurs when the size of leaves in $\T'$  is $\lambda'=\lambda+1$ (resp.\ of length $\rho'=\rho+\frac{\rho}{\ell}$).
In this case, for every $\lambda$ (resp.\ $\rho$) leaves from $\T$ that are used to construct the $\T'$ leaves, 
at least $1$  leaf is fully contained. 
Hence, when we process all $\ell$ leaves, at least $\frac{\ell^2}{|D|}$  leaves are fully contained in $\T'$. 

Hence in \hetC, in statistics computations over the leaves, 
instead of processing $|D|$ values, 
we process at most $|D|-\frac{\ell^2}{|D|}\cdot \lambda=|D|-\ell$. 
The same also holds  in \hetR, if we assume normal distribution over the values in $D$.
Therefore,  the cost for computing the leaves statistics is $O(|D|-\ell)=O(|D|-\ell'-k)$.

\clearpage   

 \pagebreak

\end{document}